\documentclass[aps,prx,preprint,onecolumn,citeautoscript,footinbib,
eqsecnum]{revtex4-1}  
\usepackage{amsmath,amssymb,bm,bbm} 
\usepackage{graphicx}  
\usepackage{float}
\usepackage[caption = false]{subfig}
\usepackage{color} 
\usepackage[dvipsnames]{xcolor}
\usepackage[papersize={8.5in,11in}]{geometry}
\usepackage[colorlinks=true]{hyperref}  
\usepackage{ulem}
\usepackage[mathscr]{euscript}
\usepackage[section]{placeins}
\usepackage{comment}
\hypersetup{
    bookmarks=true,         
    unicode=false,          
    pdftoolbar=true,        
    pdfmenubar=true,        
    pdffitwindow=false,     
    pdfstartview={FitH},    
    pdfsubject={},   
    pdfcreator={},   
    pdfproducer={}, 
    pdfkeywords={} {} {}, 
    pdfnewwindow=true,      
    colorlinks=true,       
    linkcolor=magenta, 
    citecolor=blue,        
    filecolor=magenta,      
    urlcolor=blue           
} 

\usepackage{amsfonts}
\usepackage{upgreek}
\usepackage{slashed}
\usepackage{latexsym}

\newcommand{\beq}{\begin{equation}}
\newcommand{\eeq}{\end{equation}}
\def\bea{\begin{eqnarray}}
\def\eea{\end{eqnarray}}
\newcommand{\nn}{\nonumber \\}

\newcommand{\LS}{\Lambda_{S}}

\newcommand{\lam}{\lambda} 
\newcommand{\gc}{g_0} 
\newcommand{\gcr}{g} 
\newcommand{\gam}{\gamma_0} 
\newcommand{\gamr}{\gamma} 
\newcommand{\gv}{v_{0}} 
\newcommand{\gvr}{v} 

\newcommand{\fd}{f^{\dagger}} 
\newcommand{\fa}{f} 
\newcommand{\bd}{b^{\dagger}} 
\newcommand{\ba}{b} 
\newcommand{\dd}{d^{\dagger}} 
\newcommand{\da}{d} 
\newcommand{\pa}{\phi_{a}} 

\newcommand{\zf}{Z_{f}}
\newcommand{\zb}{Z_{b}}

\newcommand{\zd}{Z_{d}}
\newcommand{\zgy}{Z_{g1}}
\newcommand{\zge}{Z_{g2}}
\newcommand{\zp}{Z_{\phi}}
\newcommand{\zgam}{Z_{\gamma}}
\newcommand{\rgs}{\mu}

\newcommand{\up}{^}

\newcommand{\betg}{\beta (g)} 
\newcommand{\betgam}{\beta (\gamma)} 
\newcommand{\betv}{\beta (v)}

\newcommand{\rb}{\bar{r}}
\newcommand{\ep}{\epsilon}
\newcommand{\epp}{\epsilon'}

\newcommand{\lo}{L_{0}}
\newcommand{\lop}{L'_{0}}
\newcommand{\lopp}{L''_{0}}
\newcommand{\lob}{\bar{L}_{0}}
\newcommand{\lobp}{\bar{L}'_{0}}

\newcommand{\Lgam}{L_{\gamma}}
\newcommand{\Lg}{L_{g}}
\newcommand{\Lv}{L_{v}}

\newcommand{\la}{L_{1}}
\newcommand{\lap}{L'_{1}}
\newcommand{\lapp}{L''_{1}}
\newcommand{\lab}{\bar{L}_{1}}
\newcommand{\labp}{\bar{L}'_{1}}

\newcommand{\lb}{L_{2}}
\newcommand{\lbp}{L'_{2}}
\newcommand{\lbpp}{L''_{2}}
\newcommand{\lbb}{\bar{L}_{2}}
\newcommand{\lbbp}{\bar{L}'_{2}}

\newcommand{\lc}{L_{3}}
\newcommand{\lcp}{L'_{3}}
\newcommand{\lcpp}{L''_{3}}
\newcommand{\lcb}{\bar{L}_{3}}
\newcommand{\lcbp}{\bar{L}'_{3}}

\newcommand{\Da}{D_{1\phi}}
\newcommand{\Db}{D_{2\phi}}
\newcommand{\Dc}{D_{3\phi}}
\newcommand{\Dab}{D_{1\zeta}}
\newcommand{\Dbb}{D_{2\zeta}}
\newcommand{\Dcb}{D_{3\zeta}}

\newcommand{\Dap}{D'_{1\psi}}
\newcommand{\Dbp}{D'_{2\psi}}
\newcommand{\Dcp}{D'_{3\psi}}

\newcommand{\Dapp}{D''_{1\psi}}
\newcommand{\Dbpp}{D''_{2\psi}}
\newcommand{\Dcpp}{D''_{3\psi}}

\newcommand{\po}{P_{0}}
\newcommand{\Pgam}{P_{\gamma}}
\newcommand{\Pg}{P_{g}}
\newcommand{\Pv}{P_{v}}

\newcommand{\Pa}{P_{1}}
\newcommand{\pap}{P'_{1}}
\newcommand{\papp}{P''_{1}}
\newcommand{\pab}{\bar{P}_{1}}
\newcommand{\pabp}{\bar{P}'_{1}}

\newcommand{\pb}{P_{2}}
\newcommand{\pbp}{P'_{2}}
\newcommand{\pbpp}{P''_{2}}
\newcommand{\pbb}{\bar{P}_{2}}
\newcommand{\pbbp}{\bar{P}'_{2}}

\newcommand{\pc}{P_{3}}
\newcommand{\pcp}{P'_{3}}
\newcommand{\pcpp}{P''_{3}}
\newcommand{\pcb}{\bar{P}_{3}}
\newcommand{\pcbp}{\bar{P}'_{3}}

\newcommand{\Ro}{R_{0}}
\newcommand{\Rgam}{R_{\gamma}}
\newcommand{\Rg}{R_{g}}
\newcommand{\Rv}{R_{v}}

\newcommand{\Ra}{R_{1}}
\newcommand{\Rap}{R'_{1}}
\newcommand{\Rapp}{R''_{1}}
\newcommand{\Rab}{\bar{R}_{1}}
\newcommand{\Rabp}{\bar{R}'_{1}}

\newcommand{\Rb}{R_{2}}
\newcommand{\Rbp}{R'_{2}}
\newcommand{\Rbpp}{R''_{2}}
\newcommand{\Rbb}{\bar{R}_{2}}
\newcommand{\Rbbp}{\bar{R}'_{2}}

\newcommand{\Rc}{R_{3}}
\newcommand{\Rcp}{R'_{3}}
\newcommand{\Rcpp}{R''_{3}}
\newcommand{\Rcb}{\bar{R}_{3}}
\newcommand{\Rcbp}{\bar{R}'_{3}}

\newcommand{\To}{T_{0}}

\newcommand{\Tv}{T_{v}}

\newcommand{\Ta}{T_{1}}
\newcommand{\Tap}{T'_{1}}
\newcommand{\Tapp}{T''_{1}}
\newcommand{\Tab}{\bar{T}_{1}}
\newcommand{\Tabp}{\bar{T}'_{1}}

\newcommand{\Tb}{T_{2}}
\newcommand{\Tbp}{T'_{2}}
\newcommand{\Tbpp}{T''_{2}}
\newcommand{\Tbb}{\bar{T}_{2}}
\newcommand{\Tbbp}{\bar{T}'_{2}}

\newcommand{\Tc}{T_{3}}
\newcommand{\Tcp}{T'_{3}}
\newcommand{\Tcpp}{T''_{3}}
\newcommand{\Tcb}{\bar{T}_{3}}
\newcommand{\Tcbp}{\bar{T}'_{3}}

\newcommand{\iw}{i\omega}
\newcommand{\inu}{i\nu}

\begin{document}

\preprint{\href{https://arxiv.org/abs/2102.01700}{arXiv:2102.01700}}
\title{Critical anomalous metals near superconductivity\\ in models with random interactions}

\author{Chenyuan Li}
\affiliation{Department of Physics, Harvard University, Cambridge MA 02138, USA}

\author{Darshan G. Joshi}
\affiliation{Department of Physics, Harvard University, Cambridge MA 02138, USA}

\author{Subir Sachdev}
\affiliation{Department of Physics, Harvard University, Cambridge MA 02138, USA}

\date{\today
\\
\vspace{0.4in}}

\begin{abstract}
Anomalous metals are observed in numerous experiments
on disordered two-dimensional systems proximate to superconductivity. A characteristic feature of an anomalous metal is that its low temperature conductivity has a weakly temperature dependent value, significantly higher than that of a disordered Fermi liquid. 
We propose a dynamical mean-field model of an anomalous metal: interacting electrons similar in structure to that of the well-studied universal Hamiltonian of mesoscopic metallic grains, but with independent random interactions between pairs of sites, involving Cooper pair hopping and spin exchange. 
We find evidence for critical anomalous phases or points between a superconducting phase and a disordered Fermi liquid phase in this model. Our results are obtained by a renormalization group analysis in a weak coupling limit, and a complementary solution at large $M$ when the spin symmetry is generalized to USp($M$). The large $M$ limit describes the anomalous metal by fractionalization of the electron into spinons, holons, and doublons, with these partons forming critical non-Fermi liquids in the Sachdev-Ye-Kitaev class. We compute the low temperature conductivity in the large $M$ limit, and find temperature-independent values moderately enhanced from that in the disordered metal.
\end{abstract}

\maketitle

\tableofcontents

\section{Introduction}

The quantum phase transition out of the superconductor in disordered two-dimensional systems has been the focus of intense theoretical and experimental study in the past few decades \cite{KKS19,Kapitulnik20,Chu20,Marcus18,Shahar18,2012Cqpt,Goldman10,Feigelman10,GRFS05,SGCS97,Finkelstein87}.
Depending upon experimental conditions, the non-superconducting phase can either be an insulator or an `anomalous' metal \cite{KKS19}. In an early proposal \cite{Finkelstein87}, it was argued that the breakdown of screening with increasing disorder could lead to a metallic phase with weak interactions and disorder. However, as argued by Kapitulnik {\it et al.\/} \cite{KKS19}, observations show that the metallic phase is anomalous, with a much larger conductivity than in the `normal' metal state at higher temperatures. They dubbed it a `failed superconductor': they surveyed the large body of experimental data, and concluded that no available model of fluctuating superconductivity in a metallic background can consistently explain the existing observations. 

Given this impasse, further advances would be greatly aided by a dynamical mean-field model which displays a quantum phase of matter analogous to an anomalous metal (a `dynamical mean-field' is a theory which is mean-field in space, but includes fluctuations in time). 
In this paper, we propose such a model with a critical metallic state which shares some characteristics with those in the Sachdev-Ye-Kitaev (SYK) models \cite{SY92,kitaev2015talk,SS15,GKST}. We consider a model of electrons with random pair-hopping and spin exchange terms, and find evidence that a critical anomalous metal phase appears either as a critical point or a critical phase between a disordered Fermi liquid and a superconductor. But unlike earlier studies of the superconductor-metal transition \cite{Finkelstein87}, the interactions do not scale to weak-coupling at the critical point, and there are large anomalous exponents. 

The infinite-range interactions (or large spatial dimensionality) in our models imply that it cannot capture aspects of the physics that are associated with low dimensionality. These include weak localization and interaction corrections associated with diffusive electrons which were key ingredients in the theory of Finkel'stein \cite{Finkelstein87}. However, these effects require coherent quasiparticle excitations, which will turn out to be absent in our SYK-like model. Indeed, from our dynamical mean-field model, our proposal is that the anomalous metal is controlled by non-quasiparticle dynamics, and so it is a useful starting point to ignore such effects which have so far been at the forefront in the theory of disordered interacting electrons. Quantum Griffiths effects associated with rare regions will also be absent in our model, but there is little experimental indication that rare regions control the observed anomalous metal. 

The critical exponents of our anomalous metal turn out to have a natural interpretation in a dynamical mean-field theory of fractionalization of the electron into spinons, holons, and doublons. 
Galitski {\it et al.\/} \cite{GRFS05} introduced the idea of fractionalizing electrons and Cooper pairs in a theory of the anomalous metal, and examined a dual theory of fermionic vortices carrying flux $h/(2e)$. It is difficult to make contact with microscopic physics in such a dual formulation, and their theory relies on a conjecture on the vanishing of an average dual field acting on the fermionized vortices. In Section~\ref{sec:largeM}, we will implement fractionalization directly on the electrons, and show that it applies in the large $M$ limit of a model with USp($M$) spin rotation symmetry. Furthermore, in Section~\ref{sec:c_rg} we will employ a renormalization group analysis to obtain very similar critical states without explicit reference to fractionalized degrees of freedom.

We will carry out our analysis on a very general class of electronic Hamiltoinans with both disorder and interations.
Consider a model of electrons ($c$) with on-site interaction $U$ and chemical potential $\mu$, and hopping and interaction terms between the sites:
\bea
H &=& - \mu \sum_{i = 1}^N  c_{i\alpha}^\dagger c_{i \alpha} + U \sum_{i = 1}^N  \left( c_{i\uparrow}^\dagger c_{i\uparrow} - \frac{1}{2} \right) \left( c_{i\downarrow}^\dagger c_{i\downarrow} - \frac{1}{2} \right) + \frac{1}{\sqrt{N}} \sum_{i \neq j=1}^N t_{ij}
 c_{i\alpha}^\dagger c_{j \alpha}^{} \nonumber \\
&~&  + \frac{1}{\sqrt{N}} \sum_{i < j=1}^N \left[ J_{ij} \vec{S}_i \cdot \vec{S}_j + K_{ij} n_i n_j \right] + \frac{1}{\sqrt{N}} \sum_{i \neq j=1}^N L_{ij} \, c_{i \uparrow}^\dagger c_{i \downarrow}^\dagger
 c_{j \downarrow} c_{j \uparrow}
 \,.
\label{eq:Ham}
\eea
(The $1/\sqrt{N}$ normalizations are for future convenience.)
The hopping between the sites is $t_{ij}$, and there are three classes of interaction terms between the sites: a spin-exchange interaction $J_{ij}$ with 
$\vec{S} = c_{\alpha}\up{\dagger} (\vec{\sigma}_{\alpha \beta}/2) c_{\beta}$ 
($\vec{\sigma}$ are the Pauli matrices), 
a Cooper-pair hopping term $L_{ij}$, and
a density-density interaction $K_{ij}$ with $n = c_{\alpha}^\dagger c_{\alpha}$; however, we note that we will only consider the $K_{ij}=0$ case here---see comments below on changes expected for non-zero $K_{ij}$. Such a class of Hamiltonian has been much studied in the context of the `universal Hamiltonian' for quantum dots \cite{KAA00,Alhassid00,Aleiner02}, which take a random set of $t_{ij}$
to obtain the random, extended, single-particle eigenstates of a Fermi liquid description of the quantum dot. Kurland {\it et al.\/}~\cite{KAA00} argued that in such a Fermi liquid phase, the effects of interactions for a sufficiently large and chaotic (in the single particle sense) quantum dot can be accounted for by taking $J_{ij}=J/\sqrt{ N}$, $K_{ij}=K/\sqrt{N}$, $L_{ij}=L/\sqrt{N}$ for some constants $J$, $K$, $L$ for all $i$ and $j$. Then the interaction terms can be written as complete squares, dependent only upon the total spin, number and pairing operator of the quantum dot. 

As in previous work \cite{Cha19,Joshi2019,Tarnopolsky20,Joshi2020,Shackleton20}, our central thesis is that Hamiltonians like (\ref{eq:Ham}) can also exhibit non-Fermi liquid phases or critical points in which the interaction terms cannot be accounted for so simply. To access these non-Fermi liquid states without quasiparticle excitations, we will consider the model in which the $J_{ij}$, $K_{ij}$, and $L_{ij}$ are independent real random numbers for each pair of sites $i$, and $j$, instead of the constant values assumed in theories of quantum dots \cite{KAA00,Aleiner02}.
So the randomness in the interaction terms is treated at an equal footing with the randomness in $t_{ij}$ (which is also an independent real random number), and we can choose
\bea
&& \overline{t_{ij}} = 0 \quad, \quad \overline{J_{ij}} = 0 \quad, \quad
\overline{K_{ij}} = 0 \quad, \quad\overline{L_{ij}} = 0 \nonumber \\
&& \overline{t_{ij}^2} = t^2 \quad,\quad \overline{J_{ij}^2} = J^2 \quad,\quad \overline{K_{ij}^2} = K^2 \quad,\quad \overline{L_{ij}^2} = L^2\,,
\label{averagetJKL}
\eea
and independent of $N$ with the normalizations in (\ref{eq:Ham}).
In previous works, Joshi {\it et al.\/} \cite{Joshi2019,Joshi2020} considered cases with the on-site, non-random interaction term $U \rightarrow \infty$. Then, double occupancy on each site is prohibited, and the $L_{ij}$ term has no effect. In this manner, they obtained a critical metal state with SYK character, and proposed it as a candidate for optimal doping criticality in the hole-doped cuprates. This critical metal appeared between a metallic spin glass phase and a Fermi liquid phase. Numerical evidence for SYK criticality between a metallic spin glass and Fermi liquid has appeared in recent work on a SU(2) model \cite{Shackleton20}. We also note earlier numerical evidence for SYK criticality in a SU(2) model at the metal-insulator transition at half-filling \cite{Cha19}.

In the present paper, we will consider  the case where $U$ is small and can take values of both signs. Then, both the $L_{ij}$ term, and a $U<0$ can favor a superconducting ground state. We find evidence for critical metallic states with SYK character between a superconducting phase and a Fermi liquid phase, and propose it as a candidate theory for the anomalous metal discussed by Kapitulnik {\it et al.} \cite{KKS19}. We will provide both a renormalization group and a large $M$ analysis of this anomalous metal. For simplicity, we will consider the case where the density-density interaction is neglected, $K=0$. We have considered the effects of non-zero $K$ for a related model in another paper \cite{Joshi2020}, and found that it did not change the basic structure of the critical phase, and mainly modified certain critical exponents. We expect a similar consequence of a non-zero $K$ here. Given the significant complexity of our analysis below, we choose to defer a full analysis of the effects of $K$ to later work.

The model which is the focus of this paper will be introduced in Section~\ref{sec:model}. As in earlier work \cite{GGS20,GeorgesRMP,Shackleton20}, we can also consider models in the limit of large spatial dimension with nearest-neighbor non-random hopping and nearest-neighbor random exchange; this leads to the same saddle-point dynamical mean-field equations, and also allows definition of the conductivity and other transport observables. 

In Section~\ref{sec:c_rg} we present a renormalization group (RG) analysis of the model of Section~\ref{sec:model} for the case at half-filling ($\mu=0$). This analysis does not require a large $M$ limit, and we retain the SU(2) spin symmetry. We use the mapping of the model of Section~\ref{sec:model} to an impurity model with a self-consistent bath. We assume that the bath correlators have a power-law decay and, following Ref.~\onlinecite{Joshi2019}, show that for suitable exponents a RG analysis is possible, analogous to the field-theoretic $\epsilon$ expansion. This RG analysis ignores the self-consistency condition on the bath; but the self-consistency is imposed {\it a posteriori\/} on the exponents of the correlators, although not the amplitudes. Our $\epsilon$ expansion is carried out at $U=0$, and finds fixed points at which $U$ is a strongly relevant perturbation about $U=0$. We can therefore identify the fixed points with critical points between a superconductor and a disordered Fermi liquid. We note that the RG analysis in Section~\ref{sec:c_rg} is carried out entirely with the electron operator and its composites, and there is no explicit reference to fractionalized degrees of freedom.
Appendices~\ref{app:fbd_rg} and~\ref{app:gi_rg} perform the RG analysis using fractionalization of the electron into spinons, holons, and doublons. The local constraint on the fractionalized fields is imposed exactly in this RG, and precisely the same RG results are obtained in these appendices as those in Section~\ref{sec:c_rg}. This acts as a strong check on our RG computations, and shows explicitly that the use of fractionalized variables is optional when the gauge constraint is imposed exactly.

Section~\ref{sec:largeM} turns to a solvable large $M$ limit of the model at general $\mu$, obtained by generalizing the spin symmetry from SU(2) to USp($M$). For our large $M$ limit, we will have to explicitly fractionalize the electron into spinons, holons an doublons. In the body of the paper we present the formulation where we make the spinons fermionic, and the holons and doublons bosonic; Appendix~\ref{app:bos_sp} considers the alternative formulation where the spinons and bosonic, and the holons and doublons are fermionic, and obtains similar results.
The large $M$ limit admits intermediate critical points or phases at variable density between the metal and the superconductor whose structure we will describe in some detail. In particular, we will show that the conductivities of these intermediate critical phases can be larger than that of the disordered Fermi liquid.  
The large $M$ solutions can be viewed as relatives of the critical points found in the RG analysis of Section~\ref{sec:c_rg}. This conclusion is supported by the agreement between the exponents of the electron, spin, and Cooper pair operators between the large $M$ computations of Section~\ref{sec:largeM}, and the RG computations of Section~\ref{sec:c_rg}, as we will review in Section~\ref{sec:conc}.


\section{Model}
\label{sec:model}

We will consider the model with all-to-all and random electron hopping, spin exchange, and Cooper-pair hopping: this is the Hamiltonian in (\ref{eq:Ham}), with $K_{ij} = 0$. Our analysis can be extended to non-zero $K_{ij}$ without significant difficulty \cite{Joshi2020}, but we choose not to do so here in the interests of simplicity; we do not expect significant changes to be induced by the $K_{ij}$. The random parameters in the Hamiltonian obey (\ref{averagetJKL}).
We can also consider a model without time-reversal symmetry with $t_{ij}$ and $L_{ij}$ complex, but the results in the non-superconducting state will remain unchanged in the all-to-all interaction limit.

Let us first discuss the possible phases of the model. For small $U>0$ the electrons can freely hop around to form a metal. Depending on the exchange interaction we may have a disordered Fermi liquid or a metallic spin glass \cite{Tarnopolsky20,Shackleton20}. For large $U>0$, doubly-occupancy is energetically costly thus restricting the electron movement to form a Mott insulator. 
This is likely to be a spin glass, although in the large $M$ limit (where $M$ corresponds to SU($M$) generalization of SU(2)) one obtains the SY state \cite{SY92}. This metal-insulator quantum phase transition at finite $U>0$ in this model has been recently studied both analytically \cite{Tarnopolsky20} and numerically \cite{Cha19}, with an evidence for a SYK-like criticality.  
On the other hand, for large $U<0$, one expects to obtain a superconductor. Thus near $U=0$ we expect a metal-superconductor quantum phase transition. This phase transition and its vicinity will be discussed in detail here.

We note that we could equally consider a model on large-dimensional lattice with non-random hopping, and nearest-neighbor random exchange. Such a model leads to the same large volume limit, with the same on-site dynamical mean-field equations \cite{GGS20,GeorgesRMP,Shackleton20}. Such a limit allows definition of transport quantities, and we will need it here to define the appropriate correlator needed to define the conductivity.

The Hilbert space of (\ref{eq:Ham}) on each site has four states: empty ($|0\rangle$), singly occupied ($\left|\uparrow \right\rangle$ or $\left|\downarrow \right\rangle$) and doubly occupied ($\left|\uparrow\downarrow \right\rangle$). As we are working at small $U$, there is no constraint we need to impose and we can perform computations directly using the electron operator $c_\alpha$. This will be our strategy in the Section~\ref{sec:c_rg}. We will introduce a fractionalized representation later in Section~\ref{sec:largeM} when we study a large $M$ limit.

\subsection{Large-volume limit}
\label{sec:largeN}

Models with random interactions often offer the possibility of simplification in the large-volume limit, i.e., $N \rightarrow \infty$. We formally introduce replica indices for the fields in the path integrals and perform a disorder average over $t_{ij}$, 
$J_{ij}$, and $L_{ij}$. Consequently we arrive at an effective single-site action. Since we are interested in the critical point the replica indices do not play a significant role. These are important when studying a spin-glass phase. The entire procedure follows the methodology developed in Refs. \cite{SY92,GPS00,GPS01}. In the end when $N \rightarrow \infty$ and dropping the replica indices for simplicity, we arrive at the effective single-site action below. (As we noted above, the same single-site action can also be obtained in the limit of large dimensionality.) 
\begin{align}
\label{eq:Zpath}
\mathcal{Z} &= \int \mathcal{D} c_{\alpha}(\tau)  e\up{-\mathcal{S}_{0} - \mathcal{S}_{1}} \,, \\    
\label{eq:SB}
\mathcal{S}_{0} &= \int d\tau \bigg[
c_{\alpha}\up{\dagger}(\tau) \frac{\partial}{\partial \tau} c_{\alpha}(\tau) - \mu c_{\alpha}\up{\dagger}(\tau)c_{\alpha}(\tau) + U c_{\uparrow}\up{\dagger}(\tau)c_{\downarrow}\up{\dagger}(\tau)c_{\downarrow}(\tau)c_{\uparrow}(\tau)
\bigg] \,, \\
\label{eq:SH}
\mathcal{S}_{1} &= \int d\tau d\tau' \bigg[
 t\up{2}  R(\tau - \tau') c_{\alpha}\up{\dagger}(\tau) c_{\alpha} (\tau') - \frac{J\up{2}}{2} Q(\tau - \tau') \vec{S}(\tau) \cdot \vec{S}(\tau') \nonumber \\
&~~~~~~~~~~~~~~~~~~~~~~~~~~~ - L^2 P(\tau - \tau')\,  c_{\uparrow}\up{\dagger}(\tau)c_{\downarrow}\up{\dagger}(\tau)c_{\downarrow}(\tau')c_{\uparrow}(\tau') \bigg] \,.
\end{align}
The fields $R$, $Q$, and $P$ have to be determined self-consistently. With respect to the above path integral we have the correlators,
\begin{eqnarray}
\label{eq:rqbar}
\bar{R}(\tau - \tau') &=& - \left\langle c_{\alpha}(\tau) c_{\alpha}\up{\dagger}(\tau') \right\rangle_{\mathcal{Z}} \nonumber \\
\bar{Q}(\tau - \tau') &=& \frac{1}{3} \left\langle 
\vec{S}(\tau) \cdot \vec{S}(\tau')
\right\rangle_{\mathcal{Z}} \nonumber \\
\bar{P}(\tau - \tau') &=& \left\langle c_{\downarrow}(\tau)c_{\uparrow}(\tau) c_{\uparrow}\up{\dagger}(\tau')c_{\downarrow}\up{\dagger}(\tau') \right\rangle_{\mathcal{Z}}\,.
\end{eqnarray}
To find the solution of the problem defined by Eq. (\ref{eq:Ham}) we have to impose the following  self-consistency conditions:
\begin{eqnarray}
\label{eq:selfcons}
R(\tau - \tau') &=& \bar{R} (\tau - \tau') \nonumber \\
Q(\tau - \tau') &=& \bar{Q} (\tau - \tau') \nonumber \\
P(\tau - \tau') &=& \bar{P} (\tau - \tau') \,.
\end{eqnarray}
This is the difficult part in obtaining the solution.
Note that since $t_{ij}$ and $L_{ij}$ are real random variables, in principle, upon disorder averaging we also obtain the anomalous (pairing) fields in Eq. (\ref{eq:SH}). However, our analyses below will be restricted to non-superconducting states where the anomalous correlators vanish, and so we have omitted them in our presentation for simplicity. If we consider $t_{ij}$ and $L_{ij}$ to be complex then the anomalous fields do not arise. So as long as we are not in the superconducting phase the analyses with real or complex $t_{ij}$ and $L_{ij}$ are same.


\section{Renormalization group analysis}
\label{sec:c_rg}

In this section we present a renormalization-group study of the model in Eq. (\ref{eq:Zpath},\ref{eq:SB},\ref{eq:SH}). 
To make progress and set-up our RG let us for the moment ignore the self-consistency conditions, Eq. (\ref{eq:selfcons}). We shall come back to it later in our analysis. Furthermore, since we are interested in criticality let us assume a  power-law time dependence for $R$, $Q$ and $P$ fields,
\begin{equation}
\label{eq:rqtau}
R(\tau) \sim \frac{\mbox{sgn}(\tau)}{|\tau|\up{r+1}} \,, ~~~
Q(\tau) \sim \frac{1}{|\tau|\up{d-1}} \,, ~~~
P(\tau) \sim \frac{1}{|\tau|\up{d'-1}} \,,
\end{equation}
where $r$, $d$ and $d'$ are arbitrary numbers. 
The idea is similar to that introduced in Ref. \cite{Joshi2019}. This allows us to decouple the quartic terms in the action, Eq. (\ref{eq:SH}), by introducing a fermionic and two bosonic bath fields. Consequently, the path integral reduces to a quantum impurity problem with the following impurity Hamiltonian: 
\begin{align}
\label{Himp_c}
H_{\rm imp} & = - \mu  \, c_{\alpha}^\dagger c_{\alpha} + U  \left( c_\uparrow^\dagger c_\uparrow - \frac{1}{2} \right) \left( c_\downarrow^\dagger c_\downarrow - \frac{1}{2} \right)   \nonumber \\
&+ g_0 \left[ c^\dagger_\alpha \, \psi_\alpha (0) + \mbox{H.c.} \right] + \gamma_0 c_\alpha^\dagger \frac{\sigma^a_{\alpha\beta}}{2} c_\beta \, \phi_a (0) +v_0\left[ c_{\downarrow} c_{\uparrow} \, \zeta(0)+\mbox{H.c.}\right]\nonumber \\
&+ \int |k|^r dk \, k \, \psi_{k\alpha}^\dagger \psi_{k \alpha} + \frac{1}{2} \int d^d x \left[ \pi_a^2 + (\partial_x \phi_a)^2 \right]+ \frac{1}{2} \int d^{d'} x \left[ \tilde{\pi}^*\tilde{\pi} + (\partial_x \zeta)(\partial_x \zeta^*) \right] \,, 
\end{align}
where $a = (x,y,z)$, $\sigma^a$ are the Pauli matrices, $\pi_a$ is canonically conjugate to the bosonic-bath field $\phi_a$, $\tilde{\pi}$ is canonically conjugate to the bosonic-bath field $\zeta$, and $\psi_{k\alpha}$ is a fermionic-bath field. Additionally, $\phi_a (0) \equiv \phi_a (x=0)$, $\zeta (0) \equiv \zeta (x=0)$ and $\psi_\alpha (0)  \equiv \int |k|^r dk \, \psi_{k \alpha}$. As discussed in Sec. \ref{sec:model}, we will perform the RG about half-filling and small $U$, and so we will set $\mu=0$ and expand perturbatively in $U$ in the following RG calculations.

Let us start by writing the tree-level scaling dimensions of the fields and couplings,
\begin{align}
&\text{dim} [c_{\alpha}] = 0 \,, ~~~~ \text{dim} [\psi_{k\alpha}] = -\frac{1+r}{2} = -\text{dim} [\psi_{\alpha} (0)]\,, ~~~~ \text{dim} [\pa] = \frac{d-1}{2} \,, ~~~~
\text{dim} [\zeta] = \frac{d'-1}{2}   \nonumber \\
&\text{dim} [g_{0}] =\frac{1-r}{2} \equiv \rb \,, ~~~~
\text{dim} [\gam] = \frac{3-d}{2} \equiv \frac{\ep}{2} \,, ~~~~ \text{dim} [v_0] = \frac{3-d'}{2} \equiv \frac{\ep'}{2} \,.
\label{eq:sca_dim_c}
\end{align}
Thus we will be using $\rb$, $\ep$, and $\epp$ as our expansion parameters in perturbative RG calculation. 
To set-up our field-theoretic RG, we introduce the following renormalization factors for the electron operator and coupling constants,
\begin{align}
\label{eq:renorm_fact_c}
c_{\alpha} = \sqrt{Z_c} c_{\alpha, R} \,, ~~~~ 
\gc = \frac{\mu^{\rb} Z'_{g} \gcr}{\sqrt{Z_c}} \,, ~~~~ 
\gam = \frac{\mu^{\ep/2} Z'_{\gamr} \gamr}{Z_c} \,, ~~~~ 
\gv = \frac{\mu^{\epp/2} Z'_{\gvr} \gvr}{Z_c} \,.
\end{align}
Note that the action in Eq. (\ref{Himp_c}) does not contain any interaction terms for the bath fields. As a result, the bath fields do not get renormalized. In the below subsections we shall calculate the electron self energy and vertex corrections at one-loop order to obtain the renormalization factors defined above. 

We note that a perturbative RG was performed for $H_{\rm imp}$ without the $\zeta$ bath field in Ref.~\cite{Tarnopolsky20} by a different method which chose field definitions so that the tree level scaling dimension of $\gamma$ was $\epsilon = 2-d$ rather than $\epsilon = 3-d$ in (\ref{eq:sca_dim_c}). That approach led to an additional boundary renormalization of the operator $\phi^2$, and a different scaling structure. Here we choose the method above because it leads to results compatible with the large $M$ analysis in Section~\ref{sec:largeM}.

\subsection{Self-energy of electron}
\label{sec:se_c}

\begin{figure}[t]
\centering
\subfloat[]{\includegraphics[width=0.2\textwidth]{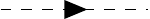}} ~~~~~
\subfloat[]{\includegraphics[width=0.2\textwidth]{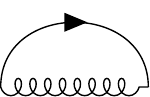}} ~~~~~
\subfloat[]{\includegraphics[width=0.2\textwidth]{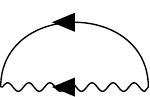}}
\caption{Feynman diagrams for the self energy of electrons. Here the solid line is electron propagator, dashed line is $\psi$ propagator, curly curve is $\phi$ propagator and wavy curve is $\zeta$ propagator. }
\label{fig:se_c}
\end{figure}

We begin with the evaluation of the electron self energy.
There are three contributions to the electron self energy, one each from the $\gc$, $\gam$, and $\gv$ vertices. These are shown in Fig. \ref{fig:se_c} and evaluate as follows:
\begin{align}
\label{eq:se_c1}
\Sigma_{\ref{fig:se_c} (a)}(i\nu) &= \gc^{2} \int dk \frac{|k|^{r}}{i\nu - k} = - \frac{\gc^{2}}{\rb} (i\nu)^{1-2\rb} 
= -\frac{\gcr^{2}}{\rb} i\nu A'_{\mu} \,, \\ 
\label{eq:se_c2} 
\Sigma_{\ref{fig:se_c} (b)} (i\nu) &= \gam^{2} \frac{3}{4} \int \frac{d^{d}k}{(2\pi)^d} \frac{1}{\beta} \sum_{i\omega} \frac{1}{\omega^2 + k^2} \frac{1}{i\nu + i\omega} =  -\gam^{2} \frac{3}{4 \ep} (i\nu)^{1-\ep}  
= -\frac{3 \gamr^{2}}{4\ep} i\nu B'_{\mu} \,, \\ 
\label{eq:se_c3} 
\Sigma_{\ref{fig:se_c} (c)} (i\nu) &= -\gv^{2} \int \frac{d^{d}k}{(2\pi)^d} \frac{1}{\beta} \sum_{i\omega} \frac{1}{\omega^2 + k^2} \frac{1}{-i\nu - i\omega} =  -\gv^{2} \frac{1}{\epp} (i\nu)^{1-\epp} 
= - \frac{\gvr^{2}}{\epp} i\nu C'_{\mu} \,,
\end{align}
where $A'_{\mu} = \mu^{2\rb} (i\nu)^{-2\rb} Z'^{2}_{g}/Z_{c}$, $B'_{\mu} = \mu^{\ep} (i\nu)^{-\ep} Z'^{2}_{\gamr}/Z^{2}_{c}$, and $C'_{\mu} = \mu^{\epp} (i\nu)^{-\epp} Z'^{2}_{\gvr}/Z^{2}_{c}$. 
Demanding the cancellation of poles at the external frequency $i\nu=\mu$, we immediately get,
\begin{equation}
\label{eq:zc_c}
Z_c = 1 - \frac{\gcr^{2}}{\rb} - \frac{3 \gamr^{2}}{4 \ep} - \frac{\gvr^{2}}{\epp} \,.
\end{equation}

\subsection{Vertex corrections} 
\label{sec:vertex_c}

\begin{figure}[t]
\centering
\subfloat[]{\includegraphics[width=0.3\textwidth]{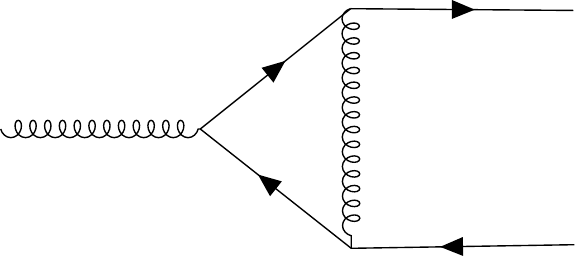}} ~~~~~
\subfloat[]{\includegraphics[width=0.3\textwidth]{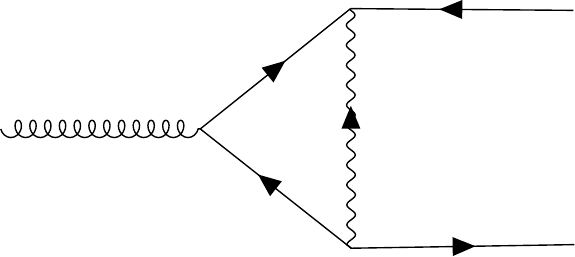}} ~~~~~
\subfloat[]{\includegraphics[width=0.3\textwidth]{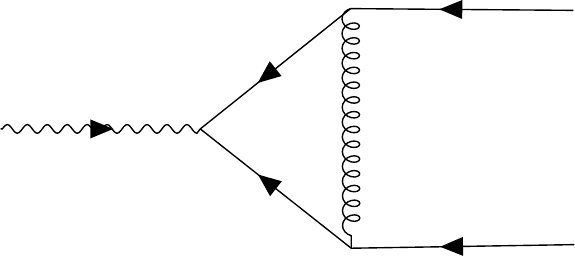}}
\caption{Feynman diagrams for the vertex corrections. Diagrams in (a) and (b) contribute to $\gam$ vertex correction, while that in (c) contribute to $\gv$ vertex correction. Conventions are the same as in Fig. \ref{fig:se_c}.  }
\label{fig:ver_c}
\end{figure}

Next we evaluate the vertex corrections. The $\gc$ vertex has no corrections and so $Z'_{g}=1$. Let us first calculate the correction to $\gam$ vertex. There are two one-loop diagrams shown in Fig. \ref{fig:ver_c} (a) and (b), which evaluate as  
\begingroup 
\allowdisplaybreaks
\begin{align}
\label{eq:ver_gam1_c}
\Gamma_{\ref{fig:ver_c} (a)} &= \gam^{3} \left(-\frac{1}{4} \right) \int \frac{d^{d} k}{(2\pi)^{d}} \frac{1}{\beta} \sum_{i\omega} \frac{1}{\omega^2 + k^2} \frac{1}{i\omega + i\Omega_{1}} \frac{1}{i\omega + i\Omega_{2}} 
= -\gam \frac{\gamr^{2}}{4 \ep} B'_{\mu} \,, \\
\label{eq:ver_gam2_c}
\Gamma_{\ref{fig:ver_c} (b)} &= \gam \gv^{2} \int \frac{d^{d} k}{(2\pi)^{d}} \frac{1}{\beta} \sum_{i\omega} \frac{1}{\omega^2 + k^2} \frac{1}{i\omega + i\Omega_{1}} \frac{1}{i\omega + i\Omega_{2}} 
= \gam \frac{\gvr^{2}}{\epp} C'_{\mu} \,.
\end{align}
\endgroup
In Eq. (\ref{eq:ver_gam1_c}), $i\Omega_1= i\Omega_2$ are external electron frequencies.
Similarly, in Eq. (\ref{eq:ver_gam2_c}) $i\Omega_1= -i\Omega_2$ are external electron frequencies.
We can compute the $\gv$ vertex correction, which has only one contribution at one-loop level (Fig. \ref{fig:ver_c} (c)),
\begin{equation}
\label{eq:ver_v1_c}
\Gamma_{\ref{fig:ver_c} (c)} = \gv \gam^{2} \frac{3}{4} \int \frac{d^{d} k}{(2\pi)^{d}} \frac{1}{\beta} \sum_{i\omega} \frac{1}{\omega^2 + k^2} \frac{1}{i\omega + i\Omega_{1}} \frac{1}{i\omega + i\Omega_{2}} 
= \gv \frac{3\gamr^{2}}{4\epp} B'_{\mu} \,.
\end{equation}
Here $i\Omega_1= -i\Omega_2$ are external electron frequencies.
The above vertex corrections then lead to following renormalization factors:
\begin{align}
\label{eq:zgam_c}
Z'_{\gamr} &= 1 + \frac{\gamr^{2}}{4\ep} - \frac{\gvr^{2}}{\epp} \,, \\
\label{eq:zv_c}
Z'_{\gvr} &= 1 - \frac{3 \gamr^{2}}{4 \ep} \,.
\end{align}

\subsection{Beta functions and fixed points}
\label{sec:beta_c}

Having obtained the renormalization factors we can now derive the RG flow equations. 
Using Eqs. (\ref{eq:renorm_fact_c}), (\ref{eq:zc_c}), (\ref{eq:zgam_c}), and (\ref{eq:zv_c}) we get the following beta functions,
\begin{align}
\label{eq:beta_g_c}
\betg &= -\rb \gcr + \gcr\up{3} + \frac{3}{8} \gcr \gamr\up{2} + \frac{\gcr \gvr^{2}}{2} \,, \\
\label{eq:beta_gam_c}
\betgam &= -\frac{\ep}{2} \gamr + \gamr\up{3} + 2 \gamr \gcr\up{2} \,, \\
\label{eq:beta_v_c}
\betv &= -\frac{\epp}{2} \gvr + \gvr\up{3} + 2 \gvr \gcr\up{2} \,.
\end{align}
The RG flow described by the beta functions found above has seven fixed points and a fixed line in the  $\gamr=0$ plane, which are located using the condition $\betg = \betgam = \betv = 0$. These fixed points/line $(g*\up{2},\gamr*\up{2},\gvr*^{2})$ are as follows:
\begingroup 
\allowdisplaybreaks 
\begin{align}
\label{eq:fp1_c}
FP_1 &: \left(0,0,0 \right) \,, \\
\label{eq:fp2_c}
FP_2 &: \left( \rb, 0,0 \right) \,, \\
\label{eq:fp3_c}
FP_3 &: \left( 0, \frac{\ep}{2}, 0 \right) \,, \\
\label{eq:fp4_c}
FP_4 &: \left( 4\rb - \frac{3\ep}{4}, 2\ep - 8\rb, 0 \right) \,, \\
\label{eq:fp5_c}
FP_5 &: \left( 0,0, \frac{\epp}{2} \right) \,, \\
\label{eq:fp6_c}
FP_6 &: \left( 0, \frac{\ep}{2}, \frac{\epp}{2} \right) \,, \\ 
\label{eq:fp7_c}
FP_7 &: \left( \bar{\gcr}^{2}, 0, \bar{\gvr}^{2} \right) \,, ~~
\text{such that $2\bar{\gcr}^{2} + \bar{\gvr}^{2}= 2\rb$ with $\epp = 4\rb$.}  \\ 
\label{eq:fp8_c}
FP_8 &: \left( -\frac{4\rb}{3} + \frac{\ep}{4} + \frac{\epp}{3}, 
\frac{8\rb}{3} - \frac{2\epp}{3}, \frac{8\rb}{3} - \frac{\ep}{2} - \frac{\epp}{6} \right) \,.
\end{align}
\endgroup 
Note that $FP_7$ is not a fixed point but a fixed line; an ellipse in the $\gamr=0$ plane described by the condition $2\bar{\gcr}^{2} + \bar{\gvr}^{2}= 2\rb$ with $\epp = 4\rb$. Thus $FP_7$ occurs only at a fine-tuned value and is generically not present. Also note that $FP_2$ and $FP_5$ lie on the same curve that describes the ellipse of $FP_7$. However, $FP_2$ and $FP_5$ do not require any fine tuning and are generically present. It so happens that if $\epp=4\rb$ then the fixed line $FP_7$ exists and it passes through both $FP_2$ and $FP_5$. 

Let us now discuss the conditions when the fixed points are real. The fixed point $FP_2$ is real for any $\rb > 0$, $FP_3$ is real for any $\ep > 0$, $FP_5$ is real for any $\epp > 0$, and $FP_6$ is real for any $\ep, \epp > 0$. The fixed point $FP_4$ is real if $16\rb/3 > \ep > 4\rb$. While the fixed point $FP_8$ is real if $(16\rb-\epp)/3 > \ep > (16\rb - 4\epp)/3$. We emphasize again that $FP_7$ exists only if $\epp=4\rb$. If $\epp=4\rb$ then $FP_8$ also lies in the $\gamr=0$ plane, and in fact it lies on the ellipse describing $FP_7$. 

We discuss in detail the stability of the fixed points in Appendix \ref{sec:fp_stab}. In particular, the non-trivial  fixed point $FP_6$, Eq. (\ref{eq:fp6_c}), is stable at the self-consistent values of $\ep, \epp$, and $\rb$, which will be determined in the next subsection.
It is interesting to note that this fixed point has $g=0$, and so charge transport occurs only via Cooper pair hopping, and not via single-particle hopping. So we may consider this as describing a critical Bose metal.

\subsection{Anomalous dimensions of spin, electron and SC order-parameter operators}
\label{sec:anom_c}

We now calculate the anomalous dimensions of the electron, spin, and SC order-parameter operators. These exponents are physical observables in scattering and spectroscopic experiments. Our aim is to evaluate the exponents corresponding to the spin correlator $\langle \vec{S}(\tau)\cdot\vec{S}(0) \rangle$, the electron correlator $\langle c_{\alpha}(\tau) c_{\alpha}\up\dagger(0) \rangle$, and the SC order-parameter correlator $\langle \Delta(\tau) \Delta\up{\dagger}(0) \rangle$. For this purpose, we define the following renormalization factors: 
$\hat{S} = \sqrt{Z_{S}} \hat{S}_{R}$ and  $\Delta=\sqrt{Z_\Delta}\Delta_R$.
To evaluate these we use the strategy of introducing source terms in the action. This is discussed in detail in Appendix \ref{sec:anom_se} for a similar RG procedure. From this analysis it is straightforward to see that 
\begin{align}
\label{eq:zs_c}
Z_{S} &= \left( \frac{Z_c}{Z'_{\gamr}} \right)\up{2}=1-\frac{2\gamma^2}{\ep}-\frac{2g^2}{\rb} \,, \\ 
\label{eq:zdel_c}
Z_\Delta &= \left(\frac{Z_c}{Z'_v} \right)^{2} = 1-\frac{2g^2}{\rb} - \frac{2\gvr^{2}}{\epp} \,.
\end{align}
We have already evaluated $Z_c$ in Eq. (\ref{eq:zc_c}). 

It is now straightforward to express the required anomalous dimensions in terms of the coupling constants, 
\begin{align}
\label{eq:etas}
\eta_{S} &= \frac{d \ln Z_{S}}{d \ln \mu} = 4g^2 + 2\gamma^2 \,, \\ 
\label{eq:etac}
\eta_{c} &= \frac{d \ln Z_{c}}{d \ln \mu} = 2g^2 + \frac{3}{4} \gamma^2 +v^2\,, \\
\label{eq:etadelta}
\eta_\Delta &=\frac{d \ln{Z_\Delta}}{d \ln{\mu}}=4g^2 + 2v^2\,.
\end{align}
These can be evaluated at the respective fixed points,
\begingroup 
\allowdisplaybreaks 
\begin{align}
\label{eq:adimcs1}
&FP_1: \eta_{S} = 0 \,,~~~~ \eta_{c} = 0  \,, ~~~~ \eta_{\Delta} = 0 \\
\label{eq:adimcs2}
&FP_2: \eta_{S} = 4\rb \,,~~~~ \eta_{c} = 2\rb \,, ~~~~ \eta_{\Delta} = 4\rb \,, \\
\label{eq:adimcs3}
&FP_3: \eta_{S} = \ep \,,~~~~ 
\eta_{c} = \frac{3\ep}{8} \,, ~~~~ \eta_{\Delta} = 0 \,, \\
\label{eq:adimcs4}
&FP_4: \eta_{S} = \ep \,,~~~~ 
\eta_{c} = 2\rb \,, ~~~~ \eta_{\Delta} = 16\rb - 3\ep  \,, \\ 
\label{eq:adimcs5}
&FP_5: \eta_{S} = 0 \,,~~~~ 
\eta_{c} = \frac{\epp}{2} \,, ~~~~ \eta_{\Delta} = \epp  \,, \\
\label{eq:adimcs6}
&FP_6: \eta_{S} = \ep \,,~~~~ 
\eta_{c} = \frac{3\ep}{8} + \frac{\epp}{2} \,, ~~~~ \eta_{\Delta} = \epp  \,, \\ 
\label{eq:adimcs7}
&FP_7: \eta_{S} = 4 \bar{g}^{2} \,,~~~~ 
\eta_{c} = 2\rb \,, ~~~~ \eta_{\Delta} = \epp  \,, \\ 
\label{eq:adimcs8}
&FP_8: \eta_{S} = \ep \,,~~~~ 
\eta_{c} = 2\rb \,, ~~~~ \eta_{\Delta} = \epp  \,. 
\end{align}
\endgroup 
Similar to Ref. \cite{Joshi2019}, we can obtain an exact result here for the spin, electron, and SC order-parameter anomalous dimensions. Using Eqs. (\ref{eq:renorm_fact_c}), and  (\ref{eq:etas}) to (\ref{eq:etadelta}) we obtain the exact result that if $g\up{*}\neq 0$ then $\eta_c = 2\rb$, if $\gamr\up{*} \neq 0$ then $\eta_S = \ep$, and if $\gvr\up{*} \neq 0$ then $\eta_{\Delta} = \epp$ at all orders in $\ep$, $\epp$ and $\rb$. In particular this means that at the non-trivial fixed $FP_{8}$ we have $\eta_c =2\rb$, $\eta_S = \ep$, and $\eta_{\Delta}=\epp$ at all orders in the perturbation theory. Note that for $K_{ij} \neq 0$ in Eq. (\ref{eq:Ham}) the exponent of density operator is similarly determined exactly at the fixed points with non-zero density coupling.

Now we are in a position to impose the self-consistency condition, Eq. (\ref{eq:selfcons}), that we had neglected so far. We shall impose the self-consistency condition at the fixed point $FP_6$, Eq. (\ref{eq:fp6_c}). This simply means equating the exponents of $Q(\tau)$ and $P(\tau)$ in Eq. (\ref{eq:rqtau}) to the anomalous dimensions calculated for the spin and SC order-parameter operators, i.e., $2-\ep = \ep$ and $2-\epp=\epp$. Solving these equations yields the self-consistent values of $\ep=\epp=1$ at the fixed point $FP_6$. This means that the spin and SC order-parameter correlators decay as $1/\tau$. And we again emphasize that this result is true at all orders in $\ep$, $\epp$ and $\rb$. Note that since $\gcr^{*}=0$ at $FP_6$ there is no self-consistency condition for $R(\tau)$ and hence the value of $\rb$ is not fixed. Remarkably, at the self-consistent values of $\ep=\epp=1$ the fixed point $FP_6$ is stable at one-loop order for $\rb<7/16$; although we can not formally trust this result at such large values of $\ep$ and $\epp$. We also note that had we decided to consider fixed point $FP_8$, we need to impose the self-consistency condition of all $P$, $Q$, and $R$, which fixes the values of $\ep=\epp=2\rb=1$. This means that spin, electron, and SC order-parameter correlators decay as $1/\tau$ at $FP_8$. However, at one-loop order $FP_8$ is always unstable for real values. But one can not rule out the possibility that at large values of our expansion parameters and at strong coupling the non-trivial fixed point $FP_8$ is stable and controls the RG flow.


\section{Fractionalization and Large-$M$ analysis}
\label{sec:largeM}

In this section, we discuss another approach to solve the model introduced in Eq. (\ref{eq:Ham}). Here we shall fractionalize the electron operator into spinons, holon and doublon. Also we will generalize the SU$(2)$ spin symmetry to USp$(M)$. As we shall see this allows us to solve a set of saddle-point equations and find a conformal solution of the electron Green's function. In particular, we shall calculate the residual conductivity. We discuss this approach in detail in the following.

With the electron operator $c_\alpha$, we represent the 4 states on each site using fermionic spinons $f_\alpha$, a bosonic holon $b$, and a bosonic doublon $d$ as follows: 
\bea
|0 \rangle &\Rightarrow & b^\dagger |v \rangle\,,\quad c^\dagger_\alpha |0 \rangle \Rightarrow  f^\dagger_\alpha |v \rangle\,,\quad c^\dagger_\uparrow \, c^\dagger_\downarrow |0 \rangle \Rightarrow  d^\dagger |v \rangle  \,,
\eea
where $|v \rangle$ is the vaccumm state. 
We therefore fractionalize the electron into spinons, holon, and doublon in the following manner:
\bea
\label{eq:c}
c_\alpha &=& f_\alpha \, b^\dagger + \varepsilon_{\alpha\beta} f^\dagger_\beta \, d \,,  \\
\label{eq:cc}
c_\alpha^\dagger c_\alpha  &=& n = 1 + d^\dagger d - b^\dagger b \,,  \\
\label{eq:S}
\vec{S} &=& \frac{1}{2} c_\alpha^\dagger {\bm \sigma}_{\alpha\beta} c_\beta = \frac{1}{2} f_\alpha^\dagger {\bm \sigma}_{\alpha\beta} f_\beta \,,  \\
\label{eq:c4}
\left( c_\uparrow^\dagger c_\uparrow - \frac{1}{2} \right) \left( c_\downarrow^\dagger c_\downarrow - \frac{1}{2} \right) &=& \frac{1}{2} \left( d^\dagger d + b^\dagger b - \frac{1}{2} \right) \,, \\
\label{eq:ccbd}
c_\downarrow c_\uparrow &=& b^\dagger d\,,
\eea
with the constraint, $f_\alpha^\dagger f_\alpha + b^\dagger b + d^\dagger d = 1$.
Such a fractionlization was studied early on by Kotliar and Ruckenstein \cite{KR86}, and applied to the Mott transition at large positive $U$. Here, we shall study the behavior at small and negative $U$, and allow for critical, gapless, SYK-like states in which the bosons don't condense.

Using Eqs. (\ref{eq:c}) - (\ref{eq:ccbd}), we can write the Hamiltonian in Eq. (\ref{eq:Ham}) in terms of the fractionalized particles, $f, b,$ and $d$,  as
\bea
H &=& \sum_{i = 1}^N \left( -\mu (d_i^\dagger d_i  - b_i^\dagger b_i) + \frac{U}{2}  (d_i^\dagger d_i  + b_i^\dagger b_i)\right) \nonumber \\
&~&~ + \frac{1}{\sqrt{N}} \sum_{i \neq j=1}^N t_{ij}
 c_{i\alpha}^\dagger c_{j \alpha}^{}  + \frac{1}{\sqrt{N}} \sum_{i < j=1}^N J_{ij} \vec{S}_i \cdot \vec{S}_j 
 + \frac{1}{\sqrt{N}} \sum_{i \neq j=1}^N L_{ij} \, b_i^\dagger d_i d_j^\dagger b_j
 \,.
 \label{HU}
\eea
Most of our analysis will be on the particle-hole symmetric case $\mu = 0$; the form of the particle-hole symmetry is briefly discussed in Sec. \ref{sec:ph}.



Next, we generalize the model in Eq. (\ref{HU}) to a model with $M$ spin indices and $M'$ orbital indices. We then study the model in the large-$M$ limit at a fixed $k\equiv M'/M$. 
As it will become clear later, this allows us to find an analytic solution for our model in the low-energy limit.
The original model has $M=2$, $M'=1$, $k=1/2$. In this generalization, the electron operator is
\beq
\label{eq:c_largeM}
c_{\ell \alpha} = f_\alpha \, b_\ell^\dagger + \mathcal{J}_{\alpha\beta} f^\dagger_\beta \, d_\ell \,,
\eeq
where $\mathcal{J}$ is the USp($M$) invariant tensor \cite{HouchesSS}, while $\alpha = 1 \ldots M$ ($M$ even), and $\ell = 1 \ldots M'$. This representation has a U(1) gauge invariance,
\beq
f_{i\alpha}\rightarrow f_{i\alpha}e^{i\phi_i(\tau)},~~~~ b_{i\ell}\rightarrow b_{i\ell}e^{i\phi_i(\tau)},~~~~ d_{i\ell}\rightarrow d_{i\ell}e^{i\phi_i(\tau)} \,.
\label{u1gauge}
\eeq
We also have the constraint fixing the U(1) gauge charge on each site,
\beq
\sum_{\alpha=1}^M f_{i \alpha}^\dagger f_{i \alpha} + \sum_{\ell=1}^{M'} \big(b_{i \ell}^\dagger b_{i \ell}+d_{i \ell}^\dagger d_{i \ell}\big) =\frac{M}{2} \,.
\label{const}
\eeq
Then the large $M, M'$ Hamiltonian is
\bea
H &=& \sum_{i,\ell} \left( - \mu\left(  d_{i \ell}^\dagger d_{i \ell} - b_{i \ell}^\dagger b_{i \ell} \right) + \frac{U}{2}  \left( b_{i \ell}^\dagger b_{i \ell} + d_{i \ell}^\dagger d_{i \ell} \right) \right)  
+ \frac{1}{\sqrt{NM}}\sum_{i,j,\ell,\alpha} t_{ij} c_{i \ell \alpha}^\dagger c_{j \ell \alpha} \nonumber \\
&~&~~+  \frac{1}{\sqrt{NM}}\sum_{i>j,\alpha\beta} J_{ij} f_{i \alpha}^\dagger f_{i \beta} f_{j \beta}^\dagger f_{j \alpha}
+  \frac{1}{\sqrt{NM}}\sum_{i \neq j,\ell \ell'} L_{ij} b_{i \ell}^\dagger d_{i \ell'} d_{j \ell'}^\dagger b_{j \ell} \,.
\label{Hrotor}
\eea

The large $M$ analysis will proceed as Refs. \cite{Fu2018} and \cite{Joshi2019}.  
We first take the $N \rightarrow \infty$ limit and perform disorder average, 
as discussed in Sec. \ref{sec:largeN} to obtain the single-site action in the large-$M$ limit,
\bea
\mathcal{Z} &=& \int \mathcal{D} f_\alpha \mathcal{D} b_\ell\mathcal{D} d_\ell \mathcal{D} \lambda
e^{- \mathcal{S}} \nn 
\label{eq:action_m}
\mathcal{S} &=& \int_0^{1/T} d \tau \Biggl[ \sum_\ell b_{\ell}^\dagger \left( \frac{\partial}{\partial \tau} + \mu + \frac{U}{2}+i \lambda \right) b_{\ell}
+ \sum_\ell d_{\ell}^\dagger \left( \frac{\partial}{\partial \tau} - \mu + \frac{U}{2} + i \lambda \right) d_{\ell} \nonumber \\
&~&~~~~~~~~~~+ \sum_{\alpha} f_{\alpha}^\dagger \left( \frac{\partial}{\partial \tau} + i \lambda \right) f_{\alpha} - i \lambda \frac{M}{2} \Biggr] +\frac{t^2}{M} \sum_{\ell\alpha} \int_{0}^{1/T} d \tau d \tau' R (\tau - \tau') c_{\ell\alpha}^\dagger(\tau)c_{\ell\alpha}(\tau') \nn 
&-&\frac{J^2}{2M} \sum_{\alpha\beta} \int_{0}^{1/T} d \tau d \tau'  Q(\tau-\tau') f_{\alpha}^\dagger (\tau) f_{\beta} (\tau) f_{\beta}^\dagger (\tau') f_{\alpha} (\tau') \nonumber \\
&-&\frac{L^2}{M} \sum_{\ell \ell'} \int_{0}^{1/T} d \tau d \tau'  P(\tau-\tau') b_{\ell}^\dagger (\tau) d_{\ell'} (\tau) d_{\ell'}^\dagger (\tau') b_{\ell} (\tau') 
\,,
\label{L}
\eea
where $T$ is the temperature. Here $\lambda (\tau)$ is the Lagrange multiplier imposing the constraint in Eq. (\ref{const}) and the self-consistency conditions read as follows:
\bea
R(\tau - \tau') &=& -\frac{1}{M M'} \sum_{\ell\alpha}\left\langle c_{\ell\alpha}(\tau)c^\dagger_{\ell\alpha}(\tau')\right\rangle_\mathcal{Z}\,,\nn
Q(\tau - \tau') &=& \frac{1}{M^2} \sum_{\alpha\beta} \left\langle f_{\alpha}^\dagger (\tau) f_{\beta} (\tau) f_{\beta}^\dagger (\tau') f_{\alpha} (\tau') \right\rangle_\mathcal{Z} \,, \nonumber \\
P(\tau - \tau') &=& \frac{1}{M^{'2}} \sum_{\ell\ell'} \left\langle b_{\ell} (\tau) d^\dagger_{\ell'} (\tau) d_{\ell'} (\tau') b_{\ell}^\dagger (\tau') \right\rangle_\mathcal{Z}\,.
\label{selfcon}
\eea

Alternatively, we can also use a representation with bosonic spinons $\mathfrak{b}_\alpha$, fermionic holons $\mathfrak{f}_\ell$ and fermionic doublons $\mathfrak{d}_\ell$. This is discussed in detail in Appendix \ref{app:bos_sp}.

\subsection{Saddle-point equations}
\label{sec:saddle}

We will now write down the saddle-point equations obtained from the action in Eq. (\ref{L}). 
Using the condensed matter notations, we first introduce the following Green's functions
\begin{align}
    G_f(\tau,\tau') &= - \frac{1}{M} \sum_\alpha \left\langle T_\tau( f_\alpha(\tau)f^\dagger_\alpha(\tau'))\right\rangle_\mathcal{Z} \,, \\
    G_b(\tau,\tau') &= - \frac{1}{M'} \sum_\ell \left\langle T_\tau( b_\ell(\tau)b^\dagger_\ell(\tau'))\right\rangle_\mathcal{Z} \,, \\
    G_d(\tau,\tau') &= -\frac{1}{M'} \sum_\ell \left\langle T_\tau( d_\ell(\tau)d^\dagger_\ell(\tau'))\right\rangle_\mathcal{Z} \,,
\end{align}
corresponding to the spinon, $f$, holon, $b$, and the doublon, $d$. 
Note that one can also construct anomalous Green's functions as well as a mixed Green's function involving $b$ and $d$. However, we take these to vanish as they break the U(1) gauge symmetry,  Eq.~(\ref{u1gauge}), down to $\mathcal{Z}_2$. 

From the action in Eq. (\ref{L}), it is straightforward to obtain the saddle point equations for the fractionalized particles:
\begin{align}
    \label{gomega1}
    \Sigma_f(\tau)&=kt^2R(-\tau)G_d(\tau)-kt^2R(\tau)G_b(\tau)+J^2Q(\tau)G_f(\tau)\,,\\
    \label{gomega1a}
    \Sigma_b(\tau)&=t^2G_f(\tau)R(-\tau)+k^2L^2G_d(\tau)P(\tau)  \,,\\
    \label{gomega5}
    \Sigma_d(\tau)&=-t^2R(\tau)G_f(\tau)+k^2L^2G_b(\tau)P(-\tau)\,,\\
    \label{gaiomega}
    G_a^{-1}(i\omega) &= i\omega-\mu_a-\Sigma_a(i\omega) \,,\quad a = f, b, d\,.
\end{align}
Here $\mu_a$ are chemical potentials, determined by $U$ and the saddle point value of $\lambda$ to satisfy
\bea
\left\langle f^\dagger f\right\rangle &=& \delta_f \,, \\
\left\langle b^\dagger b\right\rangle &=& \delta_b\,, \\
\left\langle d^\dagger d\right\rangle &=& \delta_d \,, \\ 
\frac{2}{MM'} \sum_{\ell, \alpha} \left \langle c_{\ell \alpha}^\dagger c_{\ell \alpha} \right\rangle &=& n = 1-\delta = 1 + \delta_d - \delta_b\,,
\label{allconstraints}
\eea
with the gauge-charge constraint (\ref{const}) implying
\beq 
\delta_f + k (\delta_b + \delta_d) = \frac{1}{2}\,.
\label{constdelta}
\eeq
In terms of the Green's functions, the self-consistency equations are expressed as follows:
\begin{align}
    R(\tau)&=-G_f(\tau)G_b(-\tau)+G_f(-\tau)G_d(\tau)\,,\nonumber\\
    Q(\tau)&=-G_f(\tau)G_f(-\tau)\,,\nonumber\\
    P(\tau)&=G_b(\tau)G_d(-\tau)\,.
\end{align}
Now our aim is to look for the solution of these saddle-point equations, which we discuss in the following subsections and refer to technical details  in Appendix \ref{sec:se_app}.


\subsection{Particle-hole symmetry}
\label{sec:ph}

Before we discuss the solutions of our saddle-point equations, we briefly comment on the particle-hole symmetry here. 
The form of the electron Green’s function in terms of the fractionalized particles is
\begin{equation}
    G_c(\tau)=-G_f(\tau)G_b(-\tau)+G_f(-\tau)G_d(\tau) \,.
   \label{Gctau}
\end{equation}
The model considered here has particle-hole symmetry at $\mu=0$, which exchanges the holon and doublon. The particle-hole transformation can be written as follows:
\begin{equation}
\label{eq:ph1}
b \rightarrow - d \,, ~~~ d \rightarrow b \,, ~~~ 
f_{\alpha} \rightarrow f_{\alpha} \,,
\end{equation}
Under this transformation we have $G_b(\tau)=G_d(\tau)$ and thus $\sigma_b(\Omega)=\sigma_d(\Omega)$. Importantly, $c_{\alpha} \rightarrow \epsilon_{\alpha \beta} c_{\beta}\up{\dagger}$, which means that $n \rightarrow 2 - n$. Therefore, the electron Green's function is an odd function due to particle-hole symmetry, i.e., $G_c(\tau)=-G_c(-\tau)$. Also note that the spin operator is invariant under this transformation.


\subsection{Critical solution of saddle-point equations}
\label{sec:sol_largeM}

We will now work towards finding solutions to the saddle-point equations, Eqs. (\ref{gomega1}) - (\ref{gaiomega}) assuming that the spinons, holons, and doublons are all gapless. 
The basic strategy is to find the constraints on the parameters of our low-frequency ansatz by the saddle-point equations. In terms of the imaginary time, $\tau$, such that $|\tau|\gg 1/J$ we write at $T=0$,
\bea 
G_a (\tau) = -\text{sgn}(\tau) \frac{C_a \Gamma (2 \Delta_a) \sin (\pi \Delta_a +\text{sgn}(\tau) \theta_a)}{\pi |\tau|^{2 \Delta_a}}\,, 
\label{Gftau}
\eea
where parameters $C_{a}$, $\Delta_{a}$ and $\theta_{a}$ are real, with more details explained in Appendix \ref{sec:se_app}. 
The spectral densities Eq. (\ref{Sbomega}) in comparison with Eq. \eqref{sfacOmega} immediately yields the following relations of the exponents for nonzero $t$, $J$, and $L$,
\begin{equation}
    \label{deltafbd}
    \Delta_b+\Delta_d=\frac{1}{2} \,,
\end{equation}
and a restriction on asymmetry angles
\bea
\frac{\sin (\pi \Delta_b + \theta_b)\sin (\pi \Delta_d + \theta_d)}{\sin^2(\pi \Delta_f + \theta_f)}=\frac{\sin (\pi \Delta_b - \theta_b)\sin (\pi \Delta_d - \theta_d)}{\sin^2(\pi \Delta_f - \theta_f)} \,.
\label{eq:thetafb}
\eea

We can now evaluate the anomalous dimensions of the electron, spin, and SC order-parameter operators in the present large $M$ limit. Specifically, as the electron Green’s function is given by Eqs. (\ref{Gctau}), spin correlator by $\langle {\mathbf{S}}(\tau)\cdot{\mathbf{S}}(0)\rangle\sim-G_f(\tau)G_f(-\tau)$, and SC order-parameter correlator by $\langle \Delta(\tau) \Delta^\dagger (0)\rangle\sim G_b(\tau)G_d(-\tau)$, it is straightforward to see the anomalous dimension of the corresponding operators are
\begin{align}
    \label{etac}
    \eta_c&=\text{Min}\{2(\Delta_f+\Delta_b),\, 2(\Delta_f+\Delta_d)\} \,,\\
    \label{etaS}
    \eta_S&=4\Delta_f \,,\\
    \label{etaD} 
    \eta_\Delta&=2(\Delta_b+\Delta_d)=1 \,.
\end{align}

Examining the saddle-point equations leads to several possible solutions.
Since we are interested in the general $t-J-L$ model, we will focus on the case with nonzero $t$, $J$, $L$ first and then discuss other special scenarios.


\subsubsection{$\Delta_{f} = \Delta_{b} = \Delta_{d} = 1/4$}
\label{sec:sol1}
In this case, the exponents in $t^2$, $J^2$ and $L^2$ terms in  $\Sigma_{f,b,d}$ (see Eq. (\ref{Sfomega})) are equal, and so all terms are important in our low-frequency analysis. 
Comparison of (\ref{Sfomega}) with (\ref{sfacOmega}) yields 
\bea
\label{solb}
t^2C_f^2C_b^2\cos (2\theta_f)-k^2L^2C_b^2C_d^2\cos (2\theta_d)-2t^2C_f^2C_bC_d\frac{\sin (\pi /4 - \theta_d)\sin^2 (\pi /4+ \theta_f)}{\sin (\pi /4+ \theta_b)}&=\pi\,,\\
\label{sold}
t^2C_f^2C_d^2\cos (2\theta_f)-k^2L^2C_b^2C_d^2\cos (2\theta_b)-2t^2C_f^2C_bC_d\frac{\sin (\pi /4- \theta_b)\sin^2 (\pi/4 + \theta_f)}{\sin (\pi/4 + \theta_d)}&=\pi\,,\\
\label{solf}
J^2C_f^4\cos (2\theta_f)-kt^2C_f^2\biggl[C_b^2\cos (2\theta_b)+C_d^2\cos (2\theta_d)~~~~~~~~~~~~~~~~~&~ \nonumber \\
~~~~~~~~~~~~~-4C_bC_d\frac{\sin (\pi/4 - \theta_f)\sin (\pi/4 + \theta_b)\sin (\pi /4 + \theta_d)}{\sin (\pi /4 + \theta_f)}\biggr]&=\pi
\eea
as well as the restriction on asymmetry angles Eq. \eqref{eq:thetafb}. Notice the bounds $|\theta_f|<\pi/4$, $\pi/4<\theta_b<\pi/2$ and $\pi/4<\theta_d<\pi/2$, which leads to all the coefficients on the left-hand sides of Eqs. (\ref{solb}) - (\ref{solf}) being positive. Since $C_f$, $C_b$, $C_d$ are defined to be real positive numbers, we find another constraint for $\theta_f$, $\theta_b$ and $\theta_d$:
\begin{equation}
    \label{eq:thetacut}
    -\cos{(2\theta_f)}\leq -k\big[\cos{(2\theta_b)}-\cos{(2\theta_d)}\big]\leq \cos{(2\theta_f)}\,,
\end{equation}

Therefore the parameters of the solution ($\theta_d$, $C_f$, $C_b$ and $C_d$) are fully determined by the two asymmetry angles $\theta_f$ and $\theta_b$. Recall that the values of $\theta_f$ and $\theta_b$ are related to the particle densities $\delta_b$, $\delta_d$ via the Luttinger constraints derived by techniques in Ref. \cite{GKST}: 
\begin{align}
\label{eq:lr1}
    \frac{\theta_f}{\pi} + \left( \frac{1}{2} -\Delta_f \right) \frac{\sin (2 \theta_f)}{\sin (2\pi \Delta_f)} &=\frac{1}{2} - \delta_f  \,, \\
    \label{eq:lr2}
    \frac{\theta_b}{\pi} + \left( \frac{1}{2} -\Delta_b \right) \frac{\sin (2 \theta_b)}{\sin (2\pi \Delta_b)} &= \frac{1}{2} + \delta_b \,, \\
    \label{eq:lr3}
    \frac{\theta_d}{\pi} + \left( \frac{1}{2} -\Delta_d \right) \frac{\sin (2 \theta_d)}{\sin (2\pi \Delta_d)} &= \frac{1}{2} + \delta_d \,. 
\end{align}
And from (\ref{allconstraints}), we see that at half-filling with particle-hole symmetry, we have $n=1$ and $\delta_b=\delta_d$.





It is useful to recall now all the parameters and constraint equations for the low energy behavior of the large $M$ ansatz. The low energy ansatz in (\ref{Gftau}) with $\Delta_f = \Delta_b=\Delta_d = 1/4$ is determined by the 6 parameters $C_f$, $\theta_f$, $C_b$, $\theta_b$, $C_d$, $\theta_d$. Let us also regard the densities of the fractionalized particles $\delta_f$, $\delta_b$, and $\delta_d$ as unknown quantities to be determined by the electronic doping density $\delta$. So at fixed $\delta$ we have 9 parameters specifying the low energy solution. These parameters are fully determined by $\delta$ by 9 equations: the Luttinger relations (\ref{eq:lr1}), (\ref{eq:lr2}), (\ref{eq:lr3}), the density constraints (\ref{allconstraints}), (\ref{constdelta}), and the saddle point equations (\ref{eq:thetafb}), (\ref{solb}), (\ref{sold}), (\ref{solf}). Now let us consider the solution of the saddle-point equations at all energies (this has to be done numerically, and we will not carry out the numerical analysis here; such a numerical analysis has been carried out for the $t$-$J$ model in Refs.~\cite{Tikhanovskaya:2020elb,Tikhanovskaya:2020zcw}).
The saddle point equations (\ref{gomega1}), (\ref{gomega1a}), (\ref{gomega5}) will lead to definite values of the self energies $\Sigma_f (i \omega = 0)$, $\Sigma_d (i \omega = 0)$, $\Sigma_d (i \omega = 0)$. But the gapless nature of the ansatz in (\ref{Gftau}) requires that $G_f^{-1} (i \omega = 0) = G_b^{-1} (i \omega = 0) = G_d^{-1} (i \omega = 0) = 0$, which requires cancellation of the zero frequency self energy by the chemical potential \cite{SY92}. The chemical potentials of the 3 fractionalized particles in (\ref{L}) are determined by the 3 parameters $\mu$, $\lambda$, and $U$. As $U$ is not a free parameter, we conclude that the single parameter $U$ must be tuned to realize the critical solution (\ref{Gftau}). So this large-$M$ solution describes a critical point at a fixed density $\delta$, but is present for variable $\delta$. 

From Eqs. (\ref{deltafbd}), (\ref{etac}), (\ref{etaS}), and (\ref{etaD}) we immediately see that in this critical point $\eta_c = \eta_S = \eta_{\Delta} =1$, i.e., the electron, spin, and SC order parameter correlators all decay as $1/\tau$.
In particular, the $1/\tau$ decay of the spin correlator is in stark contrast to its $1/\tau^2$ decay in Fermi liquid phase. 
Such a critical behavior of spin correlator is similar to the SYK-like criticality \cite{SY92, kitaev2015talk}.
The present scenario can be understood as a direct consequence of the fractionalization of electron into spinons, holon, and doublon, wherein the correlator of each of these particles decay as $1/\sqrt{\tau}$. 
As we shall show in the next subsection the conductivity of this critical phase is larger than that of a Fermi liquid metal. Therefore, this solution is a candidate anomalous metal. 
In Sec. \ref{sec:c_rg} we presented a weak-coupling RG analysis for the case of $M=2, M'=1$, and we found a SYK-like critical point, $FP_8$ in Eq. (\ref{eq:fp8_c}). The critical point therein is related to the critical point studied here. Our one-loop weak-coupling RG analysis found $FP_8$ to be unstable. However, as mentioned earlier we expect this fixed point to be stable at strong coupling such that it corresponds to the present large-$M$ solution.

\subsubsection{$\Delta_f>1/4$}
\label{sec:sol2}
For $\Delta_f >1/4$, we neglect the subdominant terms in our low-frequency analysis. The saddle point equations Eqs. \eqref{solb}-\eqref{solf} are now modified as 
\begin{align}
    1/2&=\Delta_f+\Delta_b\,,\\
    1/2&=\Delta_d+\Delta_b\,,\\
    1&=-\frac{k^2L^2C_b^2C_d^2\sin(\pi\Delta_b+\theta_b)\sin(\pi\Delta_b-\theta_b)}{2\pi\Delta_b\sin(2\pi\Delta_b)}\,,\\
    1&=-\frac{kt^2C_f^2C_b^2\sin(\pi\Delta_b+\theta_b)\sin(\pi\Delta_b-\theta_b)}{2\pi\Delta_b\sin(2\pi\Delta_b)}\,,\\    
    -k&=\frac{\sin (\pi \Delta_f + \theta_f)\sin (\pi \Delta_f - \theta_f)/\Delta_f}{\sin (\pi \Delta_b + \theta_b)\sin (\pi \Delta_b - \theta_b)/\Delta_b-\sin(\pi \Delta_d + \theta_d)\sin(\pi \Delta_d - \theta_d)/\Delta_d}\,,
\end{align}
or the other set of equations by making a replacement $\Delta_b\leftrightarrow\Delta_d$, $\theta_b\leftrightarrow\theta_d$, $C_b\leftrightarrow C_d$.
Since the Luttinger relations (\ref{eq:lr1}), (\ref{eq:lr2}), (\ref{eq:lr3}), the density constraints (\ref{allconstraints}) remain the same here, so this solution corresponds to a critical phase. Here the anomalous dimensions of the spin operator, $\eta_S >1$, while that of the SC order parameter, $\eta_\Delta=1$.
The saddle-point of the $J$ term vanishes in this solution, and the solution only depends upon the values of $t$ and $L$. So this solution is the analog of the fixed point $FP_7$ had only the $g$ and $v$ couplings non-zero. 


\subsubsection{Solutions at $t=0$}
\label{sec:solt0}
In this case, 
the combination of Eq. (\ref{Sfomega}) and Eq. (\ref{sfacOmega}) yields $\Delta_f=1/4$ and still satisfies the constraints Eq. \eqref{deltafbd}. We can assume $\Delta_b=\Delta>0$, $\Delta_d=1/2-\Delta>0$ and therefore Eqs. \eqref{solb}-\eqref{solf} are modified to
\begin{align}
    J^2C_f^4\cos(2\theta_f)&=\pi\,,\\
    -k^2L^2C_b^2C_d^2\frac{\cos(2\theta_d)+\cos(2\pi\Delta)}{(1-2\Delta)\sin(2\pi\Delta)}&=2\pi\,,\\
   -k^2L^2C_b^2C_d^2\frac{\cos(2\theta_b)-\cos(2\pi\Delta)}{2\Delta\sin(2\pi\Delta)}&=2\pi\,,
\end{align}
Therefore the parameters of the solution ($\theta_b$, $C_f$ and the product $C_bC_d$) are fully determined by the asymmetry angles $\theta_f$, $\theta_d$ and exponent $\Delta$. Since $C_f$, $C_b$, $C_d$ are defined to be real positive numbers, we find a constraint in analogy with \eqref{eq:thetacut} for $\theta_d$ and $\Delta$ at $t=0$:
\begin{equation}
    \label{eq:thetacutJL}
    \cos{(2\theta_d)}+\cos{(2\pi\Delta)}\leq 0\,.
\end{equation}
This solution corresponds to a critical phase with $\eta_S=\eta_\Delta=1$ and $\eta_c\leq 1$. Our weak-coupling RG in Sec. \ref{sec:c_rg} yielded a stable `Bose metal' fixed point, $FP_6$ in Eq. (\ref{eq:fp6_c}), with these properties. 


\subsubsection{Solutions at $L=0$}
\label{sec:solL0}
For $\Delta_f=\Delta_b=\Delta_d=1/4$, the saddle point equations have the same form as Eqs. \eqref{solb}-\eqref{solf} without $L$ terms, but the condition to keep all $C_f$, $C_b$, $C_d$ real and positive is different from Eq.\eqref{eq:thetacut}:
\begin{equation}
    \label{eq:thetacutl0}
    -\cos{(2\theta_f)}\leq -k\big[\cos{(2\theta_b)}+\cos{(2\theta_d)}\big]\leq \cos{(2\theta_f)}\,.
\end{equation}
This solution actually is a special case in Sec. \ref{sec:sol1}.


For $\Delta_f >1/4$ the $J^2$ term in Eq. (\ref{Sfomega}) is subdominant compared to the $t^2$ term, and so we neglect it in our low-frequency analysis. The combination of Eq. (\ref{Sfomega}) and Eq. (\ref{sfacOmega}) yields 
\begin{align}
    1/2&=\Delta_f+\Delta_b\,,\\
    1/2&=\Delta_f+\Delta_d\,,\\
    1&=t^2C_f^2C_b^2\frac{\sin (\pi\Delta_f+\theta_f)\sin (\pi\Delta_f-\theta_f)}{2\pi\Delta_f\sin (2\pi\Delta_f)}-t^2C_f^2C_bC_d\frac{\sin (\pi\Delta_d - \theta_d)\sin^2 (\pi\Delta_f+ \theta_f)}{2\pi\Delta_f\sin (2\pi\Delta_f)\sin (\pi\Delta_b+ \theta_b)}\,,\\
    1&=t^2C_f^2C_d^2\frac{\sin (\pi\Delta_f+\theta_f)\sin (\pi\Delta_f-\theta_f)}{2\pi\Delta_f\sin (2\pi\Delta_f)}-t^2C_f^2C_bC_d\frac{\sin (\pi\Delta_b- \theta_b)\sin^2 (\pi\Delta_f + \theta_f)}{2\pi\Delta_f\sin (2\pi\Delta_f)\sin (\pi\Delta_d+ \theta_d)}\,,\\
   -k&= \frac{\sin (\pi \Delta_f + \theta_f)\sin (\pi \Delta_f - \theta_f)/\Delta_f}{\sin (\pi \Delta_b + \theta_b)\sin (\pi \Delta_b - \theta_b)/\Delta_b+\sin(\pi \Delta_d + \theta_d)\sin(\pi \Delta_d - \theta_d)/\Delta_d} \,,
\end{align}
along with the restriction on the asymmetry angles Eq. \eqref{eq:thetafb}. Therefore the parameters of the solution ($\Delta_b$, $\Delta_d$, $\theta_b$, $\theta_d$, and the product $C_fC_b$, $C_fC_d$) are fully determined by the asymmetry angle $\theta_f$ and exponent $\Delta_f$. 
This solution corresponds to a critical phase with $\eta_S>1$, $\eta_\Delta<1$ and $\eta_c=1$. 

\subsection{Non-critical phases}
\label{sec:nonc}

As in Ref.~\cite{Fu2018}, we obtain non-critical phases by gapping or condensing the bosons. A solution with both $b$ and $d$ gapped is possibly only at half-filling, and this describes a gapless spin liquid of the $f$ spinons. Such a solution is only reasonable at large $U$. If we condense only one of $b$ and $d$, then we Higgs the U(1) gauge symmetry, but leave the global U(1) symmetry unbroken--such a solution is therefore a disordered Fermi liquid, as in Ref.~\cite{Joshi2019}. 
Finally, if we condense both $b$ and $d$, we break both the gauge and global U(1) symmetries, and the solution is a superconductor.

\subsection{Conductivity}
\label{sec:con}

We shall now compare the zero temperature residual conductivity between the normal state, and the critical point corresponding to the solution $\Delta_f = \Delta_b = \Delta_d =1/4$ found above. Let us denote the residual conductivities in the normal and the critical phases by $\sigma_{0}\up{N}$ and $\sigma_{0}\up{A}$ respectively. We shall show below that for a wide-range of asymmetry parameters the ratio $\mathcal{R}\equiv\sigma_{0}\up{A}/\sigma_{0}\up{N} > 1$. This suggests that the critical phase is in fact an anomalous metal phase, similar to that reported in experiments in the vicinity of a metal-SC phase transition. In this section, we will need to revert to the large dimensionality limit to properly define the expressions for the conductivity \cite{GGS20,GeorgesRMP}.

The calculation in the critical (anomalous metal) phase uses  techniques from Ref. \cite{GGS20},  which applies a more general formalism using time-reparameterization symmetry in SYK-like models. Following the steps in Ref. \cite{GGS20}, we first generalize the Greens function to finite temperature, 
\begin{equation}
    G_a(\theta)=-\frac{C_a\Gamma(2\Delta_a)}{\pi}\frac{\sin(\theta_a+\pi\Delta_a)}{|2\sin\frac{\theta}{2}|^{2\Delta_a}}e^{-\mathcal{E}_a\theta}, \quad \theta\in(0,2\pi) \,,
    \label{Gatheta}
\end{equation}
where we choose $\theta=2\pi\tau/\beta$. $\mathcal{E}_a$ is determined by
\begin{equation}
    e^{2\pi\mathcal{E}_a}=\frac{\sin(\theta_a+\pi\Delta_a)}{\sin(\theta_a-\pi\Delta_a)}\zeta_a\,,
\end{equation}
where $\zeta_b=\zeta_d=1$, $\zeta_f=-1$ indicate the periodicity of $G_a(\theta)$. Here the subscript $a = f,b, d$ is used for spinon, holon and doublon respectively.

The Kubo formula for conductivity on the Bethe lattice from electron hopping is derived in Ref. \cite{GGS20}
\begin{align}
\label{eq:kubo}
    \Re\sigma(\nu) &=\frac{M'e^2t^2a^{2-d}}{2\pi z}\int d\omega A_c(\omega)A_c(\omega+\nu) \frac{n_F(\omega)-n_F(\omega+\nu)}{\nu}\,,
\end{align}
 where $a$ is lattice constant, $z$ is coordination number, $d$ is spatial dimension and $A_c(\omega)=-2\Im G_c(\omega+i\eta)$ is electron spectral density. We can introduce the bosonic Cooper pair Green's function $G_\Delta(\theta)\equiv G_b(\theta)G_d(-\theta)$, whose contribution to conductivity is an analogy with the electron contribution after making the replacement $e\rightarrow 2e$, $A_c\rightarrow A_\Delta$, $n_F\rightarrow n_B$ and  $t^2\rightarrow kL^2$. 
The residual conductivity now yields
\begin{equation}
\label{eq:resb0}
    \sigma_{0}^A =\frac{M'e^2t^2a^{2-d}}{2\pi z}\int d\omega A_c(\omega)^2\beta n_F(\omega)n_F(-\omega)+\frac{M'(2e)^2k L^2a^{2-d}}{2\pi z}\int d\omega A_\Delta(\omega)^2\beta n_B(\omega)n_B(-\omega) \,,
\end{equation}
where 
$A_\Delta(\omega)=-2\Im G_\Delta(\omega+i\eta)$ is Cooper pair spectral density,
\begin{align}
    \label{eq:Ac_ano}
    A_c(\omega)&=2\pi C e^{-\pi\mathcal{E}}\frac{\cosh(\beta\omega/2)}{\cosh(\beta\omega/2-\pi\mathcal{E})}+2\pi C' e^{-\pi\mathcal{E'}}\frac{\cosh(\beta\omega/2)}{\cosh(\beta\omega/2-\pi\mathcal{E'})} \,,\\
    \label{eq:Ad_ano}
    A_\Delta(\omega)&=-2\pi C_\Delta e^{-\pi\mathcal{E}_\Delta}\frac{\sinh(\beta\omega/2)}{\cosh(\beta\omega/2-\pi\mathcal{E}_\Delta)} \,.
\end{align}
Here the parameters are defined as $\mathcal{E}=\mathcal{E}_f-\mathcal{E}_b$, $\mathcal{E'}=\mathcal{E}_d-\mathcal{E}_f$,  $\mathcal{E}_\Delta=\mathcal{E}_b-\mathcal{E}_d$, $C=C_fC_b\sin(\theta_f+\pi/4)\sin(\theta_b-\pi/4)/\pi$, $C'=-C_fC_d\sin(\theta_f-\pi/4)\sin(\theta_d+\pi/4)/\pi$, $C_\Delta=C_bC_d\sin(\theta_b+\pi/4)\sin(\theta_d-\pi/4)/\pi$.

We now have all the ingredients to calculate the residual conductivity using Eq. (\ref{eq:resb0}). Inserting Eqs. \eqref{eq:Ac_ano}-\eqref{eq:Ad_ano} into Eq. (\ref{eq:resb0}) we obtain,
\begin{equation}
    \sigma_{0}\up{A} 
    =2\pi\frac{M'e^2a^{2-d}}{z}\bigg[t^2\bigg( C^2e^{-2\pi\mathcal{E}}+C'^2e^{-2\pi\mathcal{E'}}-2CC'e^{-\pi(\mathcal{E}+\mathcal{E'})}\frac{\pi(\mathcal{E}-\mathcal{E'})}{\sinh[\pi(\mathcal{E}-\mathcal{E'})]}\bigg)+4kL^2C_\Delta^2 e^{-2\pi\mathcal{E}_\Delta}\bigg] \,.
\end{equation}

In the normal state with semi-circle density $A_c(\omega) =\sqrt{4t^2-(\omega+\mu)^2}/t^2$, only electron contributes to the residual conductivity. Therefore, at zero temperature and $\mu=0$ the normal state residual conductivity is
\begin{equation}
\sigma_{0}\up{N}=\frac{M'e^2t^2a^{2-d}}{2\pi z}\int d\omega A_c(\omega)^2\beta n_F(\omega)n_F(-\omega)=\frac{2M'e^2 a^{2-d}}{\pi z} \,.
\end{equation}
Now we have the ratio of these two residual conductivities, 
\begin{align}
    \mathcal{R}=\frac{\sigma_0^A}{\sigma_0^N}=\pi^2\bigg[t^2 \bigg(C^2 e^{-2\pi\mathcal{E}}+C'^2 e^{-2\pi\mathcal{E}'}+2C C' e^{-\pi(\mathcal{E}+\mathcal{E}')}\frac{\pi (\mathcal{E}-\mathcal{E}')}{\sinh{[\pi(\mathcal{E}-\mathcal{E}')]}}\bigg)+4kL^2C_\Delta^2 e^{-2\pi\mathcal{E}_\Delta}\bigg].
    \label{Ratio}
\end{align}

\begin{figure}
    \centering
    \subfloat[]{\includegraphics[height=0.32\textwidth]{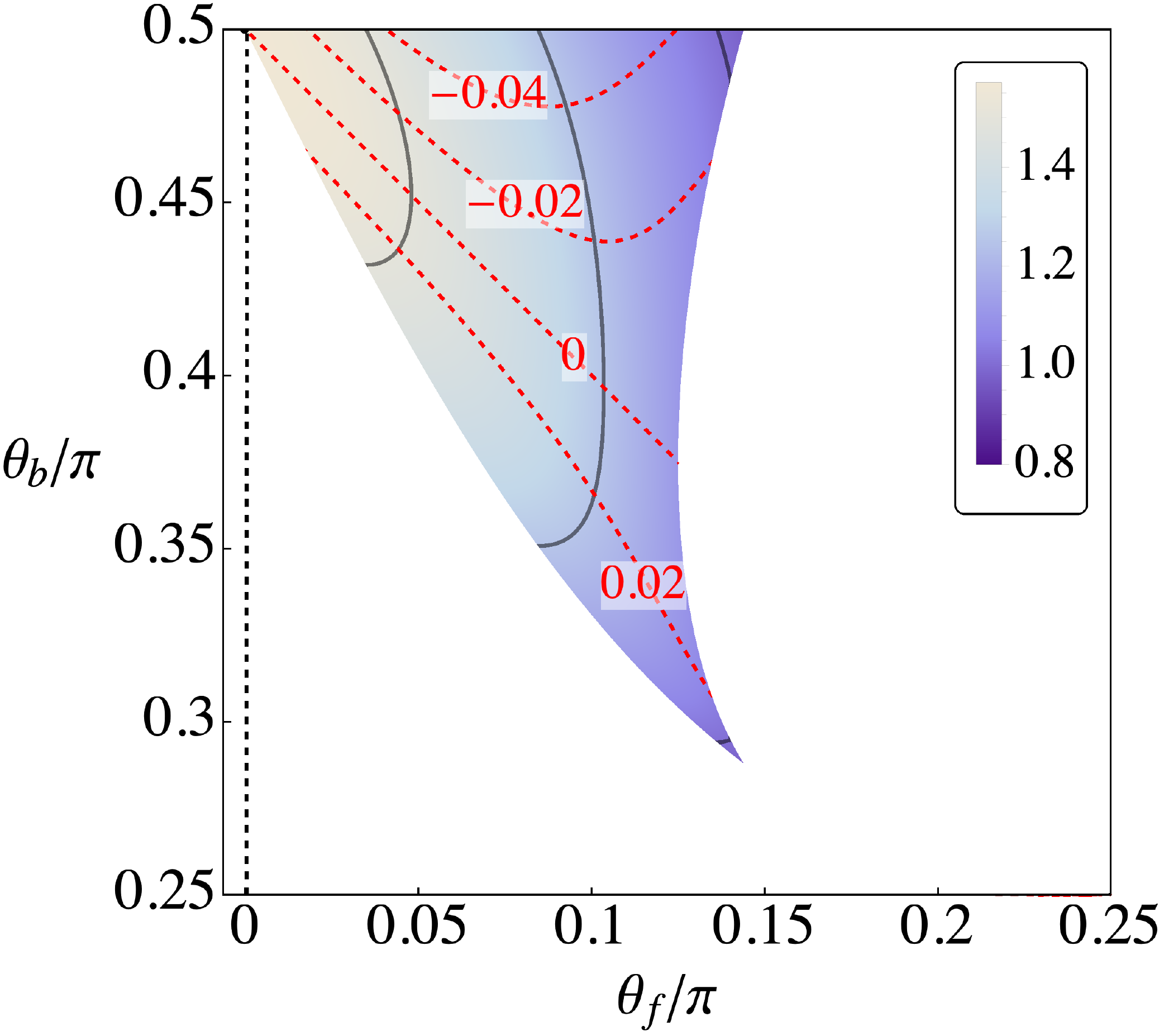}}~
    \subfloat[]{\includegraphics[height=0.32\textwidth]{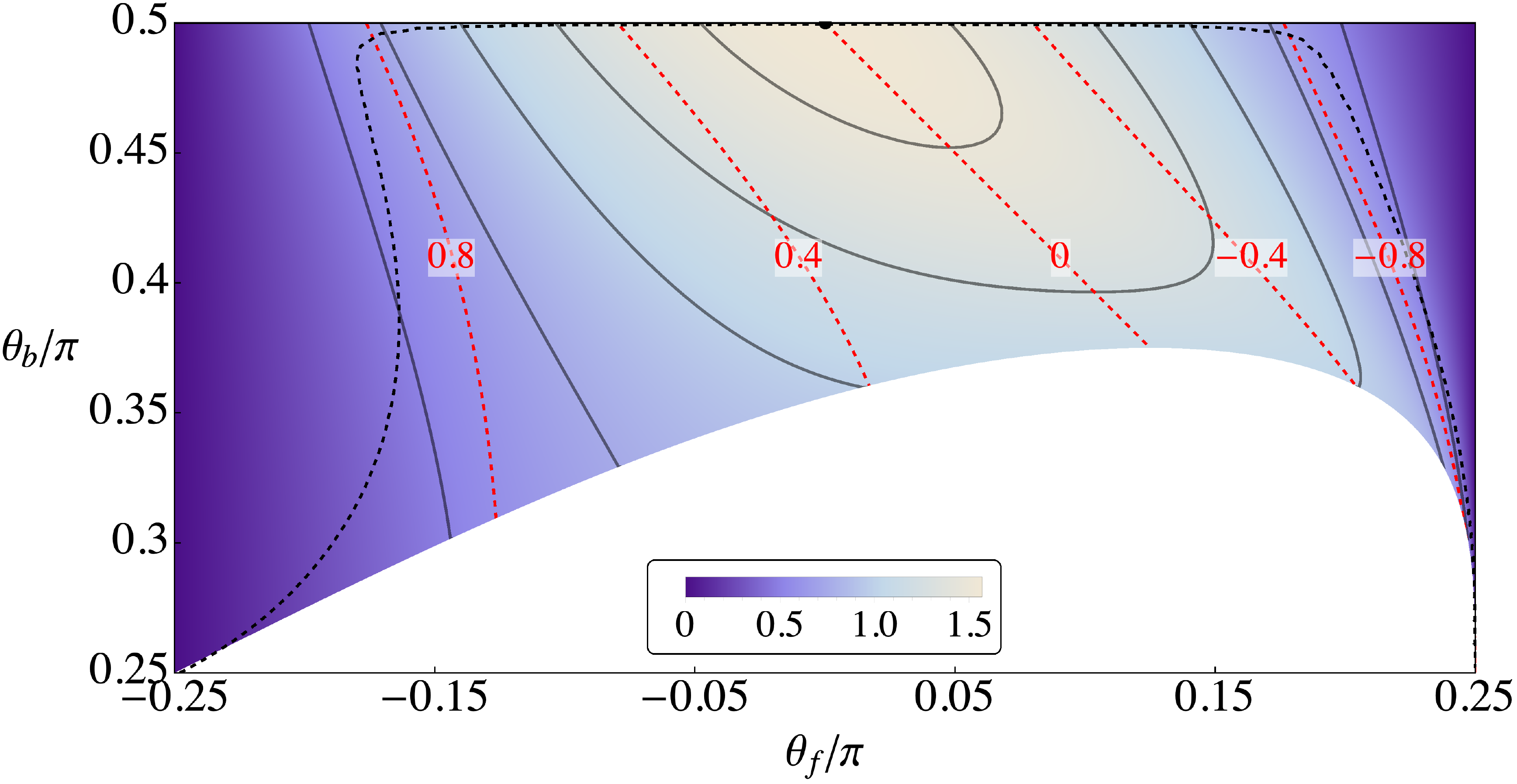}}~~~~\\
    \subfloat[]{\includegraphics[height=0.32\textwidth]{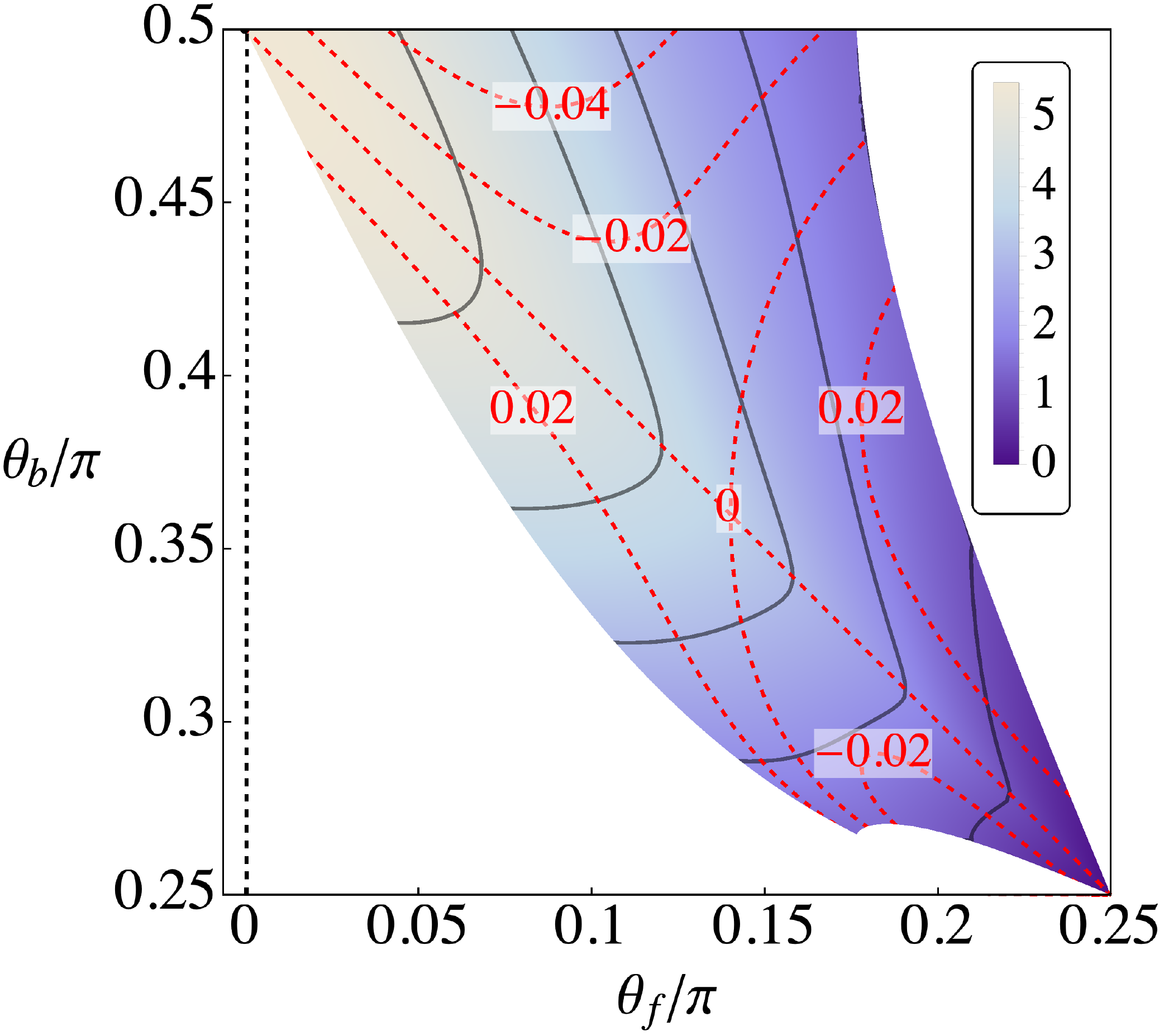}}~
    \subfloat[]{\includegraphics[height=0.32\textwidth]{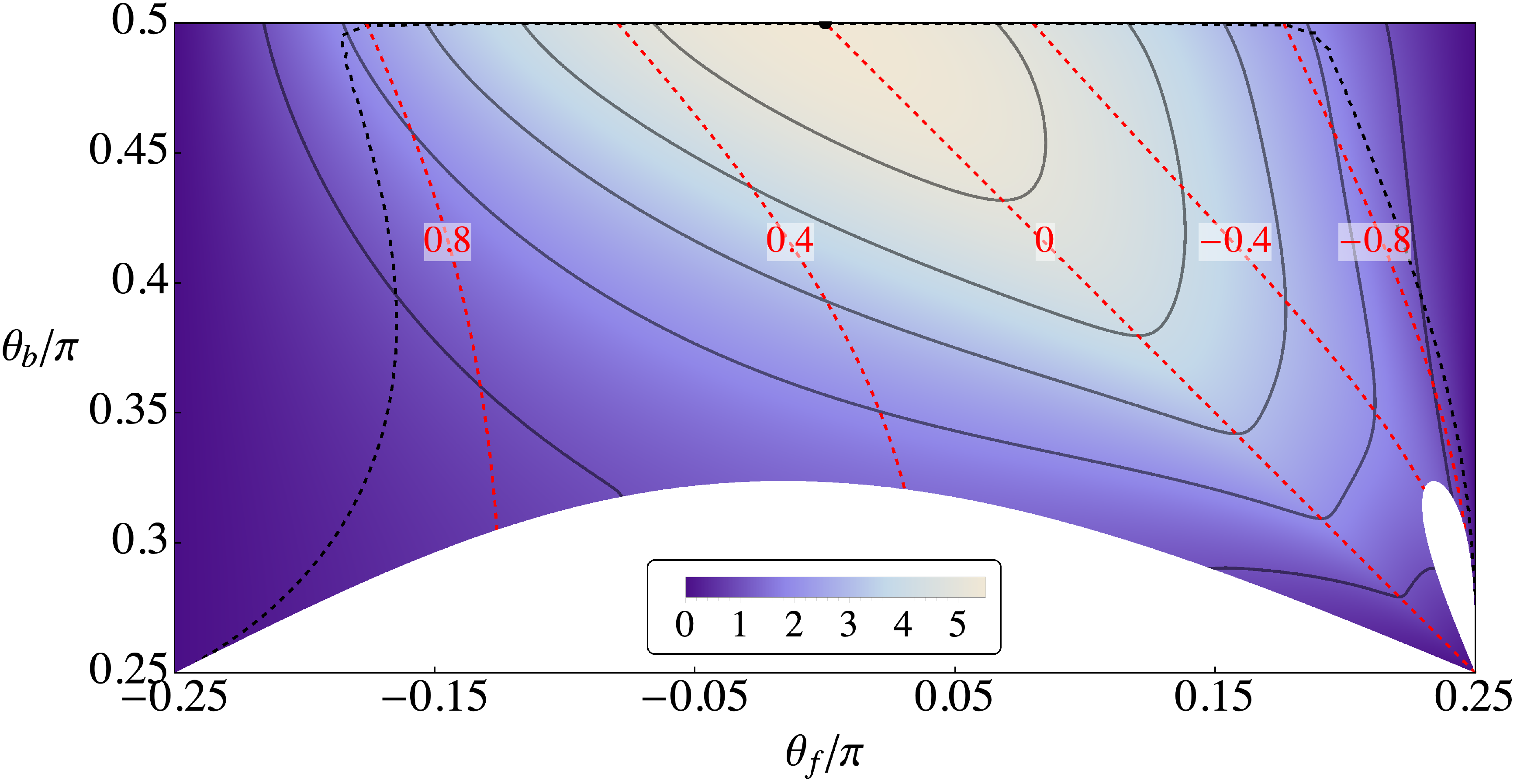}}~~~~
    \caption{The ratio of large M background conductivity to normal state solution, $\mathcal{R}=\sigma_0^A/\sigma_0^N$, in a density plot for possible asymmetry parameters $\theta_f$ and $\theta_b$ in different representations; the full black lines are contours of equal $\mathcal{R}$. The ratio is calculated at $k=1/2$, $t=1/2$, $J= 1$ and different $L$. (a) Bosonic holon scheme with $L=0$. (b) Fermionic holon scheme with $L=0$. (c) Bosonic holon scheme with $L=5$. (d) Fermionic holon scheme with $L=5$. The red dashed lines are contours of constant doping density $\delta$ obtained by solving (\ref{allconstraints},\ref{eq:lr2},\ref{eq:lr3}), and the black dashed lines shows the solution of (\ref{eq:thetafb},\ref{eq:lr1},\ref{eq:lr2},\ref{eq:lr3}). Spectral positivity and the conditions for real $C_f$, $C_b$, $C_d$ parameters restrict the range of solutions: (a) is bounded by (\ref{eq:thetafb},\ref{eq:thetacutl0},\ref{eq:spec_pos_bh}),  (b) is bounded by (\ref{eq:spec_pos_fh},\ref{thetafb2},\ref{eq:thetacut2l0}), (c) is bounded by (\ref{eq:thetafb},\ref{eq:thetacut},\ref{eq:spec_pos_bh}), and (d) is bounded by (\ref{eq:spec_pos_fh},\ref{thetafb2},\ref{eq:thetacut2}),. We could identify the ratio $\mathcal{R}$ in the two schemes (a) and (b) or (c) and (d) by $\theta_f\leftrightarrow \pi/2-\theta_b$.}
    \label{fig:sigma}
\end{figure}

As we can see from Sec. \ref{sec:sol_largeM}, the above  ratio has only two parameters, namely, $\theta_f$ and $\theta_b$. In Fig. \ref{fig:sigma} we make a density plot of $\mathcal{R}$ with contours as a function of these parameters. The ratios are calculated at a given set of parameters $k=1/2$, $t=1/2$, $J=1$, $L=0$ and 5 in two schemes: the fermionic spinon scheme in Sec. \ref{sec:largeM} and the bosonic spinon scheme discussed in Appendix \ref{app:bos_sp}. In the particle-hole symmetric case, the only solution is $\theta_f=0$, $\theta_b=\pi/2$ with $\mathcal{R}\approx 5$ at $L=5$ and $\mathcal{R}=\pi/2$ at $L=0$ for both schemes. We further explore the particle-hole symmetric solution at $k=1/2$ but arbitrary positive $t$, $J$, $L$ parameters, and obtain the ratio in the range of 
\begin{equation}
\label{eq:sigma_tJL}
\frac{\pi}{2}\leq\mathcal{R}\big|_{\theta_f=0,\,\theta_b=\pi/2}=2\pi\frac{J L+t^2}{J L+4t^2} \leq 2\pi\,.
\end{equation}
More generally, for arbitrary $k$, $t$, $J$, $L$ parameters,
\begin{equation}
\label{eq:sigma_tJLk}
\mathcal{R}\big|_{\theta_f=0,\,\theta_b=\pi/2}=\pi\frac{k J^2L^2+(1-2k)t^4+\sqrt{k^2J^2L^2t^4+(1-2k)^2t^8}}{k^2J^2L^2+2(1-2k)t^4+2\sqrt{k^2J^2L^2t^4+(1-2k)^2t^8}}=\begin{cases}
      \pi/k, & \text{if}\ JL\gg t^2 \\
      \pi/2, & \text{if}\ JL\ll t^2\,.
    \end{cases}
\end{equation}
The limit $JL\gg t^2$ likely describes a Bose metal, where incoherent pairs of electrons carry the charge current instead of bare electrons.

We now briefly discuss the conductivity of the phases corresponding to the RG fixed point $FP_7$ and $FP_6$, which are associated with the large $M$ saddle points in Sections~\ref{sec:sol2} and \ref{sec:solt0} respectively.
For $FP_7$, we assume $\Delta_f=\Delta_d=\Delta>1/4$, so $\Delta_b=1/2-\Delta<1/4$ and the ratio of these two residual conductivities yields
\begin{align}
    \mathcal{R}=\frac{\pi(1-2\Delta)\sin{\pi(1-2\Delta)}}{k}\bigg[\frac{3}{2}\cos(2\pi\Delta)-2\cos(2\theta_d)+\frac{1}{2}\cos(2\theta_f)\bigg]<\frac{5\pi}{4k}\,.
    \label{eq:Ratiofp7}
\end{align}
For $FP_6$, we consider the case with $t=0$, $\Delta_f=1/4$, $\Delta_b=\Delta$, $\Delta_d=1/2-\Delta$. The ratio becomes
\begin{align}
    \mathcal{R}=-\frac{4\pi\Delta\sin{2\pi\Delta}}{k}\bigg[\cos(2\pi\Delta)+\cos(2\theta_d)\bigg]\,.
    \label{RatioJL}
\end{align}
The maximum of $\mathcal{R}=11.4$ reaches at $\Delta=0.365$, $\theta_d=\pi/2$.

As a reminder, we also quote here the value of zero temperature residual conductivity in the critical phase found in a related model, namely the random $t$-$J$ model without double occupancy. The result obtained in Ref. \cite{GGS20} is
\begin{equation}
\label{eq:sigma_tJ}
\mathcal{R}^{tJ}=\frac{\sigma_{0}^{tJ}}{\sigma_0^N} = 2\pi C^2e^{-2\pi\mathcal{E}} \frac{M'e^2t^2a^{2-d}}{z}\bigg/\sigma_0^N =-\frac{\pi}{4}\cos(2\theta_b)\leq\frac{\pi}{4}\,. 
\end{equation}

\section{Conclusions}
\label{sec:conc}

We have examined a model of interacting electrons with random and all-to-all hopping  and interactions in (\ref{eq:Ham}): this yields a theory which is mean-field in space but includes fluctuations in time. The interactions are in one-to-one correspondence with the interactions in the universal Hamiltonian for disordered metallic grains \cite{KAA00,Alhassid00,Aleiner02}. The universal Hamiltonian has the same interaction strength acting between any pair of sites, but we instead take each interaction term to be an independent random number. An important characteristic of our analysis is that the random hopping and interaction terms are treated at an equal footing. We then employed RG and large $M$ computations to argue, as in Ref.~\cite{Joshi2019}, that such models lead to critical metallic solutions similar to those in the SYK models. For the class of models examined here, with the on-site Hubbard interaction parameter small, we have argued that such critical metals can realize a critical point or phase between a disordered Fermi liquid and a superconductor, and so are candidate dynamical mean-field descriptions of the anomalous metallic `failed superconductor' \cite{KKS19}.

The experimental observations are largely in two-dimensional systems, while our model
has either infinite-range interactions or large spatial dimensionality, a characteristic of dynamical mean-field theories. So our model does not capture effects that have been viewed as crucial to the physics of disordered interacting electrons: weak localization and interaction corrections from diffusive electrons. However, such effects require long-lived quasiparticles, and the absence of quasiparticles in our theory is reason to hope that our dynamical mean-field theory is useful starting point for a theory in two dimensions which can be applied real materials.

Another perspective on our anomalous metal is provided by an analogy with the large $U$ work of Joshi {\it et al.} \cite{Joshi2019}. That work examined the transition between a metallic spin glass and a disordered Fermi liquid, and proposed a description in terms of a 
deconfined spin liquid with doped charge carriers. Our present work is carried out at small $|U|$, and has fluctuations in the Nambu pseudospin space along with spin fluctuations. So we may consider the anomalous metal as a `deconfined pseudospin liquid' found at the transition between a supercondutor and a disordered Fermi liquid.  

Our RG computations in Section~\ref{sec:c_rg}, and our large $M$ computation in Section~\ref{sec:largeM} yielded a number of candidate solutions for the critical anomalous metal. There are interesting connections between these two classes of solutions, and we list below 3 solutions from each method that have similar physical properties and exponents. Notice, however, that the stability of the phases differs between the two methods, and so we are not able to reliably determine the optimal candidate.
\begin{center}
\begin{tabular}{c|c}
      RG & Large $M$ \\
     \hline
     Fixed point $FP_8$ & Solution in Sec.~\ref{sec:sol1} \\
     $\eta_c=\eta_S=\eta_\Delta=1$ & $\eta_c=\eta_S=\eta_\Delta=1$ \\
     Unstable critical point & Stable critical point \\
     \hline
          Fixed point $FP_7$ & Solution in Sec.~\ref{sec:sol2} \\
     $\eta_c=\eta_\Delta=1$ & $\eta_c=\eta_\Delta=1$, $\eta_S > 1$ \\
     Present only for $\epsilon' = 4 \overline{r}$~~\\
     Line of stable critical points~ & Stable critical phase \\
     \hline
          Fixed point $FP_6$ & Solution in Sec.~\ref{sec:solt0} \\
     $\eta_S=\eta_\Delta=1$ & $\eta_S=\eta_\Delta=1$, $\eta_c \leq 1$ \\
     Stable critical point & ~~Critical phase unstable to $t \neq 0$ \\
     Critical Bose metal & Critical Bose metal \\
     \hline
\end{tabular}
\end{center}
The exponents $\eta_c$, $\eta_S$, $\eta_\Delta$ determine the decay of the electron, spin, and Cooper-pair correlators in imaginary time $\sim 1/\tau^\eta$. The Bose metal transports charge primarily via Cooper pair hopping, and single electron hopping is subdominant.

The conductivities of the anomalous critical metal phases were computed in Section~\ref{sec:con} in the large $M$ theory. Here, we followed the large-$M$ formulation of the conductivity in a large dimension lattice, which was discussed in Ref.~\cite{GGS20}. It should be noted that this large $M$ approach yields residual resistivities which are significantly higher than those obtained numerically in the SU(2) solution of the same model \cite{Cha19}. So we computed the ratios of the conductivity of large $M$ solution of the critical anomalous metal to that of the disordered Fermi liquid, as an estimate of any enhancement of the conductivity in the anomalous metal in the large $M$ limit. As presented in Section~\ref{sec:con} there is a moderate, but not large, enhancement in this ratio in our computations. It would be of interest to perform numerical studies for the SU(2) case, similar to that in Ref.~\cite{Cha19}, for models like (\ref{eq:Ham}) with Cooper pair hopping, and compute their conductivities between the Fermi liquid and superconducting phases.

\section*{Acknowledgement}
\label{sec:ack}

We thank H.~Guo, A.~Kapitulnik, S.~Kivelson, and G.~Tarnopolsky for valuable discussions. 
This research was supported by the National Science Foundation under Grant No.~DMR-2002850. 
D.G.J acknowledges support from the Leopoldina fellowship by the German National Academy of Sciences through grant no. LPDS 2020-01. This work was also supported by the Simons Collaboration on Ultra-Quantum Matter, which is a grant from the Simons Foundation (651440, S.S.).



\appendix


\section{Renormalization-group analysis using fractionalization}
\label{app:fbd_rg}

This appendix will obtain the same RG results as those obtained in Section~\ref{sec:c_rg}, but by using the fractionalized degrees of freedom we defined in Section~\ref{sec:largeM}.
To this end, we rewrite the impurity Hamiltonian in (\ref{Himp_c}) using the spinon, doublon and holon operators. 
\begin{align}
\label{Himp}
H_{\rm imp} & = -\mu \left(d^\dagger d - b^\dagger b \right) + \frac{U}{2} (b^\dagger b + d^\dagger d) + \lambda \left( f_\alpha^\dagger f_\alpha + b^\dagger b + d^\dagger d \right)   \nonumber \\
&+ g_0 \left[ \left( f^\dagger_\alpha b + \varepsilon_{\alpha\beta} f_\beta d^\dagger \right)\, \psi_\alpha (0) + \mbox{H.c.} \right] + \gamma_0 f_\alpha^\dagger \frac{\sigma^a_{\alpha\beta}}{2} f_\beta \, \phi_a (0) +v_0\left[b^\dagger d\zeta(0)+\mbox{H.c.}\right]\nonumber \\
&+ \int |k|^r dk \, k \, \psi_{k\alpha}^\dagger \psi_{k \alpha} + \frac{1}{2} \int d^d x \left[ \pi_a^2 + (\partial_x \phi_a)^2 \right]+ \frac{1}{2} \int d^{d'} x \left[ \tilde{\pi}^*\tilde{\pi} + (\partial_x \zeta)(\partial_x \zeta^*) \right] \,, 
\end{align}
As discussed in Sec. \ref{sec:model}, we expand about the point $\mu=U=0$, where the propagators of $f_\alpha$, $b$ and $d$ all carry the same energy.
The constraint $f_\alpha^\dagger f_\alpha + b^\dagger b+ d^\dagger d = 1$ is imposed by taking the limit $\lambda \rightarrow \infty$ \cite{VojtaFritz04}. For a more detailed discussion about mapping to an impurity problem we refer the reader to Ref. \cite{Joshi2019}.



The impurity Hamiltonian in Eq. (\ref{Himp}) is now amenable to RG analysis similar to the problems of impurity in quantum magnets \cite{vbs, SS2001}, and Kondo models \cite{VojtaFritz04, FritzVojta04}.  
To start with, we can easily identify the tree-level scaling dimensions of the operators and vertices,
\begin{align}
&\text{dim} [\fa] = \text{dim} [\ba]= \text{dim} [d] = 0 \,; ~~~~ \text{dim} [\psi_{k\alpha}] = -\frac{1+r}{2} = -\text{dim} [\psi_{\alpha} (0)]\,; ~~~~ \text{dim} [g_{0}] =\frac{1-r}{2} \equiv \rb \,;   \nonumber \\
&\text{dim} [\gam] = \frac{3-d}{2} \equiv \frac{\ep}{2} \,; ~~~~ \text{dim} [v_0] = \frac{3-d'}{2} \equiv \frac{\ep'}{2} \,; ~~~~
\text{dim} [\pa] = \frac{d-1}{2} \,; ~~~~
\text{dim} [\zeta] = \frac{d'-1}{2} \,.
\label{eq:sca_dim}
\end{align}

Thus, $r=1$, $d=3$ and $d'=3$ are the {\it upper critical dimensions} around which we shall make a systematic expansion. More precisely, we will be using $\rb$, $\ep$ and $\ep'$ as our expansion parameters in the perturbative calculation. 

As a first step we define the renormalized fields and couplings in the following manner:
\begin{align}
&\fa_{\alpha} = \sqrt{\zf} \fa_{R \alpha} \,, ~~~~ \ba = \sqrt{\zb} \ba_{R} \,, ~~~~ d = \sqrt{Z_d} d_{R} \,,
\nonumber \\
&g_{0} = \frac{\rgs\up{\rb}\zgy }{\sqrt{\zf \zb}} g_1 \,, ~~~~ g_{0} = \frac{\rgs\up{\rb} \zge}{\sqrt{\zf Z_d}} g_2 \,, ~~~~ \gam = \frac{\rgs\up{\ep/2} \zgam}{\zf \sqrt{\tilde{S}_{d+1}}} \gamr \,, ~~~~ v_0 = \frac{\rgs\up{\ep'/2} Z_v}{ \sqrt{Z_b Z_d \tilde{S}_{d'+1}}} v \,,
\label{eq:renorm_fact}
\end{align}
where $\tilde{S}_d = \Gamma (d/2-1)/(4 \pi\up{d/2})$. 
Note that we have introduced two different renormalized couplings $g_1$ and $g_2$ corresponding to the two terms $f^\dagger b$ and $\varepsilon_{\alpha\beta} f_\beta d^\dagger $ respectively that couple to $\psi(0)$. However, in the end we shall see that these are equal as we must expect. 
Also note that the bath fields $\phi$, $\psi$, and $\zeta$ are not renormalized due to the absence of the respective interaction terms. 
In the following subsections we will evaluate the self energies and vertex corrections to one-loop order. These will determine the renormalization factors defined above. We will tune our model to the critical point by setting $U=0$ and subsequently derive the flow away from it. Also note that we work at zero temperature.


\subsection{Self energy of spinon $f$}
\label{sec:fse}

\begin{figure}[t]
\centering
\subfloat[]{\includegraphics[width=0.12\textwidth]{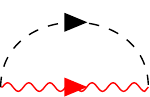}} ~~~
\subfloat[]{\includegraphics[width=0.12\textwidth]{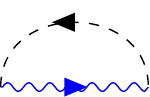}} ~~~
\subfloat[]{\includegraphics[width=0.12\textwidth]{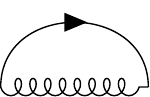}} ~~~
\subfloat[]{\includegraphics[width=0.12\textwidth]{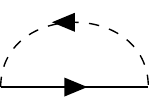}} ~~~
\subfloat[]{\includegraphics[width=0.12\textwidth]{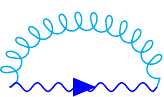}} ~~~
\subfloat[]{\includegraphics[width=0.12\textwidth]{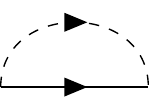}} ~~~
\subfloat[]{\includegraphics[width=0.12\textwidth]{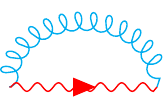}} 
\caption{One-loop self-energy diagrams. Spinon ($f$ fermion) self-energy diagrams are shown in (a)-(c), holon ($b$ boson) self energy is shown in (d-e), and doublon ($d$ boson) self energy is shown in (f-g). In these diagrams, solid line is for f propagator, dashed line is for $\psi$ propagator, red wavy is for b propagator, blue wavy is for d propagator, black spiral is for $\phi$ propagator, and cyan spiral is for $\zeta$ propagator.}
\label{fig:dia}
\end{figure}

We begin by calculating the self energy of spinon at one-loop level. There are three self-energy diagrams as shown in Fig. \ref{fig:dia}(a) - (c). These are evaluated below,
\begingroup
\allowdisplaybreaks
\begin{align}
\label{eq:se_f1}
\Sigma_{\ref{fig:dia} (a)} &= \gc\up{2}\int\frac{d\omega}{2\pi}\int_{-\infty}\up{\infty} dk \frac{|k|\up{r}}{-\iw +k} 
\frac{1}{-i(\nu - \omega) + \lam}
= -\gc\up{2} \Gamma(r)\Gamma(1-r)(-\inu+\lam)^r\nonumber\\
&= -A_{\mu1}g_1\up{2} \Gamma(r)\Gamma(1-r)(-\inu+\lam)\nonumber\\
&= A_{\mu1} g_1\up{2}\bigg[-\frac{1}{2 \rb}+1+O\left(\rb\right)\bigg] (-\inu+\lam) \hspace{0.04\textwidth} (\text{where, $A_{\mu1} = \mu\up{2\rb} (-\inu+\lam)\up{-2\rb} \frac{\zgy\up{2}}{\zf \zb}$}) \,, \\
\label{eq:se_f2}
\Sigma_{\ref{fig:dia} (b)} &= -\gc\up{2}\int\frac{d\omega}{2\pi}\int_{-\infty}\up{\infty} dk \frac{|k|\up{r}}{-\iw +k} 
\frac{1}{-i(\nu + \omega) + \lam}
= -\gc\up{2} \Gamma(r)\Gamma(1-r)(-\inu+\lam)^r\nonumber\\
&= -A_{\mu2}\gcr_2\up{2} \Gamma(r)\Gamma(1-r)(-\inu+\lam)\nonumber\\
&= -A_{\mu2} \gcr_2\up{2}\bigg[\frac{1}{2 \rb}-1+O\left(\rb\right)\bigg] (-\inu+\lam) \hspace{0.04\textwidth} (\text{where, $A_{\mu2} = \mu\up{2\rb} (-\inu+\lam)\up{-2\rb} \frac{\zge\up{2}}{\zf \zd}$}) \,, \\
\label{eq:se_f3}
\Sigma_{\ref{fig:dia} (c)} &= \gam\up{2} S(S+1) \int \frac{d\omega}{2\pi} \int \frac{d\up{d}k}{(2\pi)\up{d}} \frac{1}{\omega\up{2} +k\up{2}} 
\frac{1}{-(\inu + \iw) + \lam} = \gam\up{2} S(S+1) \frac{S_{d}}{2} \int_{0}\up{\infty} dk \frac{k\up{d-2} }{-(\inu -\lam) +k} \nonumber\\
&=  -\gam\up{2} S(S+1) \frac{S_{d}}{2} \Gamma(d)\Gamma(1-d) (\lam-\inu)\up{d-2} ~~~~~~~~~~~~~~~~~~
(\text {$d=3-\ep$}) \nonumber \\
&= B_{\mu} \gamr\up{2}S(S+1)\bigg[-\frac{1}{\ep}+N/2+O(\ep)\bigg]  (-\inu+\lam)  
\hspace{0.04\textwidth}(\text{where $B_{\mu} = \mu\up{\ep} (-\inu+\lam)\up{-\ep} \frac{\zgam\up{2}}{\zf\up{2} \zp}$}) \,.
\end{align}
\endgroup
In the momentum integral above, we have used 
$\int d\up{d}k = \frac{S_{d}}{2} \int dk k\up{d-1}$ with $S_d = 2/\Gamma(d/2) (4\pi)\up{d/2} $. Also, we used 
$\sum_\beta\epsilon_{\beta\alpha}^2=1$ in Eq. (\ref{eq:se_f2}), and the constant $N=\gamma_E-2\log[2]-\Gamma'(\frac{3}{2})/\Gamma(\frac{3}{2})=-0.8455686...$.

\subsection{Self energy of holon $b$}
\label{sec:bse}

The self energy of holon, $b$, involves two diagrams Fig. \ref{fig:dia}(d) - (e) at the one-loop level. They are evaluated here,
\begingroup
\allowdisplaybreaks
\begin{align}
\label{eq:se_f4}
\Sigma_{\ref{fig:dia} (d)} &= -\gc\up{2}\sum_\alpha \int_{-\infty}\up{\infty}\frac{dw}{2\pi}\int_{-\infty}\up{\infty} dk \frac{|k|\up{r} }{\iw -k}\frac{1}{\inu+\iw-\lam}=  -2\gc\up{2} \Gamma(r)\Gamma(1-r) (-\inu+\lam)\up{r} \nonumber\\
&= 2A_{\mu1} g_1\up{2}\bigg[-\frac{1}{2 \rb}+1+O\left(\rb\right)\bigg] (-\inu+\lam).\hspace{0.2\textwidth}\\
\label{eq:se_f5}
\Sigma_{\ref{fig:dia} (e)} &= v_0^2 \int_{-\infty}\up{\infty}\frac{dw}{2\pi}\int_{-\infty}\up{\infty} \frac{d^{d'}k}{(2\pi)^{d'}} \frac{1}{-\inu+\iw+\lam}\frac{1}{\omega^2+k^2}=  -v_0\up{2}\frac{S_d}{2} \Gamma(d')\Gamma(1-d') (-\inu+\lam)\up{d'-2} \nn
&= C_{\mu} v\up{2}\bigg[-\frac{1}{\ep'}+\gamma_E-1+O\left(\ep'\right)\bigg] (-\inu+\lam).\hspace{0.05\textwidth}(\text{where $C_{\mu} = \mu\up{\ep'} (-\inu+\lam)\up{-\ep'} \frac{Z_v\up{2}}{Z_bZ_d}$}) \,.
\end{align}
\endgroup 
Note that the self energy diagram Fig. \ref{fig:dia}(d) includes a factor of $\sum_\alpha 1=2$ due to the spin index of internal $f$-line. 


\subsection{Self energy of doublon $d$}
\label{dse}

The self energy of doublon, $d$, also involves two diagrams Fig. \ref{fig:dia}(f) - (g) at the one-loop level. They are evaluated below,
\begingroup
\allowdisplaybreaks
\begin{align}
\label{eq:se_f6}
\Sigma_{\ref{fig:dia} (f)} &= \gc\up{2}\sum_{\alpha\beta}\epsilon_{\alpha\beta}^2 \int\frac{d\omega}{2\pi}\int_{-\infty}\up{\infty} dk \frac{|k|\up{r}}{-\iw +k} 
\frac{1}{-i(\nu - \omega) + \lam}=  -2\gc\up{2} \Gamma(r)\Gamma(1-r) (-\inu+\lam)\up{r} \nonumber\\
&= 2A_{\mu2} g_2\up{2}\bigg[-\frac{1}{2 \rb}+1+O\left(\rb\right)\bigg] (-\inu+\lam).
\end{align}
\begin{align}
\label{eq:se_f7}
\Sigma_{\ref{fig:dia} (g)} &= v_0^2 \int_{-\infty}\up{\infty}\frac{dw}{2\pi}\int_{-\infty}\up{\infty} \frac{d^{d'}k}{(2\pi)^{d'}} \frac{1}{-i(\nu-\omega)+\lam}\frac{1}{\omega^2+k^2}=  -v_0\up{2}\frac{S_d}{2} \Gamma(d')\Gamma(1-d') (-\inu+\lam)\up{d'-2} \nn
&= C_{\mu} v\up{2}\bigg[-\frac{1}{\ep'}+\gamma_E-1+O\left(\ep'\right)\bigg] (-\inu+\lam).\hspace{0.2\textwidth}
\end{align}
\endgroup
Note that the self energy diagram Fig. \ref{fig:dia}(f) includes a factor of $\sum_{\alpha\beta}\epsilon_{\alpha\beta}^2 =2$ due to the spin index of internal $f$-line. 


\begin{figure}[t]
\centering
\subfloat[]{\includegraphics[width=0.25\textwidth]{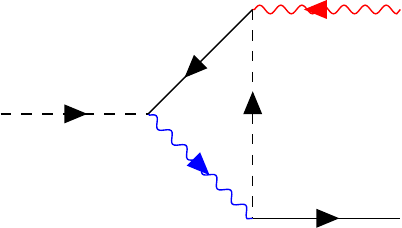}} ~~~
\subfloat[]{\includegraphics[width=0.25\textwidth]{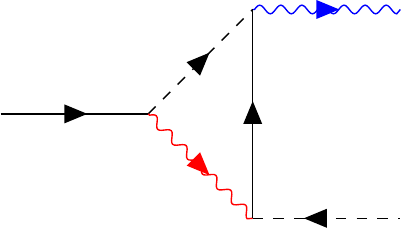}} ~~~
\subfloat[]{\includegraphics[width=0.25\textwidth]{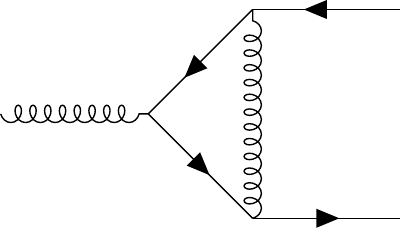}}
\caption{One-loop diagrams for vertex corrections. Vertex corrections to the fermionic bath coupling, $g_0$, are shown in (a) and (b), while that for bosonic bath coupling, $\gamma_0$, is shown in (c). The conventions are same as those used in Fig. (\ref{fig:dia}). }
\label{fig:ver}
\end{figure}

\subsection{Vertex correction}
\label{sec:vc}

Unlike in Ref. \cite{Joshi2019}, due to the presence of doublon in this case there is a one-loop correction to the vertex $\gc$ corresponding to the fermionic bath coupling. This is shown in Fig. \ref{fig:ver} (a) and (b), which is evaluated here,
\begingroup
\allowdisplaybreaks
\begin{align}
\label{eq:vertex_g1}
\Gamma_{\ref{fig:ver} (a)} &=\gc\up{3} \sum_\beta\epsilon_{\alpha\beta}\epsilon_{\beta\alpha}\int\frac{d\omega}{2\pi}\int dk|k|^r\frac{1}{-\iw_1+k}\frac{1}{-i(\omega+\omega_1+\nu)+\lambda}\frac{1}{-i(\omega_1+\nu)+\lambda }\nonumber\\
& = \gc A_{\mu2}g_2^2 (-1)[-\frac{1}{2\rb}+1+O(\rb)] \,, \\
\label{eq:vertex_g2}
\Gamma_{\ref{fig:ver} (b)} &=-\gc\up{3} \epsilon_{\alpha\beta}\int\frac{d\omega}{2\pi}\int dk|k|^r\frac{1}{-\iw_1+k}\frac{1}{-i(\omega-\omega_1+\nu)+\lambda}\frac{1}{-i(\omega-\omega_1)+\lambda }\hspace{0.06\textwidth}\nonumber\\
& = \gc\epsilon_{\beta\alpha} A_{\mu1}g_1^2 [\frac{1}{2\rb}-1+O(\rb)] \,.
\end{align}
\endgroup
The original vertex is $g_0\ep_{\beta\alpha}$.

The vertex correction to the bosonic bath coupling $\gam$, shown in Fig. \ref{fig:ver} (c), is same as that in Ref. \cite{Joshi2019} and evaluates to
\begin{align}
\Gamma_{\ref{fig:ver} (c)} &=\gam\up{3} (S\up{2} + S -1)\frac{S_d}{2}\Gamma(3-d)\Gamma(d-1)(-\inu+\lam)^{(d-3)}\hspace{0.22\textwidth}\nonumber\\
& = \gam B_{\mu} \gamr\up{2} (S\up{2} + S -1)[\frac{1}{\ep}-1-N/2+O(\ep)] \nonumber\\
&= -\gam B_{\mu} \gamr\up{2} \frac{1}{4}\bigg[\frac{1}{\ep}-1-N/2+O(\ep)\bigg]\,.
\label{eq:vertex_gam1}
\end{align}
Note that there is no vertex correction to the bosonic bath coupling $v_0$. 


\subsection{Beta functions}
\label{sec:beta_fbd}

Now that we have evaluated the self-energies and the vertex corrections we can obtain the renormalization factors defined in Eq. (\ref{eq:renorm_fact}). We choose the external frequency such that $-\inu+\lam=\mu$ and demand the cancellation of poles in the expressions for the renormalized vertices and Green's functions. Consequently we obtain,
\begin{align}
\label{eq:zf}
Z_f &= 1-\frac{g_1^2}{2\rb}-\frac{g_2^2}{2\rb}-\frac{3 \gamma^2}{4 \epsilon} \,, \\
\label{eq:zb}
Z_b&=1-\frac{g_1^2}{\rb}-\frac{v^2}{\ep'} \,, \\
\label{eq:zd}
Z_d&=1-\frac{g_2^2}{\rb}-\frac{v^2}{\ep'} \,, \\
\label{eq:zg1}
Z_{g_1}&=1-\frac{g_2^2}{2\rb} \,, \\
\label{eq:zg2}
Z_{g_2}&=1-\frac{g_1^2}{2\rb} \,, \\
\label{eq:zgam}
Z_{\gamma}&=1+\frac{\gamma^2}{4\epsilon} \,, \\
\label{eq:zv}
Z_{v} &= 1 \,.
\end{align}
Using the expressions for vertex renormalizations in Eq. (\ref{eq:renorm_fact}) and Eqs. (\ref{eq:zf}) - (\ref{eq:zv}) we find the beta functions,
\begin{align}
    \beta(g_1)&\equiv \mu \frac{d g_1}{d\mu} = -\rb g_1+\frac{1}{2}g_1^3+\frac{1}{2}g_1g_2^2+\frac{3}{8}g_1\gamr^2 +\frac{g_1\gvr^2}{2} \,, \\
    \beta(g_2)&\equiv \mu \frac{d g_2}{d\mu} = -\rb g_2+\frac{1}{2}g_1^2g_2+\frac{1}{2}g_2^3+\frac{3}{8}g_2\gamr^2 + \frac{g_2\gvr^2}{2} \,, \\
    \beta(\gamr)&\equiv \mu \frac{d \gamr}{d\mu} =-\frac{\ep}{2}\gamr+\gamr^3+g_1^2\gamr+g_2^2\gamr \,, \\
    \beta(\gvr)&\equiv \mu \frac{d \gvr}{d\mu} =-\frac{\epp}{2}\gvr+\gvr^3+g_1^2\gvr+g_2^2\gvr \,.
\end{align}
Note that the vertex corresponding to the fermionic bath is $g$ and so $g_1 = g_2 = g$. Therefore we have,
\begin{align}
\label{eq:betag}
\beta(g)  &= -\rb g +  g^3 + \frac{3}{8}g \gamr^2 + \frac{\gcr \gvr^{2}}{2}\,, \\
\label{eq:betagam}
\beta(\gamr) &= -\frac{\ep}{2}\gamr + \gamr^3 + 2 g^2 \gamr \,, \\
\label{eq:betav}
\betv &= -\frac{\epp}{2}\gvr + \gvr^3 + 2 g^2 \gvr \,.
\end{align}
These are identical to the beta functions obtained in Sec. \ref{sec:c_rg} where we used the electron operator instead of the fractionalized operators. 
The discussion of the fixed points is therefore same as in Sec. \ref{sec:beta_c}.

\subsection{Stability of fixed points}
\label{sec:fp_stab}

Here we discuss the stability of the fixed points (Eqs. (\ref{eq:fp1_c})-(\ref{eq:fp8_c})) by looking at the eigenvalues of the following stability matrix:
\begin{equation}
\label{eq_J_mat}
J \equiv 
\begin{bmatrix}
J_{1} & J_{2} & J_{3} \\
J_{4} & J_{5} & J_{6} \\
J_{7} & J_{8} & J_{9} 
\end{bmatrix} \,,
\end{equation}
where,
\begin{align}
\label{eq:Jdef}
J_{1} &\equiv \frac{\partial \betg}{\partial \gcr} = -\rb + 3 \gcr\up{2} + \frac{3}{8} \gamr\up{2} + \frac{\gvr^{2}}{2} \,, ~~~~
J_{2} \equiv \frac{\partial \betg}{\partial \gamr} = \frac{3}{4} \gcr \gamr \,, ~~~~
J_{3} \equiv \frac{\partial \betg}{\partial \gvr} = \gcr \gvr \,, \nonumber \\
J_{4} &\equiv \frac{\partial \betgam}{\partial \gcr} = 4 \gamr \gcr \,, ~~~~ 
J_{5} \equiv \frac{\partial \betgam}{\partial \gamr} = -\frac{\ep}{2} + 3 \gamr^{2} + 2 \gcr^{2} \,, ~~~~ 
J_{6} \equiv \frac{\partial \betgam}{\partial \gvr} = 0 \,, \nonumber \\ 
J_{7} &\equiv \frac{\partial \betv}{\partial \gcr} = 4 \gvr \gcr \,, ~~~~ 
J_{8} \equiv \frac{\partial \betv}{\partial \gamr} = 0 \,, ~~~~ 
J_{9} \equiv \frac{\partial \betv}{\partial \gvr} = -\frac{\epp}{2} + 3 \gvr^{2} + 2 \gcr^{2} \,.
\end{align}
We shall consider $\rb>0$, $\ep>0$ and $\epp>0$.
From the eigenvalues, 
\begin{equation}
\label{eq:ev_fp1}
E^{(1)}_{1} = -\rb \,, ~~~ E^{(1)}_{2} = -\frac{\ep}{2} \,, ~~~ E^{(2)}_{3} = -\frac{\epp}{2} \,,
\end{equation}
it is immediately clear that  the Gaussian fixed point $FP_{1}$ is unstable.
The fixed point $FP_2$ has the following eigenvalues:
\begin{equation}
\label{eq:ev_fp2}
E^{(2)}_{1} = 2\rb \,, ~~~ E^{(2)}_{2} = \frac{4\rb-\ep}{2} \,, ~~~ E^{(2)}_{3} = \frac{4\rb-\epp}{2} \,.
\end{equation}
So $FP_2$ is stable for any $\ep < 4\rb$ and $\epp < 4\rb$. The fixed point $FP_3$ is unstable and has one relevant direction as seen from the eigenvalues,
\begin{equation}
\label{eq:ev_fp3}
E^{(3)}_{1} = \ep \,, ~~~ E^{(3)}_{2} = \frac{3\ep - 16\rb}{2} \,, ~~~ E^{(3)}_{3} = -\frac{\epp}{2} \,.
\end{equation}
$FP_4$ is stable if $4\rb < \ep < 16\rb/3$ and $\epp < 16\rb - 3\ep$, which follows from its eigenvalues,
\begin{equation}
\label{eq:ev_fp4}
E^{(4)}_{1} = 5\ep - 16\rb - \Psi \,, ~~~ E^{(4)}_{2} = 5\ep - 16\rb + \Psi \,, ~~~ E^{(4)}_{3} = \frac{16\rb - 3\ep -\epp}{2} \,,
\end{equation}
where $\Psi = \sqrt{768 \rb^{2} - 384\rb \ep + 49 \ep^{2}}>0$. The fixed point $FP_5$ is unstable and has one relevant direction as is evident from the eigenvalues,
\begin{equation}
\label{eq:ev_fp5}
E^{(5)}_{1} = -\frac{\ep}{2}  \,, ~~~ E^{(5)}_{2} = \epp \,, ~~~ E^{(5)}_{3} = \frac{\epp - 4\rb}{4} \,.
\end{equation}
The fixed point $FP_6$ has the eigenvalues,
\begin{equation}
\label{eq:ev_fp6}
E^{(6)}_{1} = \ep \,, ~~~ E^{(6)}_{2} = \epp \,, ~~~ E^{(6)}_{3} = \frac{-16\rb + 3\ep + 4\epp}{16} \,.
\end{equation}
Thus it is stable if $-16\rb + 3\ep + 4\epp>0$.
We will discuss the eigenvalues at $FP_7$,
\begin{equation}
\label{eq:ev_fp7}
E^{(7)}_1 = \frac{4 \bar{g}^{2} - \ep}{2} \,, ~~~ E^{(7)}_2 = 0 \,, ~~~ 
E^{(7)}_3 = 16 \rb - 8 \bar{g}^{2} \,.
\end{equation}
It is then clear that $FP_7$ is marginally stable if $2\rb > \bar{g}^{2} > \ep/4$. The first inequality is trivially satisfied by the points on the ellipse describing $FP_7$. The second inequality is satisfied if $\ep < 4\rb$ (since $\bar{g}^2 < \rb$), which leads to atleast one stable fixed points. For lower values of $\ep$ there is a fixed line on the ellipse describing $FP_7$. 

The stability of $FP_8$ can be best studied by looking at the characteristic equation,
\begin{equation}
\label{eq:ev_fp8}
E^{3} + A E^2 + B E + C =0 \,,
\end{equation}
where $A=-J_1 - J_5 - J_9 = -2 (\gcr^{*2} + \gamr^{*2} + \gvr^{*2}) <0$, $B= J_1 J_5 + J_1 J_9 + J_5 J_9 - J_2 J_4 - J_3 J_7 = \gcr^{*2} \gamr^{*2} + 4 \gamr^{*2} \gvr^{*2} >0$, and $C= J_3 J_7 J_5 + J_2 J_4 J_9 - J_1 J_5 J_9 = 6 \gcr^{*2} \gamr^{*2} \gvr^{*2} >0$, with $\gcr^{*}, \gamr^{*}, \gvr^{*}$ being the values at $FP_8$. It is clear that $C = - E^{(8)}_{1} E^{(8)}_{2} E^{(8)}_{3}$, $B = E^{(8)}_{1} E^{(8)}_{2} + E^{(8)}_{1} E^{(8)}_{3} + E^{(8)}_{2} E^{(8)}_{3}$, and $A = - (E^{(8)}_{1} + E^{(8)}_{2} + E^{(8)}_{3})$, where $E^{(8)}_{i}$ are the eigenvalues. So, if $FP_8$ is real, one of its eigenvalue is negative and so it is an unstable fixed point. 

\subsection{Anomalous dimensions of spinon, holon and doublon operators}
\label{sec:anom_fbd}

We will now evaluate the anomalous dimensions of the fractionalized particles $f$, $b$, and $d$. Although, we must keep in mind that these are not gauge-invariant operators and these exponents are not physical observables. In terms of the renormalization factors, the anomalous dimensions are, 
\begin{equation}
\label{eq:adim2}
\eta_{f} = \left. \rgs \frac{d \ln \zf}{d \rgs} \right|_{FP} \,, 
~~~~~ \eta_{b} = \left. \rgs \frac{d \ln \zb}{d \rgs} \right|_{FP} \,,
~~~~~ \eta_{d} = \left. \rgs \frac{d \ln \zd}{d \rgs} \right|_{FP} \,.
\end{equation}
In terms of the coupling constants,
\begin{align}
\label{eq:etaf2}
\eta_{f} &= 2 \gcr^2 + \frac{3}{4} \gamr\up{2} \,, \\
\label{eq:etab2}
\eta_{b} &= 2 \gcr\up{2} + \gvr^{2} \,,\\
\label{eq:etad2}
\eta_{d} &= 2 \gcr\up{2} + \gvr^{2} \,.
\end{align}
At the fixed points, we obtain,
\begingroup 
\allowdisplaybreaks
\begin{align}
\label{eq:adim1_2}
&FP_1: \eta_{f} = 0 \,,~~~~ \eta_{b} = 0 \,,~~~~ \eta_{d} = 0 \,, \\
\label{eq:adim2_2}
&FP_2: \eta_{f} = 2\rb \,,~~~~ \eta_{b} = 2\rb \,,~~~~ \eta_{d} = 2\rb \,, \\
\label{eq:adim3_2}
&FP_3: \eta_{f} = \frac{3 \ep}{8} \,,~~~~ 
\eta_{b} = 0 \,,~~~~
\eta_{d} = 0 \,, \\
\label{eq:adim4_2}
&FP_4: \eta_{f} = 2\rb \,,~~~~ 
\eta_{b} = 8\rb -\frac{3 \ep}{2} \,,~~~~ 
\eta_{d} = 8\rb -\frac{3 \ep}{2} \,, \\
\label{eq:adim5_2}
&FP_5: \eta_{f} = 0 \,,~~~~ 
\eta_{b} = \frac{\epp}{2} \,,~~~~ 
\eta_{d} = \frac{\epp}{2} \,, \\
\label{eq:adim6_2}
&FP_6: \eta_{f} = \frac{3 \ep}{8} \,,~~~~ 
\eta_{b} = \frac{\epp}{2} \,,~~~~ 
\eta_{d} = \frac{\epp}{2} \,, \\
\label{eq:adim7_2}
&FP_7: \eta_{f} = 2 \bar{g}^{2} \,,~~~~ 
\eta_{b} = \frac{\epp}{2} \,,~~~~ 
\eta_{d} = \frac{\epp}{2} \,, \\
\label{eq:adim8_2}
&FP_8: \eta_{f} = -\frac{2\rb}{3} \,,~~~~ 
\eta_{b} = \frac{\epp}{2} \,,~~~~ 
\eta_{d} = \frac{\epp}{2} \,.
\end{align}
\endgroup 


\subsection{Anomalous dimensions of spin, electron and SC order-parameter operators}
\label{sec:anom_se}

Now we calculate the anomalous dimensions of the electron, spin, and SC order-parameter operators, which are gauge-invariant. These exponents are physical observables in scattering and spectroscopic experiments. Our aim is to evaluate the exponents corresponding to the spin correlator $\langle \vec{S}(\tau)\cdot\vec{S}(0) \rangle$, the electron correlator $\langle c_{\alpha}(\tau) c_{\alpha}\up\dagger(0) \rangle$, and the SC order-parameter correlator $\langle \Delta(\tau) \Delta\up{\dagger}(0) \rangle$. To proceed, we add a source term corresponding to these operators in the action,
\begin{equation}
\label{eq:S_comp2}
\mathcal{S}_c = \frac{1}{\beta}  \sum_{i \omega_n} \left( 
\LS \fd_{\alpha} \frac{\sigma\up{a}_{\alpha \beta}}{2} \fa_{\beta} 
+ \Lambda_{c} [\fd_{\alpha} \ba +\epsilon_{\alpha\beta}f_\beta d^\dagger+ H.c.] 
+ \Lambda_{\Delta} (\Delta + H.c.)
\right)\,.
\end{equation}
Here we have expressed the spin, electron, and SC order-parameter operators in terms of the fractionalized operators: $\hat{S} = \fd_{\alpha} ({\sigma\up{a}_{\alpha \beta}}/{2}) \fa_{\beta}$, 
$c_{\alpha}\up{\dagger} = \fd_{\alpha} \ba + \epsilon_{\alpha\beta}f_\beta d^\dagger$ and $\Delta=d^{\dagger} b$. In terms of these operators the spin, electron, and SC order-parameter are composite operators. 
Within the field-theoretic RG scheme, we have
\begin{equation}
\label{eq:lam_renorm2}
\LS = \frac{Z_{ff} \Lambda_{S,R}}{\zf} \,,~~~~~ \Lambda_{c} = \frac{Z_{fb} \Lambda_{c,R}}{\sqrt{\zf \zb}}= \frac{Z_{fd} \Lambda_{c,R}}{\sqrt{\zf \zd}} \,, ~~~~~ 
\Lambda_\Delta=\frac{Z_D}{\sqrt{Z_bZ_d}}\Lambda_{\Delta R} \,.
\end{equation}
The composite operators $\hat{S} = \fd_{\alpha} ({\sigma\up{a}_{\alpha \beta}}/{2}) \fa_{\beta}$,  
$c_{\alpha}\up{\dagger} = \fd_{\alpha} \ba + \epsilon_{\alpha\beta}f_\beta d^\dagger$, and $\Delta=d^{\dagger} b$ are renormalized as follows:
\begin{equation}
\label{eq:sc_renorm}
\hat{S} = \sqrt{Z_{S}} \hat{S}_{R} \,,~~~~~ c = \sqrt{Z_{c}} c_{R}  \,, ~~~~~ 
\Delta=\sqrt{Z_\Delta}\Delta_R \,.
\end{equation}
Just like in Ref. \cite{Joshi2019}, even in this case the diagrams involved in the above vertex corrections are exactly the same as those involved in the vertex corrections of $g$, $\gamr$, and $\gvr$ vertices evaluated in Sec. \ref{sec:vc}. Note that this is true at all orders in $\ep$, $\rb$, and $\epp$ and not just at one-loop level.
Setting $g_1=g_2=g$, we thus have,
\begin{align}
\label{eq:zs}
Z_{S} &= \left( \frac{\zf}{\zgam} \right)\up{2}=1-\frac{2\gamma^2}{\ep}-\frac{2g^2}{\rb} \,, \\ 
\label{eq:zc}
Z_{c} &= \frac{\zf \zb}{\zgy\up{2}}  = \frac{\zf \zd}{\zge\up{2}}=1-\frac{3\gamma^2}{4\ep}-\frac{g^2}{\rb} - \frac{\gvr^{2}}{\epp} \,, \\
\label{eq:zdel}
Z_\Delta &=\frac{Z_bZ_d}{Z_v^2} = 1-\frac{2g^2}{\rb} - \frac{2\gvr^{2}}{\epp} \,.
\end{align}
These renormalization factors are exactly the same as evaluated in Sec. \ref{sec:anom_c}. The consequent discussion of anomalous dimension is thus identical to that discussed in Sec. \ref{sec:anom_c}. 


\subsection{Density correlator}
\label{sec:anom_den}

Next, we shall now evaluate the correlator of the density $n = 1 + \dd \da - \bd \ba$. This quantity has been of recent interest and can be measured in a M-EELS experiment. The strategy to evaluate this quantity is the same as in the last two subsections via introducing a source term with vertex renormalization $Z_{nb}$ in the action. 
We find that at the one-loop level there are no vertex corrections to the correlator $\langle n \bd \ba \rangle$ to order $\gcr^{2}$, but there is a one-loop correction to order $\gvr^{2}$. Hence it is straightforward to see that 
\begin{equation}
Z_{n} = \left(\frac{\zb}{Z_{nb}}\right)\up{2} = \left( \frac{\zd}{Z_{nb}} \right)\up{2} = 1 - \frac{2\gcr\up{2}}{\rb} - \frac{4\gvr^{2}}{\epp} \,.
\end{equation}
Using this we can obtain the anomalous dimension for the density correlator,
\begin{equation}
\label{eq:etan}
\eta_{n} = \frac{d \ln Z_{n}}{d \ln \mu} = 4 \gcr\up{2} + 4 \gvr^{2} \,.
\end{equation}
Note that higher-order corrections in $\ep$, $\epp$ and $\rb$ will change the value of this exponent.
At the fixed points we have,
\begingroup
\allowdisplaybreaks
\begin{align}
&FP_1 : \eta_{n} = 0 \,,  \\
&FP_2 : \eta_{n} = 4\rb \,,  \\
&FP_3 : \eta_{n} = 0 \,,  \\
&FP_4 : \eta_{n} = 16\rb - 3\ep \,, \\
&FP_5 : \eta_{n} = 2\epp \,, \\
&FP_6 : \eta_{n} = 2\epp \,, \\ 
&FP_7 : \eta_{n} = 4\bar{\gcr}^{2} + 4 \bar{\gvr}^{2} \,, \\
&FP_8 : \eta_{n} = \frac{16\rb}{3} - \ep + \frac{2\epp}{3} \,.
\end{align}
\endgroup
At the fixed points $FP_6$ (Eq. (\ref{eq:fp6_c})) and $FP_8$ (Eq. (\ref{eq:fp8_c})), we have $\eta_n=2$ and $\eta_n = 7/3$, using the self-consistent values of $\ep=\epp=1$ and $\ep=\epp=2\rb=1$ respectively. There will be corrections to these values though.

\section{Solving the saddle point equations in large-$M$ analysis}
\label{sec:se_app}
We proceed by introducing a low-frequency ansatz for the Green's functions defined earlier. In terms of the imaginary time, $\tau$, such that $|\tau|\gg 1/J$ we write at $T=0$,
\bea 
G_a (\tau) = -\text{sgn}(\tau) \frac{C_a \Gamma (2 \Delta_a) \sin (\pi \Delta_a +\text{sgn}(\tau) \theta_a)}{\pi |\tau|^{2 \Delta_a}}\,, 
\label{Gftauf}
\eea
where parameters $C_{a}$, $\Delta_{a}$ and $\theta_{a}$ are real. To simplify the expression, we define 
\beq
C_a^\pm=\frac{C_a\Gamma(2\Delta_a)\sin(\pi\Delta_a\pm\text{sgn}(\tau)\theta_a)}{\pi} \,.
\eeq
Positivity constraints on the spectral densities impose restrictions on asymmetry angles:
\begin{equation}
    \label{eq:spec_pos_bh}
    -\pi\Delta_f<\theta_f<\pi\Delta_f\,,\quad\pi\Delta_b<\theta_b<\pi/2\,,\quad\pi\Delta_d<\theta_d<\pi/2\,.
\end{equation}
We can now collect the corresponding expressions for self energies at $|\tau|\gg 1/J$ and $T=0$ by plugging (\ref{Gftauf}) into (\ref{gomega1})-(\ref{gomega5}) and obtain,
\bea
\label{Sbtau}
\Sigma_b(\tau) &=& 
\displaystyle  -\text{sgn}(\tau)t^2\bigg( 
 \frac{C_b^+C_f^+C_f^-}{|\tau|^{4 \Delta_f + 2 \Delta_b}}- \frac{C_d^-{C_f^+}^2}{|\tau|^{4 \Delta_f + 2 \Delta_d}}\bigg)+\text{sgn}(\tau)k^2L^2\frac{ C_b^+C_d^+C_d^-}{|\tau|^{2\Delta_b+4\Delta_d}}\,, \\
\label{Sdtau}
\Sigma_d(\tau) &=& 
\displaystyle  -\text{sgn}(\tau)t^2 \bigg(
 \frac{C_d^+C_f^+C_f^-}{|\tau|^{4 \Delta_f + 2 \Delta_d}}- \frac{C_b^-{C_f^+}^2}{|\tau|^{4 \Delta_f + 2 \Delta_b}}\bigg)+\text{sgn}(\tau)k^2L^2\frac{ C_b^+C_b^-C_d^+}{|\tau|^{2\Delta_d+4\Delta_b}} \,, \\
 \label{Sftau}
\Sigma_f(\tau) &=& 
\displaystyle  \text{sgn}(\tau)kt^2\bigg(
 \frac{C_f^+C_b^+C_b^-}{|\tau|^{2 \Delta_f + 4 \Delta_b}}- \frac{2C_f^-C_b^+C_d^+}{|\tau|^{2 \Delta_f + 2 \Delta_b+ 2 \Delta_d}}+\frac{C_f^+C_d^+C_d^-}{|\tau|^{2 \Delta_f + 4 \Delta_d}}\bigg)-\text{sgn}(\tau)J^2 \frac{{C_f^+}^2C_f^-}{|\tau|^{6 \Delta_f}} \,.
\eea

Using the spectral representation of the self energies,
\beq
\Sigma (z) = \int_{-\infty}^{\infty} \frac{d \Omega}{\pi} \frac{\sigma (\Omega)}{z - \Omega} \,. \label{sspec}
\eeq
we perform the Laplace transforms at $T=0$, $|\Omega|\ll J$ and obtain\bea 
 \label{Sbomega}
\sigma_b(\Omega)=
\displaystyle  &~& \frac{\pi  t^2}{\Gamma(2 \Delta_b + 4 \Delta_f)}
 \frac{C_b^+C_f^+C_f^-}{|\Omega|^{1-2 \Delta_b - 4 \Delta_f}}-\frac{\pi  t^2}{\Gamma(2 \Delta_d + 4 \Delta_f)}
 \frac{C_d^-{C_f^+}^2}{|\Omega|^{1-2 \Delta_d - 4 \Delta_f}}-\frac{\pi  k^2L^2}{\Gamma(2 \Delta_b + 4 \Delta_d)}
 \frac{C_b^+C_d^+C_d^-}{|\Omega|^{1-2 \Delta_b - 4 \Delta_d}}\,,\nonumber\\
 \label{Sdomega}
 \sigma_d(\Omega)=
\displaystyle  &~& \frac{\pi  t^2}{\Gamma(2 \Delta_d + 4 \Delta_f)}
 \frac{C_d^+C_f^+C_f^-}{|\Omega|^{1-2 \Delta_d - 4 \Delta_f}}-\frac{\pi  t^2}{\Gamma(2 \Delta_b + 4 \Delta_f)}
 \frac{C_b^-{C_f^+}^2}{|\Omega|^{1-2 \Delta_b - 4 \Delta_f}}-\frac{\pi  k^2L^2}{\Gamma(2 \Delta_d + 4 \Delta_b)}
 \frac{C_b^+C_b^-C_d^+}{|\Omega|^{1-2 \Delta_d - 4 \Delta_b}}\,,\nonumber\\
 \label{Sfomega}
  \sigma_f(\Omega)=
\displaystyle  &~& -\frac{\pi k t^2}{\Gamma(2 \Delta_f + 4 \Delta_b)}
 \frac{C_f^+C_b^+C_b^-}{|\Omega|^{1-2 \Delta_f - 4 \Delta_b}}-\frac{\pi k t^2}{\Gamma(2 \Delta_f + 4 \Delta_d)}
 \frac{C_f^+C_d^+C_d^-}{|\Omega|^{1-2 \Delta_f - 4 \Delta_d}}\nonumber\\
 &~&+\frac{\pi k t^2}{\Gamma(2 \Delta_f + 2\Delta_b+2 \Delta_d)}
 \frac{2C_f^-C_b^+C_d^+}{|\Omega|^{1-2 \Delta_f - 2 \Delta_b-2\Delta_d}}+\frac{\pi J^2}{\Gamma(6 \Delta_f )}
 \frac{{C_f^+}^2C_f^-}{|\Omega|^{1-6 \Delta_f}} \,, 
\eea
where
\beq
C_a^\pm=\frac{C_a\Gamma(2\Delta_a)\sin(\pi\Delta_a\pm\text{sgn}(\Omega)\theta_a)}{\pi} \,.
\eeq

On the other hand, we can rewrite the saddle point equations \eqref{gaiomega} at $|\omega|\ll J$ and perform the analytic continuation to obtain
\bea 
 \label{sfacOmega}
\sigma_a(\Omega)=
\displaystyle  &~&\frac{1}{C_a}\frac{\sin(\pi\Delta_a+\text{sgn}(\Omega)\theta_a)}{|\Omega|^{2\Delta_a-1}}\,,\quad a = f,\, b,\, d\,.
 \eea
Comparison of \eqref{sfacOmega} with \eqref{Sbomega} for both positive and negative $\Omega$ leads to the solutions to the saddle point equations in Sec. \ref{sec:sol_largeM}.

\section{Large $M$ analysis with bosonic spinon + fermionic holon and doublon}
\label{app:bos_sp}

We now discuss the other fractionalization scheme, consisting of bosonic spinons $\mathfrak{b}_\alpha$ and fermionic holons $\mathfrak{f}_\ell$ and doublons $\mathfrak{d}_\ell$, 
\beq
c_{\ell \alpha} = \mathfrak{b}_\alpha \, \mathfrak{f}_\ell^\dagger + \mathcal{J}_{\alpha\beta} \mathfrak{b}^\dagger_\beta \, \mathfrak{d}_\ell\,,\quad 
\eeq
where $\mathcal{J}$ is the USp($M$) invariant tensor (see Section 2.2 of Ref.~\onlinecite{HouchesSS}), while the spin index $\alpha = 1 \ldots M$ ($M$ even) and the orbital index $\ell = 1 \ldots M'$. This representation has a U(1) gauge invariance,
\beq
\mathfrak{f}_{i\alpha}\rightarrow \mathfrak{f}_{i\alpha}e^{i\phi_i(\tau)},~~~~ \mathfrak{b}_{i\ell}\rightarrow \mathfrak{b}_{i\ell}e^{i\phi_i(\tau)},~~~~ \mathfrak{d}_{i\ell}\rightarrow \mathfrak{d}_{i\ell}e^{i\phi_i(\tau)} \,.
\label{u1gauge2}
\eeq
We also have the constraint fixing the U(1) gauge charge on each site,
\beq
\sum_{\alpha=1}^M \mathfrak{b}_{i \alpha}^\dagger \mathfrak{b}_{i \alpha} + \sum_{\ell=1}^{M'} \big(\mathfrak{f}_{i \ell}^\dagger \mathfrak{f}_{i \ell}+\mathfrak{d}_{i \ell}^\dagger \mathfrak{d}_{i \ell}\big) = \frac{M}{2}\,.
\label{constf}
\eeq

The large $M$ analysis will proceed as Sec. \ref{sec:largeM}, Refs. \cite{Fu2018} and \cite{Joshi2019}.  
We first take the $N \rightarrow \infty$ limit and perform disorder average, 
as discussed in Sec. \ref{sec:largeN} to obtain the single-site action in the large-$M$ limit,
\bea
\mathcal{Z} &=& \int \mathcal{D} \mathfrak{b}_\alpha \mathcal{D} \mathfrak{f}_\ell\mathcal{D} \mathfrak{d}_\ell \mathcal{D} \lambda
e^{- \mathcal{S}} \nn 
\label{eq:action_m2}
\mathcal{S} &=& \int_0^{1/T} d \tau \Biggl[ \sum_\ell \mathfrak{f}_{\ell}^\dagger \left( \frac{\partial}{\partial \tau} + \mu + \frac{U}{2}+i \lambda \right) \mathfrak{f}_{\ell}
+ \sum_\ell \mathfrak{d}_{\ell}^\dagger \left( \frac{\partial}{\partial \tau} - \mu + \frac{U}{2} + i \lambda \right) \mathfrak{d}_{\ell} \nonumber \\
&~&~~~~~~~~~~+ \sum_{\alpha} \mathfrak{b}_{\alpha}^\dagger \left( \frac{\partial}{\partial \tau} + i \lambda \right) \mathfrak{b}_{\alpha} - i \lambda \frac{M}{2} \Biggr] \nn 
&+&\frac{t^2}{M} \sum_{\ell,\alpha} \int_{0}^{1/T} d \tau d \tau' R (\tau - \tau') c_{\ell\alpha}^\dagger(\tau)c_{\ell\alpha}(\tau') \nn 
&-&\frac{J^2}{2M} \sum_{\alpha,\beta} \int_{0}^{1/T} d \tau d \tau'  Q(\tau-\tau') \mathfrak{b}_{\alpha}^\dagger (\tau) \mathfrak{b}_{\beta} (\tau) \mathfrak{b}_{\beta}^\dagger (\tau') \mathfrak{b}_{\alpha} (\tau') \nonumber \\
&-&\frac{L^2}{M'} \sum_{\ell \ell'} \int_{0}^{1/T} d \tau d \tau'  P(\tau-\tau') \mathfrak{f}_{\ell}^\dagger (\tau) \mathfrak{d}_{\ell'} (\tau) \mathfrak{d}_{\ell'}^\dagger (\tau') \mathfrak{f}_{\ell} (\tau') 
\,,
\label{Lp}
\eea
where $T$ is the temperature. 
In the large-$M$ limit the Lagrange multiplier, $\lambda (\tau)$ imposes the constraint in Eq. (\ref{constf}) and the self-consistency condition reads as follows:
\bea
R(\tau - \tau') &=& -\frac{1}{M M'} \sum_{\ell,\alpha}\left\langle c_{\ell\alpha}(\tau)c_{\ell\alpha}^\dagger(\tau')\right\rangle_\mathcal{Z}\,,\nn
Q(\tau - \tau') &=& \frac{1}{M^2} \sum_{\alpha,\beta} \left\langle \mathfrak{b}_{\alpha}^\dagger (\tau) \mathfrak{b}_{\beta} (\tau) \mathfrak{b}_{\beta}^\dagger (\tau') \mathfrak{b}_{\alpha} (\tau') \right\rangle_\mathcal{Z} \nonumber \\
P(\tau - \tau') &=& \frac{1}{M^{'2}} \sum_{\ell,\ell'} \left\langle \mathfrak{f}_{\ell} (\tau) \mathfrak{d}_{\ell'}^\dagger (\tau) \mathfrak{d}_{\ell'} (\tau') \mathfrak{f}_{\ell}^\dagger (\tau') \right\rangle_\mathcal{Z}\,.
\label{selfcon2}
\eea

\subsection{Saddle point equations}
Introducing bilocal fields 
\begin{align}
    G_\mathfrak{b}(\tau,\tau')=-\frac{1}{M}\sum_\alpha\mathfrak{b}_\alpha(\tau)\mathfrak{b}^\dagger_\alpha(\tau')\,,~~ G_\mathfrak{f}(\tau,\tau')=-\frac{1}{M'}\sum_\ell\mathfrak{f}_\ell(\tau)\mathfrak{f}^\dagger_\ell(\tau')\,,~~ G_\mathfrak{d}(\tau,\tau')=-\frac{1}{M'}\sum_\ell\mathfrak{d}_\ell(\tau)\mathfrak{d}^\dagger_\ell(\tau')
\end{align}
and the corresponding self-energies, we can obtain the saddle point equations:
\begin{align}
    \label{gfiomega2}
    G_\mathfrak{a}(i\omega)&=\frac{1}{i\omega-\mu_\mathfrak{a}-\Sigma_\mathfrak{a}(i\omega)}\,,\quad \mathfrak{a}=\mathfrak{f}, \mathfrak{b}, \mathfrak{d}\\
    \Sigma_\mathfrak{b}(\tau)&=kt^2R(-\tau)G_\mathfrak{d}(\tau)-kt^2R(\tau)G_\mathfrak{f}(\tau)+J^2Q(\tau)G_\mathfrak{b}(\tau)\,,\label{gomega12}\\
    \Sigma_\mathfrak{f}(\tau)&=t^2G_\mathfrak{b}(\tau)R(-\tau)+k^2L^2G_\mathfrak{d}(\tau)P(\tau)\,,\\
    \Sigma_\mathfrak{d}(\tau)&=-t^2R(\tau)G_\mathfrak{b}(\tau)+k^2L^2G_\mathfrak{f}(\tau)P(-\tau)\,.
    \label{gomega52}
\end{align}
Here $\mu_\mathfrak{a}$ are chemical potentials, determined by $U$ and the saddle point value of $\lambda$ to satisfy
\bea
\left\langle \mathfrak{f}^\dagger \mathfrak{f}\right\rangle &=& \delta_\mathfrak{f} \,, \\
\left\langle \mathfrak{b}^\dagger \mathfrak{b}\right\rangle &=& \delta_\mathfrak{b}\,, \\
\left\langle \mathfrak{d}^\dagger \mathfrak{d}\right\rangle &=& \delta_\mathfrak{d} \,, \\ 
\frac{2}{MM'} \sum_{\ell, \alpha} \left \langle c_{\ell \alpha}^\dagger c_{\ell \alpha} \right\rangle &=& n = 1-\delta = 1 + \delta_\mathfrak{d} - \delta_\mathfrak{f}\,,
\label{allconstraints2}
\eea
with the gauge-charge constraint (\ref{constf}) implying
\beq 
\delta_\mathfrak{b} + k (\delta_\mathfrak{f} + \delta_\mathfrak{d}) = \frac{1}{2}\,.
\label{constdelta2}
\eeq
The self-consistency equations are
\begin{align}
    R(\tau)&=G_\mathfrak{b}(\tau)G_\mathfrak{f}(-\tau)-G_\mathfrak{b}(-\tau)G_\mathfrak{d}(\tau)\,,\\
    Q(\tau)&=G_\mathfrak{b}(\tau)G_\mathfrak{b}(-\tau)\,,\\
    P(\tau)&=-G_\mathfrak{f}(\tau)G_\mathfrak{d}(-\tau)\,.
\end{align}

We solve Eqs. \eqref{gfiomega2}-\eqref{gomega52} in a similar procedure as in Sec. \ref{sec:largeM} and Appendix \ref{sec:se_app}. We introduce a low-frequency ansatz for the Green's functions in terms of the imaginary time $|\tau|\gg 1/J$ at $T=0$,
\bea 
G_a (\tau) = -\text{sgn}(\tau) \frac{C_a \Gamma (2 \Delta_a) \sin (\pi \Delta_a +\text{sgn}(\tau) \theta_a)}{\pi |\tau|^{2 \Delta_a}}\,, ~~~~a=\mathfrak{f},\,\mathfrak{b},\,\mathfrak{d}
\label{Gftau2}
\eea
where parameters $C_{a}$, $\Delta_{a}$ and $\theta_{a}$ are real and positivity constraints on the spectral densities yields  
\begin{equation}
    \label{eq:spec_pos_fh}
    -\pi\Delta_\mathfrak{f}<\theta_\mathfrak{f}<\pi\Delta_\mathfrak{f}\,,\quad\pi\Delta_\mathfrak{b}<\theta_\mathfrak{b}<\pi/2\,,\quad-\pi\Delta_\mathfrak{d}<\theta_\mathfrak{d}<\pi\Delta_\mathfrak{d}\,.
\end{equation}
Assuming $\Delta_{f} = \Delta_{b} = \Delta_{d} = 1/4$, the exponents in $t^2$, $J^2$ and $L^2$ terms in Eqs. \eqref{gomega12}-\eqref{gomega52} are equal, and so all terms are important in our low-frequency analysis. 
Combination of Eqs. \eqref{gfiomega2}-\eqref{gomega52} yields 
\bea
\label{solb1}
-t^2C_\mathfrak{b}^2\bigg[C_\mathfrak{f}^2\cos (2\theta_\mathfrak{b})-2C_\mathfrak{f}C_\mathfrak{d}\frac{\sin (\pi /4 - \theta_\mathfrak{d})\sin^2 (\pi /4+ \theta_\mathfrak{b})}{\sin (\pi /4+ \theta_\mathfrak{f})}\bigg]+k^2L^2C_\mathfrak{f}^2C_\mathfrak{d}^2\cos(2\theta_\mathfrak{d})&=\pi\,,\\
\label{sold1}
-t^2C_\mathfrak{b}^2\bigg[C_\mathfrak{d}^2\cos (2\theta_\mathfrak{b})-2C_\mathfrak{f}C_\mathfrak{d}\frac{\sin (\pi /4 - \theta_\mathfrak{f})\sin^2 (\pi /4+ \theta_\mathfrak{b})}{\sin (\pi /4+ \theta_\mathfrak{d})}\bigg]+k^2L^2C_\mathfrak{f}^2C_\mathfrak{d}^2\cos(2\theta_\mathfrak{f})&=\pi\,,
\eea
\bea
\label{solf1}
&& -J^2C_\mathfrak{b}^4\cos (2\theta_\mathfrak{b})+kt^2C_\mathfrak{b}^2\biggl[C_\mathfrak{f}^2\cos (2\theta_\mathfrak{f})+C_\mathfrak{d}^2\cos (2\theta_\mathfrak{d}) \nonumber \\
&&~~~~~~~~~~~~~-4C_\mathfrak{f}C_\mathfrak{d}\frac{\sin (\pi/4 - \theta_\mathfrak{b})\sin (\pi/4 + \theta_\mathfrak{f})\sin (\pi /4 + \theta_\mathfrak{d})}{\sin (\pi /4 + \theta_\mathfrak{b})}\biggr]=\pi\,,
\eea
and a restriction on asymmetry angles
\bea
\label{thetafb2}
\frac{\sin(\pi\Delta+\theta_{\mathfrak{f}})}{\sin(\pi\Delta-\theta_{\mathfrak{f}})}\frac{\sin(\pi\Delta-\theta_{\mathfrak{b}})}{\sin(\pi\Delta+\theta_{\mathfrak{b}})}=\frac{\sin(\pi\Delta+\theta_{\mathfrak{b}})}{\sin(\pi\Delta-\theta_{\mathfrak{b}})}\frac{\sin(\pi\Delta-\theta_{\mathfrak{d}})}{\sin(\pi\Delta+\theta_{\mathfrak{d}})}\,.
\eea
Notice the bounds $|\theta_\mathfrak{f}|<\pi/4$, $\pi/4<\theta_\mathfrak{b}<\pi/2$ and $|\theta_\mathfrak{d}|<\pi/4$, which leads to all the coefficients on the left-hand sides of Eqs. (\ref{solb1}) - (\ref{solf1}) being positive. Since $C_\mathfrak{f}$, $C_\mathfrak{b}$, $C_\mathfrak{d}$ are defined to be real positive numbers, we find another constraint for $\theta_\mathfrak{f}$, $\theta_\mathfrak{b}$ and $\theta_\mathfrak{d}$ at nonzero $t$, $J$, $L$:
\begin{equation}
    \label{eq:thetacut2}
    \cos{(2\theta_\mathfrak{b})}\leq -k\big[\cos{(2\theta_\mathfrak{f})}-\cos{(2\theta_\mathfrak{d})}\big]\leq -\cos{(2\theta_\mathfrak{b})}
\end{equation}
and at $L=0$:
\begin{equation}
    \label{eq:thetacut2l0}
    \cos{(2\theta_\mathfrak{b})}\leq -k\big[\cos{(2\theta_\mathfrak{f})}+\cos{(2\theta_\mathfrak{d})}\big]\leq -\cos{(2\theta_\mathfrak{b})}
\end{equation}

Therefore the parameters of the solution ($\theta_\mathfrak{d}$, $C_\mathfrak{f}$, $C_\mathfrak{b}$ and $C_\mathfrak{d}$) are fully determined by the two asymmetry angles $\theta_\mathfrak{f}$ and $\theta_\mathfrak{b}$. Note that the values of $\theta_\mathfrak{f}$ and $\theta_\mathfrak{b}$ are now related to the particle densities $\delta_\mathfrak{f}$, $\delta_\mathfrak{d}$ via the Luttinger constraints in analogy with Eqs. \eqref{eq:lr1}-\eqref{eq:lr3} of the other represention: 
\begin{align}
\label{eq:lr11}
    \frac{\theta_\mathfrak{f}}{\pi} + \left( \frac{1}{2} -\Delta_\mathfrak{f} \right) \frac{\sin (2 \theta_\mathfrak{f})}{\sin (2\pi \Delta_\mathfrak{f})} &=\frac{1}{2}- \delta_\mathfrak{f} \,, \\
    \label{eq:lr22}
    \frac{\theta_\mathfrak{b}}{\pi} + \left( \frac{1}{2} -\Delta_\mathfrak{b} \right) \frac{\sin (2 \theta_\mathfrak{b})}{\sin (2\pi \Delta_\mathfrak{b})} &= \frac{1}{2} + \delta_\mathfrak{b} \,, \\
    \label{eq:lr33}
    \frac{\theta_\mathfrak{d}}{\pi} + \left( \frac{1}{2} -\Delta_\mathfrak{d} \right) \frac{\sin (2 \theta_\mathfrak{d})}{\sin (2\pi \Delta_\mathfrak{d})} &= \frac{1}{2}- \delta_\mathfrak{d} \,. 
\end{align}
At half-filling with particle-hole symmetry, we have $n=1$ and $\delta_\mathfrak{f}=\delta_\mathfrak{d}$. Consequently, all parameters of the solution are fully determined at fixed doping density. So this large-$M$ theory also describes a critical point. 

\subsection{Conductivity}
The calculation follows the steps in Ref. \cite{GGS20}. We first generalize the Greens function to finite temperature similar to Eq. \eqref{Gatheta} in Sec. \ref{sec:con} with $\zeta_\mathfrak{b}=1$, $\zeta_\mathfrak{f}=\zeta_\mathfrak{d}=-1$. Here the subscript $a = \mathfrak{b},\mathfrak{f}, \mathfrak{d}$ is used for spinon, holon and doublon respectively.

We assume the system living on a Bethe lattice and consider both electron's and cooper pair's contribution to the residual resistivity:
\begin{equation}
\label{eq:resf0}
    \sigma_{0}^A =\frac{M'e^2t^2a^{2-d}}{2\pi z}\int d\omega A_c(\omega)^2\beta n_F(\omega)n_F(-\omega)+\frac{M'(2e)^2k L^2a^{2-d}}{2\pi z}\int d\omega A_\Delta(\omega)^2\beta n_B(\omega)n_B(-\omega) \,,
\end{equation}
where $a$ is lattice constant, $z$ is coordination number, $A_c(\omega)$ is electron spectral density and $A_\Delta(\omega)$ is Cooper pair spectral density,
\begin{align}
    \label{eq:Ac_anof}
    A_c(\omega)&=2\pi C e^{-\pi\mathcal{E}}\frac{\cosh(\beta\omega/2)}{\cosh(\beta\omega/2-\pi\mathcal{E})}+2\pi C' e^{-\pi\mathcal{E'}}\frac{\cosh(\beta\omega/2)}{\cosh(\beta\omega/2-\pi\mathcal{E'})} \,,\\
    \label{eq:Ad_anof}
    A_\Delta(\omega)&=-2\pi C_\Delta e^{-\pi\mathcal{E}_\Delta}\frac{\sinh(\beta\omega/2)}{\cosh(\beta\omega/2-\pi\mathcal{E}_\Delta)} \,.
\end{align}
Here the parameters are defined as $\mathcal{E}=\mathcal{E}_\mathfrak{b}-\mathcal{E}_\mathfrak{f}$, $\mathcal{E'}=\mathcal{E}_d-\mathcal{E}_\mathfrak{b}$,  $\mathcal{E}_\Delta=\mathcal{E}_\mathfrak{f}-\mathcal{E}_\mathfrak{d}$, $C=C_fC_b\sin(\theta_\mathfrak{f}-\pi/4)\sin(\theta_\mathfrak{b}+\pi/4)/\pi$, $C'=-C_\mathfrak{b}C_\mathfrak{d}\sin(\theta_\mathfrak{b}-\pi/4)\sin(\theta_\mathfrak{d}+\pi/4)/\pi$, $C_\Delta=-C_\mathfrak{f}C_\mathfrak{d}\sin(\theta_\mathfrak{d}-\pi/4)\sin(\theta_\mathfrak{f}+\pi/4)/\pi$.
The ratio in this fermionic holon representation then becomes
\begin{align}
    \mathcal{R}=\frac{\sigma_0^A}{\sigma_0^N}=\pi^2\bigg[t^2 \bigg(C^2 e^{-2\pi\mathcal{E}}+C'^2 e^{-2\pi\mathcal{E}'}+2C C' e^{-\pi(\mathcal{E}+\mathcal{E}')}\frac{\pi (\mathcal{E}-\mathcal{E}')}{\sinh{[\pi(\mathcal{E}-\mathcal{E}')]}}\bigg)+4kL^2C_\Delta^2 e^{-2\pi\mathcal{E}_\Delta}\bigg].
    \label{Ratio2}
\end{align}

We could transfer the bosonic holon scheme in Sec. \ref{sec:largeM} into the fermionic holon scheme by
\begin{equation}
    \theta_\mathfrak{f}=\frac{\pi}{2}-\theta_b\,,~~~ \theta_\mathfrak{b}=\frac{\pi}{2}-\theta_f\,.
\end{equation}
Therefore both schemes give the same ratio $\mathcal{R}$ at the particle-hole symmetric solution $\theta_f=0$, $\theta_b=\pi/2$.



\section{Gauge-invariant RG}
\label{app:gi_rg}

There is a third way to perform the renormalization group analysis of the impurity Hamiltonians, Eq. (\ref{Himp_c}) or (\ref{Himp}): this gives us yet another derivation of the same RG equations as those obtained in Section~\ref{sec:c_rg} and Appendix~\ref{app:fbd_rg}.

In this appendix, we proceed directly using the gauge-invariant operators \cite{vbs, Whitsitt17}. We will follow the strategy from Ref. \cite{vbs, Whitsitt17, Joshi2019}, which relies on explicit evaluation of operator traces rather than the Wick's theorem. More technical details can be found in Refs. \cite{vbs, Whitsitt17, Joshi2019}.  

Let us begin by introducing the following renormalization factors for the operators and couplings,
\begin{align}
\label{eq:renorm_fact_m}
S\up{a}&=\sqrt{Z_{S}} S\up{a}_{R}\,, ~~~ c_{\alpha}=\sqrt{Z_{c}}c_{R,\alpha} \,, ~~~ \Delta=\sqrt{Z_{\Delta}} \Delta_{R} \,, \nonumber \\  
\gam &= \frac{\mu\up{\ep/2} \tilde{Z}_{\gamma}}{\sqrt{Z_{S} \tilde{S}_{d+1}}} \gamr \,, ~~~
\gc= \frac{\mu\up{\rb} \tilde{Z}_{g}}{\sqrt{Z_{c} \Gamma(r+1)}} \gcr \,, ~~~ 
\gv= \frac{\mu\up{\epp/2} \tilde{Z}_{v}}{\sqrt{Z_{\Delta} \tilde{S}_{d'+1}}} \gvr \,.
\end{align}
We perform the calculation on the same lines as in Ref. \cite{Joshi2019}. Similar to the one used in Ref. \cite{Joshi2019}, we introduce the following notation that will be used in the calculation below. 
\begin{equation}\label{Iabc}
\mathcal{I}_{a,b,c} = \langle (\fd_\alpha \fa_\alpha)\up{a} (\bd \ba)\up{b} (\dd \da)\up{c} \rangle \,.
\end{equation}
For $M=2, M'=1$, it turns out that for all $a,b,c \geq 1$, $\mathcal{I}_{a,0,0}=1/2$, $\mathcal{I}_{0,b,0}=1/4$, $\mathcal{I}_{0,0,c}=1/4$, $\mathcal{I}_{a,b,0}=0$, $\mathcal{I}_{a,0,c}=0$, $\mathcal{I}_{0,b,c}=0$, and $\mathcal{I}_{a,b,c}=0$. We also note that the operators in (\ref{Iabc}) can be expressed in terms of the gauge-invariant electrons via
\begin{eqnarray}
\fd_\alpha f_\alpha &=& n_\uparrow + n_\downarrow - 2 n_\uparrow n_\downarrow\nonumber \\
\bd \ba &=& 1 - \left( n_\uparrow + n_\downarrow \right) + n_\uparrow n_\downarrow \nonumber \\
\dd \da &=& n_\uparrow n_\downarrow
\end{eqnarray}
where $n_\alpha = c_{\alpha}^\dagger c_\alpha$.

\subsection{Spin correlator}
\label{sec:spin_m}

Here we calculate the spin correlator, $\langle O_{1} \rangle \equiv \langle S\up{a}(\tau) S\up{a}(0) \rangle$, which will give us $Z_{S}$. As mentioned above we will not use the Wick's theorem and instead evaluate the numerator and the denominator in $\langle O_{1} \rangle = N_{1}/D$ separately. The numerator and denominator in $\langle O_{1} \rangle$ are 
\begin{align}
\label{eq:d}
D &= Tr \mathbbm{1} + \gam\up{2} \lo \left( \Da + \Db + \Dc \right) + \gc\up{2} \lop \left( \Dap + \Dbp + \Dcp \right) \nonumber \\
&~~~~~~~~~~+ \gc\up{2} \lopp \left( \Dapp + \Dbpp + \Dcpp \right) 
+\gv\up{2} (\lob + \lobp) \left( \Dab + \Dbb + \Dcb \right) \,, \\
\label{eq:n1}
N_{1} &= \lo + \gam\up{2} \left( \la\Da + \lb\Db + \lc\Dc \right) 
+ \gc\up{2} \left( \lap\Dap + \lbp\Dbp + \lcp\Dcp \right) \nonumber \\
&+ \gc\up{2} \left( \lapp\Dapp + \lbpp\Dbpp + \lcpp\Dcpp \right) 
+ \gv\up{2} \left( (\lab + \labp)\Dab + (\lbb+\lbbp)\Dbb + (\lcb+\lcbp)\Dcb \right)  \,,
\end{align}


In the above expressions,
\begingroup
\allowdisplaybreaks
\begin{align}
\lo &= \langle S\up{a} S\up{a} \rangle = \frac{3}{4} (2\mathcal{I}_{1,0,0} - \mathcal{I}_{2,0,0}) \,, \\
\lop &= \langle c_{\ell \alpha} c\up{\dagger}_{\ell \alpha} \rangle = - \mathcal{I}_{1,1,0} + 2 \mathcal{I}_{0,1,0}
+ \mathcal{I}_{1,0,0} + \mathcal{I}_{1,0,1}  \,, \\
\lopp &= \langle c\up{\dagger}_{\ell \alpha} c_{\ell \alpha} \rangle = \mathcal{I}_{1,0,0} + \mathcal{I}_{1,1,0} 
+ 2\mathcal{I}_{0,0,1} - \mathcal{I}_{1,0,1} \,, \\
\la &= \langle S\up{a} S\up{b} S\up{b} S\up{a} \rangle = \frac{9}{16}(4 \mathcal{I}_{2,0,0} 
+ \mathcal{I}_{4,0,0} - 4 \mathcal{I}_{3,0,0}) \,, \\
\lb &= \langle S\up{a} S\up{a} S\up{b} S\up{b} \rangle = \frac{9}{16}(4 \mathcal{I}_{2,0,0} 
+ \mathcal{I}_{4,0,0} - 4 \mathcal{I}_{3,0,0}) \,, \\
\lc &= \langle S\up{a} S\up{b} S\up{a} S\up{b} \rangle = \frac{3}{16}(-8\mathcal{I}_{1,0,0} + 16\mathcal{I}_{2,0,0} 
-12\mathcal{I}_{3,0,0} + 3\mathcal{I}_{4,0,0}) \,,\\
\lap &= \langle S\up{a} c_{\ell \alpha} c_{\ell \alpha}\up{\dagger} S\up{a}\rangle = -\frac{3}{4}(4\mathcal{I}_{2,1,0} 
- 4 \mathcal{I}_{1,1,0} - \mathcal{I}_{3,1,0} - 2\mathcal{I}_{2,0,1} + \mathcal{I}_{3,0,1} - 2\mathcal{I}_{2,0,0} 
+ \mathcal{I}_{3,0,0}) \,, \\
\lbp &= \langle S\up{a} S\up{a} c_{\ell \alpha} c\up{\dagger}_{\ell \alpha}\rangle = -\frac{3}{4}(4\mathcal{I}_{2,1,0} 
- 4 \mathcal{I}_{1,1,0} - \mathcal{I}_{3,1,0} - 2\mathcal{I}_{2,0,1} + \mathcal{I}_{3,0,1} - 2\mathcal{I}_{2,0,0} 
+ \mathcal{I}_{3,0,0}) \,, \\
\lcp &= \langle S\up{a} c_{\ell \alpha} S\up{a} c\up{\dagger}_{\ell \alpha} \rangle = -\frac{3}{4}(3\mathcal{I}_{2,1,0} 
- 2 \mathcal{I}_{1,1,0} - \mathcal{I}_{3,1,0} - 3\mathcal{I}_{2,0,0} + \mathcal{I}_{3,0,0} + 2\mathcal{I}_{1,0,0} 
- 3\mathcal{I}_{2,0,1}  \nonumber \\ 
&~~~~~~~~~~~~~~~~~~~~~~~~~~~+ \mathcal{I}_{3,0,1} + 2\mathcal{I}_{1,0,1}) \,, \\
\lapp &= \langle S\up{a} c\up{\dagger}_{\ell \alpha} c_{\ell \alpha} S\up{a} \rangle = \frac{3}{4}(2 \mathcal{I}_{2,0,0} 
- \mathcal{I}_{3,0,0} + 2\mathcal{I}_{2,1,0} - \mathcal{I}_{3,1,0} 
+ 4\mathcal{I}_{1,0,1} + \mathcal{I}_{3,0,1} - 4\mathcal{I}_{2,0,1}) \,, \\
\lbpp &= \langle S\up{a} S\up{a} c\up{\dagger}_{\ell \alpha} c_{\ell \alpha}\rangle = \frac{3}{4}(2 \mathcal{I}_{2,0,0} 
- \mathcal{I}_{3,0,0} + 2\mathcal{I}_{2,1,0} - \mathcal{I}_{3,1,0} 
+ 4\mathcal{I}_{1,0,1} + \mathcal{I}_{3,0,1} - 4\mathcal{I}_{2,0,1}) \,, \\
\lcpp &= \langle S\up{a} c\up{\dagger}_{\ell \alpha} S\up{a} c_{\ell \alpha} \rangle = \frac{3}{4}(3 \mathcal{I}_{2,0,0} 
- 2\mathcal{I}_{1,0,0} - \mathcal{I}_{3,0,0} + 3\mathcal{I}_{2,1,0} - 2\mathcal{I}_{1,1,0} -\mathcal{I}_{3,1,0} 
 + 2\mathcal{I}_{1,0,1}  \nonumber \\
 &~~~~~~~~~~~~~~~~~~~~~~~~~~~- 3\mathcal{I}_{2,0,1} + \mathcal{I}_{3,0,1} ) \,, \\
\lab &= \langle S^{a} \Delta \Delta^{\dagger} S^{a} \rangle = \frac{3}{4} (2 \mathcal{I}_{1,0,1} + 2 \mathcal{I}_{1,1,1} - \mathcal{I}_{2,0,1} - \mathcal{I}_{2,1,1} ) \,, \\
\labp &= \langle S^{a} \Delta^{\dagger} \Delta S^{a} \rangle = \frac{3}{4} (2 \mathcal{I}_{1,1,0} + 2 \mathcal{I}_{1,1,1} - \mathcal{I}_{2,1,0} - \mathcal{I}_{2,1,1} ) \,, \\ 
\lbb &= \langle S^{a} S^{a} \Delta \Delta^{\dagger} \rangle = \lab \,, \\
\lbbp &= \langle S^{a} S^{a} \Delta^{\dagger} \Delta \rangle = \labp \,, \\
\lcb &= \langle S^{a} \Delta S^{a} \Delta^{\dagger} \rangle = \lab \,, \\ 
\lcbp &= \langle S^{a} \Delta^{\dagger} S^{a} \Delta \rangle = \labp \,.
\end{align}
\endgroup


Also,
\begingroup
\allowdisplaybreaks
\begin{align}
\Da &= \int_{0}\up{\tau} d\tau_{1} \int_{\tau_{1}}\up{\tau} d\tau_{2} G_{\phi} (\tau_{1} - \tau_{2})  
= - \frac{\widetilde{S}_{d+1} \tau\up{\ep}}{\ep (1-\ep)} \,, \\
\Db &= \int_{\tau}\up{\beta} d\tau_{1} \int_{\tau_{1}}\up{\beta} d\tau_{2} G_{\phi} (\tau_{1} - \tau_{2}) 
= - \frac{\widetilde{S}_{d+1} \tau\up{\ep}}{\ep (1-\ep)} \,, \\
\Dc &= \int_{0}\up{\tau} d\tau_{1} \int_{\tau}\up{\beta} d\tau_{2} G_{\phi} (\tau_{1} - \tau_{2})  
= \frac{2 \widetilde{S}_{d+1} \tau\up{\ep}}{\ep (1-\ep)} \,, \\
%
\Dap &= \int_{0}\up{\tau} d\tau_{1} \int_{\tau_{1}}\up{\tau} d\tau_{2} G_{\psi} (\tau_{2} - \tau_{1})  
= - \frac{\Gamma(r+1) \tau\up{2\rb}}{2\rb (1-2\rb)} \,, \\
\Dbp &= \int_{\tau}\up{\beta} d\tau_{1} \int_{\tau_{1}}\up{\beta} d\tau_{2} G_{\psi} (\tau_{2} - \tau_{1}) 
= - \frac{\Gamma(r+1) \tau\up{2\rb}}{2\rb (1-2\rb)} \,, \\
\Dcp &= \int_{0}\up{\tau} d\tau_{1} \int_{\tau}\up{\beta} d\tau_{2} G_{\psi} (\tau_{2} - \tau_{1}) 
= \frac{2 \Gamma(r+1) \tau\up{2\rb}}{2\rb (1-2\rb)} \,, \\
\Dapp &= -\int_{0}\up{\tau} d\tau_{1} \int_{\tau_{1}}\up{\tau} d\tau_{2} G_{\psi} (\tau_{1} - \tau_{2})  
= - \frac{\Gamma(r+1) \tau\up{2\rb}}{2\rb (1-2\rb)} \,, \\
\Dbpp &= -\int_{\tau}\up{\beta} d\tau_{1} \int_{\tau_{1}}\up{\beta} d\tau_{2} G_{\psi} (\tau_{1} - \tau_{2}) 
= - \frac{\Gamma(r+1) \tau\up{2\rb}}{2\rb (1-2\rb)} \,, \\
\Dcpp &= -\int_{0}\up{\tau} d\tau_{1} \int_{\tau}\up{\beta} d\tau_{2} G_{\psi} (\tau_{1} - \tau_{2}) 
=  \frac{2\Gamma(r+1) \tau\up{2\rb}}{2\rb (1-2\rb)} \,, \\
\Dab &= \int_{0}\up{\tau} d\tau_{1} \int_{\tau_{1}}\up{\tau} d\tau_{2} G_{\zeta} (\tau_{1} - \tau_{2})  
= - \frac{\widetilde{S}_{d'+1} \tau\up{\epp}}{\epp (1-\epp)} \,, \\
\Dbb &= \int_{\tau}\up{\beta} d\tau_{1} \int_{\tau_{1}}\up{\beta} d\tau_{2} G_{\zeta} (\tau_{1} - \tau_{2}) 
= - \frac{\widetilde{S}_{d'+1} \tau\up{\epp}}{\epp (1-\epp)} \,, \\
\Dcb &= \int_{0}\up{\tau} d\tau_{1} \int_{\tau}\up{\beta} d\tau_{2} G_{\zeta} (\tau_{1} - \tau_{2})  
= \frac{2 \widetilde{S}_{d'+1} \tau\up{\epp}}{\epp (1-\epp)} \,,
\end{align}
\endgroup 
Note that we have evaluated the above integrals at $T=0$, by extending the integrals appropriately as explained in Ref. \cite{vbs}. Here,
\begin{equation}
G_{\phi}(\tau) = \int \frac{d\up{d}k}{(2\pi)\up{d}} \frac{d\omega}{2\pi} \frac{e\up{-i\omega \tau}}{k\up{2} + \omega\up{2}} 
= \frac{\widetilde{S}_{d+1}}{|\tau|\up{d-1}} \,.
\end{equation}
Similarly,
\begin{align}
G_{\psi}(\tau) &= \int dk |k|\up{r} \int \frac{d\omega}{2\pi} \frac{e\up{-i\omega \tau}}{i\omega - k} \nonumber \\
&= \int dk |k|\up{r} \left[ -e\up{-k\tau} \left( \theta(k) \theta(\tau) - \theta(-k) \theta(-\tau) \right)  \right] \nonumber \\
&= \frac{\Gamma(1+r)}{|\tau|\up{1+r}} \left[ \theta(-\tau) - \theta(\tau) \right] \,.
\end{align}

From Eqs. \ref{eq:d} and \ref{eq:n1} we obtain,
\begin{align}
\langle O_{1} \rangle = \frac{N_{1}}{D} &= \lo \bigg \lbrace 
1 + \gam\up{2} \left[ \left( \frac{\la}{\lo} - \lo \right) \Da 
+ \left( \frac{\lb}{\lo} - \lo \right) \Db 
+ \left( \frac{\lc}{\lo} - \lo \right) \Dc \right]  \nonumber \\
&+ \gc\up{2} \left[ \left( \frac{\lap}{\lo} - \lop \right) \Dap 
+ \left( \frac{\lbp}{\lo} - \lop \right) \Dbp 
+ \left( \frac{\lcp}{\lo} - \lop \right) \Dcp \right] \nonumber \\
&+ \gc\up{2} \left[ \left( \frac{\lapp}{\lo} - \lopp \right) \Dapp 
+ \left( \frac{\lbpp}{\lo} - \lopp \right) \Dbpp 
+ \left( \frac{\lcpp}{\lo} - \lopp \right) \Dcpp \right] \nonumber \\
&+ \gv\up{2} \left[ \left( \frac{\lab + \labp}{\lo} - \lob - \lobp \right) \Dab 
+ \left( \frac{\lbb + \lbbp}{\lo} - \lob - \lobp \right) \Dbb 
+ \left( \frac{\lcb + \lcbp}{\lo} - \lob - \lobp \right) \Dcb \right]
\bigg \rbrace \,.
\end{align}
It is then straightforward to identify, 
\begin{equation}
\label{eq:zs_m}
Z_{S} = 1 - \frac{\gamr\up{2}}{\ep} \Lgam - \frac{\gcr\up{2}}{2\rb} \Lg  
- \frac{\gvr^{2}}{\epp} \Lv \,,
\end{equation}
where,
\begin{align}
\Lgam &= \frac{\la + \lb -2\lc}{\lo} =2 \,, \\
\Lg &= \frac{\lap+\lapp+\lbp+\lbpp-2\lcp-2\lcpp}{\lo} = 4\,, \\
\Lv &= \frac{\lab+\labp+\lbb+\lbbp-2\lcb-2\lcbp}{\lo} = 0 \,.
\end{align}
We thus have,
\begin{equation}
Z_{S} = 1 - \frac{2 \gamr\up{2}}{\ep} - \frac{2 \gcr\up{2}}{\rb}  \,.
\end{equation}
This is exactly as obtained in Sec. \ref{sec:anom_se}. 

\subsection{Electron correlator}
\label{sec:elec_m}

Next we evaluate the electron correlation, $\langle O_{2} \rangle \equiv \langle c(\tau) c\up{\dagger}(0) \rangle = N_{2}/D$. 
We have, 

\begin{align}
\label{eq:n2}
N_{2} &= \po + \gam\up{2} \left( \Pa\Da + \pb\Db + \pc\Dc \right) 
+ \gc\up{2} \left( \pap\Dap + \pbp\Dbp + \pcp\Dcp \right) \nonumber \\
&~~~~~~~+ \gc\up{2} \left( \papp\Dapp + \pbpp\Dbpp + \pcpp\Dcpp \right) 
+ \gv\up{2} \left( (\pab+\pabp)\Dab + (\pbb+\pbbp)\Dbb + (\pcb+\pcbp)\Dcb \right)
\,, 
\end{align}
where,
\begingroup
\allowdisplaybreaks
\begin{align}
\po &= \langle c\up{\dagger}_{\ell \alpha} c_{\ell \alpha} \rangle = \mathcal{I}_{1,0,0} + \mathcal{I}_{1,1,0} + 2 \mathcal{I}_{0,0,1} - \mathcal{I}_{1,0,1} \,,\\
\Pa &= \langle c\up{\dagger}_{\ell \alpha} S\up{a} S\up{a} c_{\ell \alpha} \rangle = \frac{3}{4}(4\mathcal{I}_{2,0,0} 
-3\mathcal{I}_{1,0,0} - \mathcal{I}_{3,0,0} + 4\mathcal{I}_{2,1,0} - 3\mathcal{I}_{1,1,0} - \mathcal{I}_{3,1,0} \nonumber \\ 
&~~~~~~~~~~~~~~~~~~~~~~~~~~~ + 2\mathcal{I}_{0,0,1} - \mathcal{I}_{1,0,1} - 2 \mathcal{I}_{2,0,1} + \mathcal{I}_{3,0,1}) \,, \\
\pb &= \langle c\up{\dagger}_{\ell \alpha} c_{\ell \alpha} S\up{a} S\up{a}\rangle = \frac{3}{4}(2\mathcal{I}_{2,0,0} 
- \mathcal{I}_{3,0,0} + 2\mathcal{I}_{2,1,0} -\mathcal{I}_{3,1,0} 
+ 4\mathcal{I}_{1,0,1} - 4\mathcal{I}_{2,0,1} + \mathcal{I}_{3,0,1} ) \,, \\
\pc &= \langle c\up{\dagger}_{\ell \alpha} S\up{a} c_{\ell \alpha} S\up{a} \rangle = \frac{3}{4}(3\mathcal{I}_{2,0,0} 
- 2\mathcal{I}_{1,0,0} - \mathcal{I}_{3,0,0} + 3\mathcal{I}_{2,1,0} - 2\mathcal{I}_{1,1,0} -\mathcal{I}_{3,1,0} 
+ 2\mathcal{I}_{1,0,1} \nonumber \\ 
&~~~~~~~~~~~~~~~~~~~~~~~~~~~- 3\mathcal{I}_{2,0,1} + \mathcal{I}_{3,0,1} ) \,, \\
\pap &= \langle c\up{\dagger}_{\ell \alpha} c_{\ell'\beta} c_{\ell'\beta}\up{\dagger} c_{\ell\alpha}\rangle =
3 \mathcal{I}_{1,0,0} 
- \mathcal{I}_{2,0,0} + 6 \mathcal{I}_{1,1,0} + 3 \mathcal{I}_{1,2,0} - 2 \mathcal{I}_{2,1,0} - \mathcal{I}_{2,2,0} 
+4 \mathcal{I}_{1,0,1} - 2 \mathcal{I}_{2,0,1}   \nonumber \\
&~~~~~~~~~~~~~~~~~~~~~~~+ 4 \mathcal{I}_{1,1,1} - 2 \mathcal{I}_{2,1,1} + \mathcal{I}_{2,0,0} + \mathcal{I}_{2,1,0} + \mathcal{I}_{2,0,1} + \mathcal{I}_{2,1,1} - \mathcal{I}_{1,0,0} - \mathcal{I}_{1,1,0} - \mathcal{I}_{1,0,1}   \nonumber \\
&~~~~~~~~~~~~~~~~~~~~~~~- \mathcal{I}_{1,1,1} 
+ 2 \mathcal{I}_{0,1,1} - 3 \mathcal{I}_{1,1,1} + \mathcal{I}_{2,1,1} + 4 \mathcal{I}_{1,0,1} + 4 \mathcal{I}_{1,1,1} - 2 \mathcal{I}_{2,0,1} - 2 \mathcal{I}_{2,1,1} \nonumber \\ 
&~~~~~~~~~~~~~~~~~~~~~~~+ 2 \mathcal{I}_{0,0,2} + \mathcal{I}_{1,0,2} - \mathcal{I}_{2,0,2} \,, \\
\pbp &= \langle c\up{\dagger}_{\ell\alpha} c_{\ell\alpha} c_{\ell'\beta} c\up{\dagger}_{\ell'\beta} \rangle =  2 \mathcal{I}_{1,1,0} 
+ 2 \mathcal{I}_{1,2,0} - \mathcal{I}_{2,1,0} - \mathcal{I}_{2,2,0} + \mathcal{I}_{2,0,0} + \mathcal{I}_{2,1,0} + \mathcal{I}_{2,0,1} + \mathcal{I}_{2,1,1}  \nonumber \\ 
&~~~~~~~~~~~~~~~~~~~~~~~+ 2 \mathcal{I}_{1,0,0} + 2 \mathcal{I}_{1,1,0} + 2 \mathcal{I}_{1,0,1} + 2 \mathcal{I}_{1,1,1} - 2 \mathcal{I}_{2,0,0} - 2 \mathcal{I}_{2,1,0} - 2 \mathcal{I}_{2,0,1} - 2 \mathcal{I}_{2,1,1}  \nonumber \\ &~~~~~~~~~~~~~~~~~~~~~~~-4 \mathcal{I}_{0,1,1} + 6 \mathcal{I}_{1,1,1} - 2 \mathcal{I}_{2,1,1} + \mathcal{I}_{2,1,1} - 4 \mathcal{I}_{1,1,1} + 4 \mathcal{I}_{0,1,1} + 2 \mathcal{I}_{1,0,2} -  \mathcal{I}_{2,0,2}  \,, \\
\pcp &= -\langle c\up{\dagger}_{\ell\alpha} c_{\ell'\beta} c_{\ell\alpha} c\up{\dagger}_{\ell'\beta}\rangle  = 2 \mathcal{I}_{1,1,0} 
+ 2 \mathcal{I}_{1,2,0} - \mathcal{I}_{2,1,0} - \mathcal{I}_{2,2,0} + \mathcal{I}_{2,0,0} + \mathcal{I}_{2,1,0} + \mathcal{I}_{2,0,1} + \mathcal{I}_{2,1,1} + 2 \mathcal{I}_{1,0,0}  \nonumber \\ 
&~~~~~~~~~~~~~~~~~~~~~~~+ 2 \mathcal{I}_{1,1,0} + 2 \mathcal{I}_{1,0,1} + 2 \mathcal{I}_{1,1,1} - 2 \mathcal{I}_{2,0,0} - 2 \mathcal{I}_{2,1,0} - 2 \mathcal{I}_{2,0,1} - 2 \mathcal{I}_{2,1,1}  -4 \mathcal{I}_{0,1,1} \nonumber \\ &~~~~~~~~~~~~~~~~~~~~~~~+ 6 \mathcal{I}_{1,1,1} - 2 \mathcal{I}_{2,1,1} + \mathcal{I}_{2,1,1} - 4 \mathcal{I}_{1,1,1} + 4 \mathcal{I}_{0,1,1} + 2 \mathcal{I}_{1,0,2} -  \mathcal{I}_{2,0,2} \,, \\
\papp &= \langle c\up{\dagger}_{\ell\alpha} c_{\ell'\beta}\up{\dagger} c_{\ell'\beta} c_{\ell\alpha}\rangle = 
2(\mathcal{I}_{2,0,0}-\mathcal{I}_{1,0,0}) + 3 (\mathcal{I}_{2,1,0} -\mathcal{I}_{1,1,0}) - \mathcal{I}_{1,2,0} + \mathcal{I}_{2,2,0}  \nonumber \\
&~~~~~~~~~~~~~~~~~~~~~~~+ 2 \mathcal{I}_{2,0,1} + 2 \mathcal{I}_{2,1,1} - 4 \mathcal{I}_{1,0,1} - 4 \mathcal{I}_{1,1,1} + 3 \mathcal{I}_{1,0,1} + 3 \mathcal{I}_{1,1,1} - \mathcal{I}_{2,0,1} - \mathcal{I}_{2,1,1} \nonumber \\
&~~~~~~~~~~~~~~~~~~~~~~~+ 2 \mathcal{I}_{0,0,1} + 2 \mathcal{I}_{0,1,1} + \mathcal{I}_{1,0,1} + \mathcal{I}_{1,1,1} - \mathcal{I}_{2,0,1} - \mathcal{I}_{2,1,1} \nonumber \\ 
&~~~~~~~~~~~~~~~~~~~~~~~+2 \mathcal{I}_{2,0,1} + 2 \mathcal{I}_{2,1,1} - 4 \mathcal{I}_{1,0,1} - 4 \mathcal{I}_{1,1,1} \nonumber \\ 
&~~~~~~~~~~~~~~~~~~~~~~~-2 \mathcal{I}_{0,0,1} + 2 \mathcal{I}_{0,0,2} + 3 \mathcal{I}_{1,0,1} - 3 \mathcal{I}_{1,0,2} - \mathcal{I}_{2,0,1} + \mathcal{I}_{2,0,2}  \,, \\
\pbpp &= \langle c\up{\dagger}_{\ell\alpha} c_{\ell\alpha} c\up{\dagger}_{\ell'\beta} c_{\ell'\beta} \rangle = 
\mathcal{I}_{2,0,0} + 2 \mathcal{I}_{2,1,0} + \mathcal{I}_{2,2,0} + 2 \mathcal{I}_{1,0,1} + 2 \mathcal{I}_{1,1,1} - \mathcal{I}_{2,0,1} - \mathcal{I}_{2,1,1}  \nonumber \\ 
&~~~~~~~~~~~~~~~~~~~~~~~+ 2 \mathcal{I}_{2,0,0} + 2 \mathcal{I}_{2,1,0} + 2 \mathcal{I}_{2,0,1} + 2 \mathcal{I}_{2,1,1} - 2 \mathcal{I}_{1,0,0} - 2 \mathcal{I}_{1,1,0} - 2 \mathcal{I}_{1,0,1} - 2 \mathcal{I}_{1,1,1} \nonumber \\
&~~~~~~~~~~~~~~~~~~~~~~~+ 4 \mathcal{I}_{0,1,1} - 6 \mathcal{I}_{1,1,1} + 2 \mathcal{I}_{2,1,1} + 2 \mathcal{I}_{1,0,1} + 2 \mathcal{I}_{1,1,1} - \mathcal{I}_{2,0,1} - \mathcal{I}_{2,1,1} \nonumber \\ 
&~~~~~~~~~~~~~~~~~~~~~~~+ \mathcal{I}_{2,0,2} - 4 \mathcal{I}_{1,0,2} + 4 \mathcal{I}_{0,0,2} \,, \\
\pcpp &= -\langle c\up{\dagger}_{\ell\alpha} c\up{\dagger}_{\ell'\beta} c_{\ell\alpha} c_{\ell'\beta} \rangle = 
2(\mathcal{I}_{2,0,0}-\mathcal{I}_{1,0,0}) + 3 (\mathcal{I}_{2,1,0} -\mathcal{I}_{1,1,0}) - \mathcal{I}_{1,2,0} + \mathcal{I}_{2,2,0}  \nonumber \\
&~~~~~~~~~~~~~~~~~~~~~~~+ 2 \mathcal{I}_{2,0,1} + 2 \mathcal{I}_{2,1,1} - 4 \mathcal{I}_{1,0,1} - 4 \mathcal{I}_{1,1,1} + 3 \mathcal{I}_{1,0,1} + 3 \mathcal{I}_{1,1,1} - \mathcal{I}_{2,0,1} - \mathcal{I}_{2,1,1} \nonumber \\
&~~~~~~~~~~~~~~~~~~~~~~~+ 2 \mathcal{I}_{0,0,1} + 2 \mathcal{I}_{0,1,1} + \mathcal{I}_{1,0,1} + \mathcal{I}_{1,1,1} - \mathcal{I}_{2,0,1} - \mathcal{I}_{2,1,1} \nonumber \\ 
&~~~~~~~~~~~~~~~~~~~~~~~+2 \mathcal{I}_{2,0,1} + 2 \mathcal{I}_{2,1,1} - 4 \mathcal{I}_{1,0,1} - 4 \mathcal{I}_{1,1,1} \nonumber \\ 
&~~~~~~~~~~~~~~~~~~~~~~~-2 \mathcal{I}_{0,0,1} + 2 \mathcal{I}_{0,0,2} + 3 \mathcal{I}_{1,0,1} - 3 \mathcal{I}_{1,0,2} - \mathcal{I}_{2,0,1} + \mathcal{I}_{2,0,2} \,, \\
\pab &= \langle c_{\ell \alpha}^{\dagger} \Delta \Delta^{\dagger} c_{\ell \alpha} \rangle = 2 \mathcal{I}_{1,0,1} + 3 \mathcal{I}_{1,1,1} + \mathcal{I}_{1,2,1} + 2 \mathcal{I}_{0,0,2} - 2 \mathcal{I}_{0,0,1} + 2\mathcal{I}_{0,1,2} - 2 \mathcal{I}_{0,1,1} - \mathcal{I}_{1,0,2} \nonumber \\ 
&~~~~~~~~~~~~~~~~~~~~~~~~+ \mathcal{I}_{1,0,1} - \mathcal{I}_{1,1,2} + \mathcal{I}_{1,1,1} \,, \\
\pabp &= \langle c_{\ell \alpha}^{\dagger} \Delta^{\dagger} \Delta c_{\ell \alpha} \rangle = \mathcal{I}_{1,0,0} + 2 \mathcal{I}_{1,1,0} + \mathcal{I}_{1,2,0} + \mathcal{I}_{1,0,1} + 2 \mathcal{I}_{1,1,1} + \mathcal{I}_{1,2,1} + 2 \mathcal{I}_{0,1,2} - \mathcal{I}_{1,1,2} \,, \\
\pbb &= \langle c_{\ell \alpha}^{\dagger} c_{\ell \alpha} \Delta \Delta^{\dagger} \rangle = \mathcal{I}_{1,0,1} + 2 \mathcal{I}_{1,1,1} + \mathcal{I}_{1,2,1} + 2 \mathcal{I}_{0,0,2} + 2 \mathcal{I}_{0,1,2} - \mathcal{I}_{1,0,2} - \mathcal{I}_{1,1,2} \,, \\
\pbbp &= \langle c_{\ell \alpha}^{\dagger} c_{\ell \alpha} \Delta^{\dagger} \Delta \rangle = \mathcal{I}_{1,1,0} + \mathcal{I}_{1,2,0} + \mathcal{I}_{1,1,1} + \mathcal{I}_{1,2,1} + 2 \mathcal{I}_{0,1,1} + 2 \mathcal{I}_{0,1,2} - \mathcal{I}_{1,1,1} - \mathcal{I}_{1,1,2} \,, \\
\pcb &= \langle c_{\ell \alpha}^{\dagger} \Delta c_{\ell \alpha} \Delta^{\dagger} \rangle = 2 \mathcal{I}_{1,0,1} + 3 \mathcal{I}_{1,1,1} + \mathcal{I}_{1,2,1} + 2 \mathcal{I}_{0,0,2} - 2 \mathcal{I}_{0,0,1} + 2 \mathcal{I}_{0,1,2} - 2 \mathcal{I}_{0,1,1} - \mathcal{I}_{1,0,2} \nonumber \\
&~~~~~~~~~~~~~~~~~~~~~~~~+ \mathcal{I}_{1,0,1} - \mathcal{I}_{1,1,2} + \mathcal{I}_{1,1,1} \,, \\ 
\pcbp &= \langle c_{\ell \alpha}^{\dagger} \Delta^{\dagger} c_{\ell \alpha} \Delta \rangle = \mathcal{I}_{1,1,0} + \mathcal{I}_{1,1,1} + \mathcal{I}_{1,2,0} + \mathcal{I}_{1,2,1} + 2 \mathcal{I}_{0,1,1} + 2 \mathcal{I}_{0,1,2} - \mathcal{I}_{1,1,1} - \mathcal{I}_{1,1,2} \,.
\end{align}
\endgroup
Note that there is a minus sign in $\pcp$ and $\pcpp$ because we need to move $\psi$ across an odd number of $c$ operators to contract it with $\psi^\dagger$. This minus sign was missed in Refs. \cite{Joshi2019,Joshi2020} in the corresponding terms of electron correlator (Eqs. (B39) and (B42) in Ref. \cite{Joshi2019}, which were also used in Ref. \cite{Joshi2020}). However, in Refs. \cite{Joshi2019,Joshi2020} these terms vanished and hence this minus sign does not influence the results therein. 
For the electron correlator we thus have,
\begin{align}
\langle O_{2} \rangle = \frac{N_{2}}{D} &= \po \bigg \lbrace 
1 + \gam\up{2} \left[ \left( \frac{\Pa}{\po} - \lo \right) \Da 
+ \left( \frac{\pb}{\po} - \lo \right) \Db 
+\left( \frac{\pc}{\po} - \lo \right) \Dc \right]  \nonumber \\
&+ \gc\up{2} \left[ \left( \frac{\pap}{\po} - \lop \right) \Dap 
+ \left( \frac{\pbp}{\po} - \lop \right) \Dbp 
+\left( \frac{\pcp}{\po} - \lop \right) \Dcp \right] \nonumber \\
&+ \gc\up{2} \left[ \left( \frac{\papp}{\po} - \lopp \right) \Dapp 
+ \left( \frac{\pbpp}{\po} - \lopp \right) \Dbpp 
+\left( \frac{\pcpp}{\po} - \lopp \right) \Dcpp \right] \nonumber \\
&+ \gv\up{2} \left[ \left( \frac{\pab+\pabp}{\po} - \lob - \lobp \right) \Dab 
+ \left( \frac{\pbb + \pbbp}{\po} - \lob - \lobp \right) \Dbb 
+\left( \frac{\pcb + \pcbp}{\po} - \lob - \lobp \right) \Dcb \right]
\bigg \rbrace \,.
\end{align}
Similarly, it is then straightforward to write, 
\begin{equation}
\label{eq:zc_m}
Z_{c} = 1 - \frac{\gamr\up{2}}{\ep} P_\gamma - \frac{\gcr\up{2}}{2\rb} P_g - \frac{\gvr\up{2}}{\epp} \Pv \,,
\end{equation}
where
\begin{align}
P_\gamma &= \frac{P_1+P_2 -2P_3}{P_0} = \frac{3}{4} \,, \\
P_g &= \frac{P_1'+P_2'-2P_3'+P_1''+P_2''-2P_3''}{P_0} = 2 \,, \\
\Pv &= \frac{\pab + \pabp + \pbb + \pbbp - 2 \pcb - 2 \pcbp }{\po} = 1 \,.
\end{align}
Therefore we have,
\begin{equation}
Z_{c} = 1 - \frac{3 \gamr\up{2}}{4 \ep} - \frac{ \gcr\up{2}}{\rb} - \frac{\gvr\up{2}}{\epp} \,,
\end{equation}
which is same as obtained earlier in Sec. \ref{sec:anom_se}.


\subsection{SC order-parameter correlator}
\label{sec:sc_m}

We define superconductor order parameter, $\Delta = c_{\uparrow}\up{\dagger} c_{\downarrow}\up{\dagger} = b d\up{\dagger}$. 
We will now evaluate the correlator of $\Delta$, $\langle O_{3} \rangle \equiv \langle \Delta(\tau) \Delta\up{\dagger}(0) \rangle = N_{3}/D$. 
Here we have,
\begin{align}
\label{eq:n3}
N_{3} &= \Ro + \gam\up{2} \left( \Ra\Da + \Rb\Db + \Rc\Dc \right) 
+ \gc\up{2} \left( \Rap\Dap + \Rbp\Dbp + \Rcp\Dcp \right) \nonumber \\
&+ \gc\up{2} \left( \Rapp\Dapp + \Rbpp\Dbpp + \Rcpp\Dcpp \right) 
+ \gv\up{2} \left( (\Rab+\Rabp)\Dab + (\Rbb+\Rbbp)\Dbb + (\Rcb+\Rcbp)\Dcb \right) \,, 
\end{align}
where,
\begingroup
\allowdisplaybreaks
\begin{align}
\Ro &= \langle \Delta\up{\dagger} \Delta \rangle = \mathcal{I}_{0,1,0} + \mathcal{I}_{0,1,1} \,,\\
\Ra &= \langle \Delta\up{\dagger} S\up{a} S\up{a} \Delta \rangle = \frac{3}{4}(2\mathcal{I}_{1,1,0} 
+ 2\mathcal{I}_{1,1,1} - \mathcal{I}_{2,1,0} - \mathcal{I}_{2,1,1} ) \,, \\
\Rb &= \langle \Delta\up{\dagger} \Delta S\up{a} S\up{a} \rangle = \Ra \,, \\
\Rc &= \langle \Delta\up{\dagger} S\up{a} \Delta S\up{a} \rangle = \Ra \,, \\
\Rap &= \langle \Delta\up{\dagger} c_{\ell'\beta} c_{\ell'\beta}\up{\dagger} \Delta \rangle = 
\mathcal{I}_{0,2,0} - \mathcal{I}_{0,1,0} - \mathcal{I}_{1,2,0} + \mathcal{I}_{1,1,0} + \mathcal{I}_{0,2,1} - \mathcal{I}_{0,1,1} 
- \mathcal{I}_{1,2,1} + 4 \mathcal{I}_{1,1,1} \nonumber \\ 
&~~~~~~~~~~~~~~~~~~~~~~+ 2 \mathcal{I}_{1,1,0} + \mathcal{I}_{1,1,2}  \,, \\
\Rbp &= \langle \Delta\up{\dagger} \Delta c_{\ell'\beta} c\up{\dagger}_{\ell'\beta} \rangle =  
2\mathcal{I}_{0,2,0} + 2\mathcal{I}_{0,2,1} - \mathcal{I}_{1,2,0} - \mathcal{I}_{1,2,1} + \mathcal{I}_{1,1,0} + 2 \mathcal{I}_{1,1,1} + \mathcal{I}_{1,1,2}  \,, \\
\Rcp &= \langle \Delta\up{\dagger} c_{\ell'\beta} \Delta c\up{\dagger}_{\ell'\beta}\rangle  = 
2\mathcal{I}_{0,2,0} - 2\mathcal{I}_{0,1,0} + 2\mathcal{I}_{0,2,1} - 2\mathcal{I}_{0,1,1} - \mathcal{I}_{1,2,0} + \mathcal{I}_{1,1,0} - \mathcal{I}_{1,2,1} + 4\mathcal{I}_{1,1,1}  \nonumber \\ 
&~~~~~~~~~~~~~~~~~~~~~~+ 2 \mathcal{I}_{1,1,0} + \mathcal{I}_{1,1,2}  \,, \\
\Rapp &= \langle \Delta\up{\dagger} c_{\ell'\beta}\up{\dagger} c_{\ell'\beta} \Delta \rangle = 
\mathcal{I}_{1,2,0} + \mathcal{I}_{1,2,1} + 2\mathcal{I}_{0,1,0} + 4 \mathcal{I}_{0,1,1} + 2\mathcal{I}_{0,1,2} - \mathcal{I}_{1,1,0}  - 2 \mathcal{I}_{1,1,1} - \mathcal{I}_{1,1,2}  \,, \\
\Rbpp &= \langle \Delta\up{\dagger} \Delta c\up{\dagger}_{\ell'\beta} c_{\ell'\beta} \rangle = 
\mathcal{I}_{1,1,0} + \mathcal{I}_{1,1,1} + \mathcal{I}_{1,2,0} + \mathcal{I}_{1,2,1} + 2\mathcal{I}_{0,1,1} + 2\mathcal{I}_{0,1,2} - \mathcal{I}_{1,1,1}  - \mathcal{I}_{1,1,2} \,, \\
\Rcpp &= \langle \Delta\up{\dagger} c\up{\dagger}_{\ell'\beta} \Delta c_{\ell'\beta} \rangle = \Rbpp  \,, \\
\Rab &= \langle \Delta^{\dagger} \Delta \Delta^{\dagger} \Delta \rangle = \mathcal{I}_{0,2,0} + 2 \mathcal{I}_{0,2,1} + \mathcal{I}_{0,2,2} \,, \\ 
\Rabp &= \langle \Delta^{\dagger} \Delta^{\dagger} \Delta \Delta \rangle = 
2 \mathcal{I}_{0,2,0} - 2 \mathcal{I}_{0,1,0} + 3 \mathcal{I}_{0,2,1} - 3 \mathcal{I}_{0,1,1} + \mathcal{I}_{0,2,2} - \mathcal{I}_{0,1,2} \,, \\ 
\Rbb &= \langle \Delta^{\dagger} \Delta \Delta \Delta^{\dagger} \rangle = 
\mathcal{I}_{0,1,1} + \mathcal{I}_{0,2,1} + \mathcal{I}_{0,1,2} + \mathcal{I}_{0,2,2} \,, \\ 
\Rbbp &= \langle \Delta^{\dagger} \Delta \Delta^{\dagger} \Delta \rangle = \Rab \,, \\
\Rcb &= \langle \Delta^{\dagger} \Delta \Delta \Delta^{\dagger} \rangle = \Rbb \,, \\ 
\Rcbp &= \langle \Delta^{\dagger} \Delta^{\dagger} \Delta \Delta \rangle = \Rabp \,.
\end{align}
\endgroup
Thus we have,
\begin{align}
\langle O_{3} \rangle = \frac{N_{3}}{D} &= \Ro \bigg \lbrace 
1 + \gam\up{2} \left[ \left( \frac{\Ra}{\Ro} - \lo \right) \Da 
+ \left( \frac{\Rb}{\Ro} - \lo \right) \Db 
+\left( \frac{\Rc}{\Ro} - \lo \right) \Dc \right]  \nonumber \\
&+ \gc\up{2} \left[ \left( \frac{\Rap}{\Ro} - \lop \right) \Dap 
+ \left( \frac{\Rbp}{\Ro} - \lop \right) \Dbp 
+\left( \frac{\Rcp}{\Ro} - \lop \right) \Dcp \right] \nonumber \\
&+ \gc\up{2} \left[ \left( \frac{\Rapp}{\Ro} - \lopp \right) \Dapp 
+ \left( \frac{\Rbpp}{\Ro} - \lopp \right) \Dbpp 
+\left( \frac{\Rcpp}{\Ro} - \lopp \right) \Dcpp \right] \nonumber \\ 
&+ \gv\up{2} \left[ \left( \frac{\Rab+\Rabp}{\Ro} - \lob - \lobp \right) \Dab 
+ \left( \frac{\Rbb+\Rbbp}{\Ro} - \lob - \lobp \right) \Dbb 
+\left( \frac{\Rcb+\Rcbp}{\Ro} - \lob - \lobp \right) \Dcb \right]
\bigg \rbrace \,.
\end{align}
Similarly, it is the straightforward to write, 
\begin{equation}
\label{eq:zd_m}
Z_{\Delta} = 1 - \frac{\gamr\up{2}}{\ep} R_\gamma - \frac{\gcr\up{2}}{2\rb} R_g - \frac{\gvr\up{2}}{\epp} \Rv \,,
\end{equation}
where
\begin{align}
R_\gamma &= \frac{R_1+R_2 -2R_3}{R_0} = 0 \,, \\
R_g &= \frac{R_1'+R_2'-2R_3'+R_1''+R_2''-2R_3''}{R_0} = 4 \,, \\
\Rv &= \frac{\Rab + \Rabp + \Rbb + \Rbbp - 2 \Rcb -2 \Rcbp}{\Ro} = 2 \,.
\end{align}
Therefore,
\begin{equation}
Z_{\Delta} = 1 - \frac{2\gcr\up{2}}{\rb} - \frac{2 \gvr^{2}}{\epp} \,,
\end{equation}
same as obtained in Sec. \ref{sec:anom_se}.


\subsection{Density correlator}

We also evaluate the correlator of density $n = 1 + \dd \da - \bd \ba$. The constant $1$ is not renormalized and we ignore it here and consider $n = \dd \da - \bd \ba$ for renormalization purpose. We have $\langle O_{4} \rangle \equiv \langle n(\tau) n(0) \rangle = N_{4}/D$ with
\begin{align}
\label{eq:n4}
N_{4} &= \To + \gam\up{2} \left( \Ta\Da + \Tb\Db + \Tc\Dc \right) 
+ \gc\up{2} \left( \Tap\Dap + \Tbp\Dbp + \Tcp\Dcp \right) \nonumber \\
&+ \gc\up{2} \left( \Tapp\Dapp + \Tbpp\Dbpp + \Tcpp\Dcpp \right) 
+ \gv\up{2} \left( (\Tab+\Tabp)\Dab + (\Tbb+\Tbbp)\Dbb + (\Tcb+\Tcbp)\Dcb \right) \,, 
\end{align}
where,
\begingroup
\allowdisplaybreaks
\begin{align}
\To &= \langle n n \rangle = \mathcal{I}_{0,2,0} + \mathcal{I}_{0,0,2} - 2 \mathcal{I}_{0,1,1} \,,\\
\Ta &= \langle n S\up{a} S\up{a} n \rangle = \frac{3}{4}(2\mathcal{I}_{1,2,0} 
+ 2\mathcal{I}_{1,0,2} - 4\mathcal{I}_{1,1,1} - \mathcal{I}_{2,2,0} - \mathcal{I}_{2,0,2} + 2\mathcal{I}_{2,1,1}) \,, \\
\Tb &= \langle n n S\up{a} S\up{a} \rangle = \Ta \,, \\
\Tc &= \langle n S\up{a} n S\up{a} \rangle = \Ta \,, \\
\Tap &= \langle n c_{\ell'\beta} c_{\ell'\beta}\up{\dagger} n \rangle = 
2\mathcal{I}_{0,1,2} - \mathcal{I}_{1,1,2} - 2\mathcal{I}_{0,2,1} + \mathcal{I}_{1,2,1} + \mathcal{I}_{1,0,2} + \mathcal{I}_{1,0,3} 
- \mathcal{I}_{1,1,1} - \mathcal{I}_{1,1,2} - 2 \mathcal{I}_{0,2,1} \nonumber \\
&~~~~~~~~~~~~~~~~~~~+ \mathcal{I}_{1,2,1} + 2\mathcal{I}_{0,3,0} - \mathcal{I}_{1,3,0}  - \mathcal{I}_{1,1,1} - \mathcal{I}_{1,1,2} + \mathcal{I}_{1,2,0} + \mathcal{I}_{1,2,1} \,, \\
\Tbp &= \langle n n c_{\ell'\beta} c\up{\dagger}_{\ell'\beta} \rangle =  
2\mathcal{I}_{0,1,2} - \mathcal{I}_{1,1,2} - 2\mathcal{I}_{0,2,1} + \mathcal{I}_{1,2,1} + \mathcal{I}_{1,0,2} + \mathcal{I}_{1,0,3} 
- \mathcal{I}_{1,1,1} - \mathcal{I}_{1,1,2} - 2 \mathcal{I}_{0,2,1} \nonumber \\
&~~~~~~~~~~~~~~~~~~~+ \mathcal{I}_{1,2,1} + 2\mathcal{I}_{0,3,0} - \mathcal{I}_{1,3,0}  - \mathcal{I}_{1,1,1} - \mathcal{I}_{1,1,2} + \mathcal{I}_{1,2,0} + \mathcal{I}_{1,2,1} \,, \\
\Tcp &= \langle n c_{\ell'\beta} n c\up{\dagger}_{\ell'\beta}\rangle  = 
2\mathcal{I}_{0,1,2} - \mathcal{I}_{1,1,2} - 2\mathcal{I}_{0,2,1} + 2\mathcal{I}_{0,1,1} + \mathcal{I}_{1,2,1} - \mathcal{I}_{1,1,1} + \mathcal{I}_{1,0,1} + 2\mathcal{I}_{1,0,2} + \mathcal{I}_{1,0,3}  \nonumber \\ 
&~~~~~~~~~~~~~~~~~~~- \mathcal{I}_{1,1,1} - \mathcal{I}_{1,1,2} - 2\mathcal{I}_{0,2,1} + \mathcal{I}_{1,2,1} + 2\mathcal{I}_{0,3,0} - 2\mathcal{I}_{0,2,0} - \mathcal{I}_{1,3,0} + \mathcal{I}_{1,2,0}    \nonumber \\
&~~~~~~~~~~~~~~~~~~~- \mathcal{I}_{1,1,0} - 2\mathcal{I}_{1,1,1} - \mathcal{I}_{1,1,2} + \mathcal{I}_{1,2,0} + \mathcal{I}_{1,2,1}   \,, \\
\Tapp &= \langle n c_{\ell'\beta}\up{\dagger} c_{\ell'\beta} n \rangle = 
\mathcal{I}_{1,0,2} + \mathcal{I}_{1,1,2} - \mathcal{I}_{1,1,1} - \mathcal{I}_{1,2,1} + 2\mathcal{I}_{0,0,3} - \mathcal{I}_{1,0,3}  - 2 \mathcal{I}_{0,1,2} + \mathcal{I}_{1,1,2}  \nonumber \\
&~~~~~~~~~~~~~~~~~~~- \mathcal{I}_{1,1,1} - \mathcal{I}_{1,2,1} + \mathcal{I}_{1,2,0} + \mathcal{I}_{1,3,0} - 2\mathcal{I}_{0,1,2} + \mathcal{I}_{1,1,2}  + 2 \mathcal{I}_{0,2,1} - \mathcal{I}_{1,2,1}  \,, \\
\Tbpp &= \langle n n c\up{\dagger}_{\ell'\beta} c_{\ell'\beta} \rangle = 
\mathcal{I}_{1,0,2} + \mathcal{I}_{1,1,2} - \mathcal{I}_{1,1,1} - \mathcal{I}_{1,2,1} + 2\mathcal{I}_{0,0,3} - \mathcal{I}_{1,0,3}  - 2 \mathcal{I}_{0,1,2} + \mathcal{I}_{1,1,2}  \nonumber \\
&~~~~~~~~~~~~~~~~~~~- \mathcal{I}_{1,1,1} - \mathcal{I}_{1,2,1} + \mathcal{I}_{1,2,0} + \mathcal{I}_{1,3,0} - 2\mathcal{I}_{0,1,2} + \mathcal{I}_{1,1,2}  + 2 \mathcal{I}_{0,2,1} - \mathcal{I}_{1,2,1}  \,, \\
\Tcpp &= \langle n c\up{\dagger}_{\ell'\beta} n c_{\ell'\beta} \rangle = 
\mathcal{I}_{1,0,2} + \mathcal{I}_{1,1,2} - \mathcal{I}_{1,0,1} - 2\mathcal{I}_{1,1,1} - \mathcal{I}_{1,2,1} + 2\mathcal{I}_{0,0,3} - 2 \mathcal{I}_{0,0,2} - \mathcal{I}_{1,0,3} + \mathcal{I}_{1,0,2} \nonumber \\
&~~~~~~~~~~~~~~~~~~~- \mathcal{I}_{1,1,1} - \mathcal{I}_{1,2,1} + \mathcal{I}_{1,1,0} + 2 \mathcal{I}_{1,2,0} + \mathcal{I}_{1,3,0} - 2\mathcal{I}_{0,1,2} + 2\mathcal{I}_{0,1,1} + \mathcal{I}_{1,1,2} - \mathcal{I}_{1,1,1} \nonumber \\
&~~~~~~~~~~~~~~~~~~~+ 2\mathcal{I}_{0,2,1} - \mathcal{I}_{1,2,1}   \,, \\
\Tab &= \langle n \Delta \Delta^{\dagger} n \rangle = \mathcal{I}_{0,0,3} + \mathcal{I}_{0,1,3} - 2 \mathcal{I}_{0,1,2} - 2 \mathcal{I}_{0,2,2} + \mathcal{I}_{0,2,1} + \mathcal{I}_{0,3,1} \,, \\ 
\Tabp &= \langle n \Delta^{\dagger} \Delta n \rangle = \mathcal{I}_{0,1,2} + \mathcal{I}_{0,1,3} - 2 \mathcal{I}_{0,2,1} - 2 \mathcal{I}_{0,2,2} + \mathcal{I}_{0,3,0} + \mathcal{I}_{0,3,1} \,, \\ 
\Tbb &= \langle n n \Delta \Delta^{\dagger} \rangle = \mathcal{I}_{0,0,3} + \mathcal{I}_{0,1,3} - 2 \mathcal{I}_{0,1,2} - 2 \mathcal{I}_{0,2,2} + \mathcal{I}_{0,2,1} + \mathcal{I}_{0,3,1} \,, \\
\Tbbp &= \langle n n \Delta^{\dagger} \Delta \rangle = \mathcal{I}_{0,1,2} + \mathcal{I}_{0,1,3} - 2 \mathcal{I}_{0,2,1} - 2 \mathcal{I}_{0,2,2} + \mathcal{I}_{0,3,0} + \mathcal{I}_{0,3,1} \,, \\
\Tcb &= \langle n \Delta n \Delta^{\dagger} \rangle = \mathcal{I}_{0,0,3} - \mathcal{I}_{0,0,2} + \mathcal{I}_{0,1,3} - \mathcal{I}_{0,1,2} - \mathcal{I}_{0,0,2} - 2 \mathcal{I}_{0,1,2} - \mathcal{I}_{0,2,2} \nonumber \\ 
&~~~~~~~~~~~~~~~~ - \mathcal{I}_{0,1,2} + \mathcal{I}_{0,1,1} - \mathcal{I}_{0,2,2} + \mathcal{I}_{0,2,1} + \mathcal{I}_{0,1,1} + 2 \mathcal{I}_{0,2,1} + \mathcal{I}_{0,3,1} \,, \\
\Tcbp &= \langle n \Delta^{\dagger} n \Delta \rangle =  \mathcal{I}_{0,1,1} + 2 \mathcal{I}_{0,1,2} + \mathcal{I}_{0,1,3} - \mathcal{I}_{0,2,1} + \mathcal{I}_{0,1,1} - \mathcal{I}_{0,2,2} + \mathcal{I}_{0,1,2} \nonumber \\ 
&~~~~~~~~~~~~~~~~ - \mathcal{I}_{0,2,0} - 2 \mathcal{I}_{0,2,1} - \mathcal{I}_{0,2,2} + \mathcal{I}_{0,3,0} + \mathcal{I}_{0,3,1} - \mathcal{I}_{0,2,0} - \mathcal{I}_{0,2,1} \,.
\end{align}
\endgroup
Thus we have,
\begin{align}
\langle O_{4} \rangle = \frac{N_{4}}{D} &= \To \bigg \lbrace 
1 + \gam\up{2} \left[ \left( \frac{\Ta}{\To} - \lo \right) \Da 
+ \left( \frac{\Tb}{\To} - \lo \right) \Db 
+\left( \frac{\Tc}{\To} - \lo \right) \Dc \right]  \nonumber \\
&+ \gc\up{2} \left[ \left( \frac{\Tap}{\To} - \lop \right) \Dap 
+ \left( \frac{\Tbp}{\To} - \lop \right) \Dbp 
+\left( \frac{\Tcp}{\To} - \lop \right) \Dcp \right] \nonumber \\
&+ \gc\up{2} \left[ \left( \frac{\Tapp}{\To} - \lopp \right) \Dapp 
+ \left( \frac{\Tbpp}{\To} - \lopp \right) \Dbpp 
+\left( \frac{\Tcpp}{\To} - \lopp \right) \Dcpp \right] \nonumber \\
&+ \gv\up{2} \left[ \left( \frac{\Tab+\Tabp}{\To} - \lob - \lobp \right) \Dab 
+ \left( \frac{\Tbb+\Tbbp}{\To} - \lob - \lobp \right) \Dbb 
+\left( \frac{\Tcb+\Tcbp}{\To} - \lob - \lobp \right) \Dcb \right] 
\bigg \rbrace \,.
\end{align}
Similarly, it is then straightforward to write, 
\begin{equation}
\label{eq:zn_m}
Z_{n} = 1 - \frac{\gamr\up{2}}{\ep} T_\gamma - \frac{\gcr\up{2}}{2\rb} T_g - \frac{\gvr\up{2}}{\epp} \Tv \,,
\end{equation}
where
\begin{align}
T_\gamma &= \frac{T_1+T_2 -2T_3}{T_0} = 0 \,, \\
T_g &= \frac{T_1'+T_2'-2T_3'+T_1''+T_2''-2T_3''}{T_0} = 4 \,, \\
\Tv &= \frac{\Tab + \Tabp + \Tbb + \Tbbp - 2\Tcb -2\Tcbp}{\To} = 4 \,.
\end{align}
Therefore,
\begin{equation}
Z_{n} = 1 - \frac{2\gcr\up{2}}{\rb} - \frac{4\gvr^{2}}{\epp} \,,
\end{equation}
same as evaluated in Sec. \ref{sec:anom_den}.


\subsection{Beta functions}
\label{sec:beta_m}

We are now in a position to write the beta functions for the coupling constants. Using Eq. \ref{eq:renorm_fact_m} we find three equations,
\begin{align}
\label{eq:beta1_m}
\frac{\ep}{2}\gamr Z_{S} + \left[ Z_{S} - \frac{\gamr}{2} \frac{\partial Z_{S}}{\partial \gamr} \right] \betgam 
- \frac{\gamr}{2} \frac{\partial Z_{S}}{\partial \gcr} \betg 
- \frac{\gamr}{2} \frac{\partial Z_{S}}{\partial \gvr} \betv &= 0 \,, \\
\label{eq:beta2_m}
\rb\gcr Z_{c} + \left[ Z_{c} - \frac{\gcr}{2} \frac{\partial Z_{c}}{\partial \gcr} \right] \betg 
- \frac{\gcr}{2} \frac{\partial Z_{c}}{\partial \gamr} \betgam 
- \frac{\gcr}{2} \frac{\partial Z_{c}}{\partial \gvr} \betv &= 0 \,, \\ 
\label{eq:beta3_m}
\frac{\epp}{2}\gvr Z_{\Delta} + \left[ Z_{\Delta} - \frac{\gvr}{2} \frac{\partial Z_{\Delta}}{\partial \gvr} \right] \betv 
- \frac{\gvr}{2} \frac{\partial Z_{\Delta}}{\partial \gcr} \betg 
- \frac{\gvr}{2} \frac{\partial Z_{\Delta}}{\partial \gamr} \betgam &= 0 \,.
\end{align}
We have used the exact relations $\widetilde{Z}_{\gcr}=\widetilde{Z}_{\gamr}=\widetilde{Z}_{\gvr}=1$ in obtaining these equations. 
Solving these three equations and using the expressions for the renormalization factors found above we obtain the following one-loop beta functions,
\begin{align}
\label{eq:beta_g_m}
\betg &= -\rb \gcr + \frac{\Pg}{2} \gcr\up{3} + \frac{\Pgam}{2} \gcr \gamr\up{2} + \frac{\Pv}{2} \gcr \gvr\up{2} 
= -\rb \gcr + \gcr\up{3} + \frac{3}{8} \gcr \gamr\up{2} + \frac{\gcr \gvr^{2}}{2} \,, \\
\label{eq:beta_gam_m}
\betgam &= -\frac{\ep}{2} \gamr + \frac{\Lgam}{2} \gamr\up{3} + \frac{\Lg}{2} \gamr \gcr\up{2} + \frac{\Lv}{2} \gamr \gvr\up{2}  
= -\frac{\ep}{2} \gamr + \gamr\up{3} + 2 \gamr \gcr\up{2} \,, \\
\label{eq:beta_v_m}
\betv &= -\frac{\epp}{2} \gvr + \frac{\Rv}{2} \gvr\up{3} + \frac{\Rg}{2} \gvr \gcr\up{2} + \frac{\Rgam}{2} \gvr \gamr\up{2}  
= -\frac{\epp}{2} \gvr + \gvr\up{3} + 2 \gvr \gcr\up{2} \,.
\end{align}
These are exactly the same as Eqs. (\ref{eq:beta_g_c}) - (\ref{eq:beta_v_c}) derived in Sec. \ref{sec:beta_c} using diagrammatic RG with electron operator, as well as Eqs. (\ref{eq:betag}) - (\ref{eq:betav}) found using the fractionalized RG method in Sec. \ref{sec:beta_fbd}. 
Therefore the fixed points are also the same as obtained earlier as well as all the exponents. Hence we do not discuss them here again.

\bibliography{scmt}

\begin{thebibliography}{37}%
\makeatletter
\providecommand \@ifxundefined [1]{%
 \@ifx{#1\undefined}
}%
\providecommand \@ifnum [1]{%
 \ifnum #1\expandafter \@firstoftwo
 \else \expandafter \@secondoftwo
 \fi
}%
\providecommand \@ifx [1]{%
 \ifx #1\expandafter \@firstoftwo
 \else \expandafter \@secondoftwo
 \fi
}%
\providecommand \natexlab [1]{#1}%
\providecommand \enquote  [1]{``#1''}%
\providecommand \bibnamefont  [1]{#1}%
\providecommand \bibfnamefont [1]{#1}%
\providecommand \citenamefont [1]{#1}%
\providecommand \href@noop [0]{\@secondoftwo}%
\providecommand \href [0]{\begingroup \@sanitize@url \@href}%
\providecommand \@href[1]{\@@startlink{#1}\@@href}%
\providecommand \@@href[1]{\endgroup#1\@@endlink}%
\providecommand \@sanitize@url [0]{\catcode `\\12\catcode `\$12\catcode
  `\&12\catcode `\#12\catcode `\^12\catcode `\_12\catcode `\%12\relax}%
\providecommand \@@startlink[1]{}%
\providecommand \@@endlink[0]{}%
\providecommand \url  [0]{\begingroup\@sanitize@url \@url }%
\providecommand \@url [1]{\endgroup\@href {#1}{\urlprefix }}%
\providecommand \urlprefix  [0]{URL }%
\providecommand \Eprint [0]{\href }%
\providecommand \doibase [0]{http://dx.doi.org/}%
\providecommand \selectlanguage [0]{\@gobble}%
\providecommand \bibinfo  [0]{\@secondoftwo}%
\providecommand \bibfield  [0]{\@secondoftwo}%
\providecommand \translation [1]{[#1]}%
\providecommand \BibitemOpen [0]{}%
\providecommand \bibitemStop [0]{}%
\providecommand \bibitemNoStop [0]{.\EOS\space}%
\providecommand \EOS [0]{\spacefactor3000\relax}%
\providecommand \BibitemShut  [1]{\csname bibitem#1\endcsname}%
\let\auto@bib@innerbib\@empty
\bibitem [{\citenamefont {{Kapitulnik}}\ \emph {et~al.}(2019)\citenamefont
  {{Kapitulnik}}, \citenamefont {{Kivelson}},\ and\ \citenamefont
  {{Spivak}}}]{KKS19}%
  \BibitemOpen
  \bibfield  {author} {\bibinfo {author} {\bibfnamefont {A.}~\bibnamefont
  {{Kapitulnik}}}, \bibinfo {author} {\bibfnamefont {S.~A.}\ \bibnamefont
  {{Kivelson}}}, \ and\ \bibinfo {author} {\bibfnamefont {B.}~\bibnamefont
  {{Spivak}}},\ }\bibfield  {title} {\enquote {\bibinfo {title} {{Anomalous
  metals -- failed superconductors}},}\ }\href {\doibase
  10.1103/RevModPhys.91.011002} {\bibfield  {journal} {\bibinfo  {journal}
  {Rev. Mod. Phys.}\ }\textbf {\bibinfo {volume} {91}},\ \bibinfo {pages}
  {011002} (\bibinfo {year} {2019})},\ \Eprint
  {http://arxiv.org/abs/1712.07215} {arXiv:1712.07215 [cond-mat.supr-con]}
  \BibitemShut {NoStop}%
\bibitem [{\citenamefont {{Zhang}}\ \emph {et~al.}(2020)\citenamefont
  {{Zhang}}, \citenamefont {{Hen}}, \citenamefont {{Palevski}},\ and\
  \citenamefont {{Kapitulnik}}}]{Kapitulnik20}%
  \BibitemOpen
  \bibfield  {author} {\bibinfo {author} {\bibfnamefont {X.}~\bibnamefont
  {{Zhang}}}, \bibinfo {author} {\bibfnamefont {B.}~\bibnamefont {{Hen}}},
  \bibinfo {author} {\bibfnamefont {A.}~\bibnamefont {{Palevski}}}, \ and\
  \bibinfo {author} {\bibfnamefont {A.}~\bibnamefont {{Kapitulnik}}},\
  }\bibfield  {title} {\enquote {\bibinfo {title} {{Robust anomalous metallic
  states and vestiges of self duality in two-dimensional granular In
  composites}},}\ }\href@noop {} {\  (\bibinfo {year} {2020})},\ \Eprint
  {http://arxiv.org/abs/2008.09325} {arXiv:2008.09325 [cond-mat.supr-con]}
  \BibitemShut {NoStop}%
\bibitem [{\citenamefont {{Malinowski}}\ \emph {et~al.}(2020)\citenamefont
  {{Malinowski}}, \citenamefont {{Jiang}}, \citenamefont {{Sanchez}},
  \citenamefont {{Mutch}}, \citenamefont {{Liu}}, \citenamefont {{Went}},
  \citenamefont {{Liu}}, \citenamefont {{Ryan}}, \citenamefont {{Kim}},\ and\
  \citenamefont {{Chu}}}]{Chu20}%
  \BibitemOpen
  \bibfield  {author} {\bibinfo {author} {\bibfnamefont {P.}~\bibnamefont
  {{Malinowski}}}, \bibinfo {author} {\bibfnamefont {Q.}~\bibnamefont
  {{Jiang}}}, \bibinfo {author} {\bibfnamefont {J.~J.}\ \bibnamefont
  {{Sanchez}}}, \bibinfo {author} {\bibfnamefont {J.}~\bibnamefont {{Mutch}}},
  \bibinfo {author} {\bibfnamefont {Z.}~\bibnamefont {{Liu}}}, \bibinfo
  {author} {\bibfnamefont {P.}~\bibnamefont {{Went}}}, \bibinfo {author}
  {\bibfnamefont {J.}~\bibnamefont {{Liu}}}, \bibinfo {author} {\bibfnamefont
  {P.~J.}\ \bibnamefont {{Ryan}}}, \bibinfo {author} {\bibfnamefont {J.-W.}\
  \bibnamefont {{Kim}}}, \ and\ \bibinfo {author} {\bibfnamefont {J.-H.}\
  \bibnamefont {{Chu}}},\ }\bibfield  {title} {\enquote {\bibinfo {title}
  {{Suppression of superconductivity by anisotropic strain near a nematic
  quantum critical point}},}\ }\href {\doibase 10.1038/s41567-020-0983-9}
  {\bibfield  {journal} {\bibinfo  {journal} {Nature Physics}\ }\textbf
  {\bibinfo {volume} {16}},\ \bibinfo {pages} {1189} (\bibinfo {year}
  {2020})},\ \Eprint {http://arxiv.org/abs/1911.03390} {arXiv:1911.03390
  [cond-mat.supr-con]} \BibitemShut {NoStop}%
\bibitem [{\citenamefont {{B{\o}ttcher}}\ \emph {et~al.}(2018)\citenamefont
  {{B{\o}ttcher}}, \citenamefont {{Nichele}}, \citenamefont {{Kjaergaard}},
  \citenamefont {{Suominen}}, \citenamefont {{Shabani}}, \citenamefont
  {{Palmstr{\o}m}},\ and\ \citenamefont {{Marcus}}}]{Marcus18}%
  \BibitemOpen
  \bibfield  {author} {\bibinfo {author} {\bibfnamefont {C.~G.~L.}\
  \bibnamefont {{B{\o}ttcher}}}, \bibinfo {author} {\bibfnamefont
  {F.}~\bibnamefont {{Nichele}}}, \bibinfo {author} {\bibfnamefont
  {M.}~\bibnamefont {{Kjaergaard}}}, \bibinfo {author} {\bibfnamefont {H.~J.}\
  \bibnamefont {{Suominen}}}, \bibinfo {author} {\bibfnamefont
  {J.}~\bibnamefont {{Shabani}}}, \bibinfo {author} {\bibfnamefont {C.~J.}\
  \bibnamefont {{Palmstr{\o}m}}}, \ and\ \bibinfo {author} {\bibfnamefont
  {C.~M.}\ \bibnamefont {{Marcus}}},\ }\bibfield  {title} {\enquote {\bibinfo
  {title} {{Superconducting, insulating and anomalous metallic regimes in a
  gated two-dimensional semiconductor-superconductor array}},}\ }\href
  {\doibase 10.1038/s41567-018-0259-9} {\bibfield  {journal} {\bibinfo
  {journal} {Nature Physics}\ }\textbf {\bibinfo {volume} {14}},\ \bibinfo
  {pages} {1138} (\bibinfo {year} {2018})},\ \Eprint
  {http://arxiv.org/abs/1711.01451} {arXiv:1711.01451 [cond-mat.mes-hall]}
  \BibitemShut {NoStop}%
\bibitem [{\citenamefont {{Dubouchet}}\ \emph {et~al.}(2018)\citenamefont
  {{Dubouchet}}, \citenamefont {{Sac{\'e}p{\'e}}}, \citenamefont {{Seidemann}},
  \citenamefont {{Shahar}}, \citenamefont {{Sanquer}},\ and\ \citenamefont
  {{Chapelier}}}]{Shahar18}%
  \BibitemOpen
  \bibfield  {author} {\bibinfo {author} {\bibfnamefont {T.}~\bibnamefont
  {{Dubouchet}}}, \bibinfo {author} {\bibfnamefont {B.}~\bibnamefont
  {{Sac{\'e}p{\'e}}}}, \bibinfo {author} {\bibfnamefont {J.}~\bibnamefont
  {{Seidemann}}}, \bibinfo {author} {\bibfnamefont {D.}~\bibnamefont
  {{Shahar}}}, \bibinfo {author} {\bibfnamefont {M.}~\bibnamefont {{Sanquer}}},
  \ and\ \bibinfo {author} {\bibfnamefont {C.}~\bibnamefont {{Chapelier}}},\
  }\bibfield  {title} {\enquote {\bibinfo {title} {{Collective energy gap of
  preformed Cooper pairs in disordered superconductors}},}\ }\href {\doibase
  10.1038/s41567-018-0365-8} {\bibfield  {journal} {\bibinfo  {journal} {Nature
  Physics}\ }\textbf {\bibinfo {volume} {15}},\ \bibinfo {pages} {233}
  (\bibinfo {year} {2018})},\ \Eprint {http://arxiv.org/abs/1806.00323}
  {arXiv:1806.00323 [cond-mat.supr-con]} \BibitemShut {NoStop}%
\bibitem [{\citenamefont {{Dobrosavljevic}}\ \emph {et~al.}(2012)\citenamefont
  {{Dobrosavljevic}}, \citenamefont {{Trivedi}},\ and\ \citenamefont
  {{Valles}}}]{2012Cqpt}%
  \BibitemOpen
  \bibinfo {editor} {\bibfnamefont {V.}~\bibnamefont {{Dobrosavljevic}}},
  \bibinfo {editor} {\bibfnamefont {N.}~\bibnamefont {{Trivedi}}}, \ and\
  \bibinfo {editor} {\bibfnamefont {J.~M.}\ \bibnamefont {{Valles}}},\ eds.,\
  \href {\doibase 10.1093/acprof:oso/9780199592593.001.0001} {\emph {\bibinfo
  {title} {Conductor-Insulator Quantum Phase Transitions}}}\ (\bibinfo
  {publisher} {Oxford University Press},\ \bibinfo {address} {Oxford},\
  \bibinfo {year} {2012})\BibitemShut {NoStop}%
\bibitem [{\citenamefont {{Goldman}}(2010)}]{Goldman10}%
  \BibitemOpen
  \bibfield  {author} {\bibinfo {author} {\bibfnamefont {A.~M.}\ \bibnamefont
  {{Goldman}}},\ }\bibfield  {title} {\enquote {\bibinfo {title}
  {{Superconductor-Insulator Transitions}},}\ }\href {\doibase
  10.1142/S0217979210056451} {\bibfield  {journal} {\bibinfo  {journal}
  {International Journal of Modern Physics B}\ }\textbf {\bibinfo {volume}
  {24}},\ \bibinfo {pages} {4081} (\bibinfo {year} {2010})}\BibitemShut
  {NoStop}%
\bibitem [{\citenamefont {{Feigel'man}}\ \emph {et~al.}(2010)\citenamefont
  {{Feigel'man}}, \citenamefont {{Ioffe}}, \citenamefont {{Kravtsov}},\ and\
  \citenamefont {{Cuevas}}}]{Feigelman10}%
  \BibitemOpen
  \bibfield  {author} {\bibinfo {author} {\bibfnamefont {M.~V.}\ \bibnamefont
  {{Feigel'man}}}, \bibinfo {author} {\bibfnamefont {L.~B.}\ \bibnamefont
  {{Ioffe}}}, \bibinfo {author} {\bibfnamefont {V.~E.}\ \bibnamefont
  {{Kravtsov}}}, \ and\ \bibinfo {author} {\bibfnamefont {E.}~\bibnamefont
  {{Cuevas}}},\ }\bibfield  {title} {\enquote {\bibinfo {title} {{Fractal
  superconductivity near localization threshold}},}\ }\href {\doibase
  10.1016/j.aop.2010.04.001} {\bibfield  {journal} {\bibinfo  {journal} {Annals
  of Physics}\ }\textbf {\bibinfo {volume} {325}},\ \bibinfo {pages} {1390}
  (\bibinfo {year} {2010})},\ \Eprint {http://arxiv.org/abs/1002.0859}
  {arXiv:1002.0859 [cond-mat.supr-con]} \BibitemShut {NoStop}%
\bibitem [{\citenamefont {{Galitski}}\ \emph {et~al.}(2005)\citenamefont
  {{Galitski}}, \citenamefont {{Refael}}, \citenamefont {{Fisher}},\ and\
  \citenamefont {{Senthil}}}]{GRFS05}%
  \BibitemOpen
  \bibfield  {author} {\bibinfo {author} {\bibfnamefont {V.~M.}\ \bibnamefont
  {{Galitski}}}, \bibinfo {author} {\bibfnamefont {G.}~\bibnamefont
  {{Refael}}}, \bibinfo {author} {\bibfnamefont {M.~P.}\ \bibnamefont
  {{Fisher}}}, \ and\ \bibinfo {author} {\bibfnamefont {T.}~\bibnamefont
  {{Senthil}}},\ }\bibfield  {title} {\enquote {\bibinfo {title} {{Vortices and
  Quasiparticles near the Superconductor-Insulator Transition in Thin
  Films}},}\ }\href {\doibase 10.1103/PhysRevLett.95.077002} {\bibfield
  {journal} {\bibinfo  {journal} {Phys. Rev. Lett.}\ }\textbf {\bibinfo
  {volume} {95}},\ \bibinfo {eid} {077002} (\bibinfo {year} {2005})},\ \Eprint
  {http://arxiv.org/abs/cond-mat/0504745} {arXiv:cond-mat/0504745
  [cond-mat.supr-con]} \BibitemShut {NoStop}%
\bibitem [{\citenamefont {{Sondhi}}\ \emph {et~al.}(1997)\citenamefont
  {{Sondhi}}, \citenamefont {{Girvin}}, \citenamefont {{Carini}},\ and\
  \citenamefont {{Shahar}}}]{SGCS97}%
  \BibitemOpen
  \bibfield  {author} {\bibinfo {author} {\bibfnamefont {S.~L.}\ \bibnamefont
  {{Sondhi}}}, \bibinfo {author} {\bibfnamefont {S.~M.}\ \bibnamefont
  {{Girvin}}}, \bibinfo {author} {\bibfnamefont {J.~P.}\ \bibnamefont
  {{Carini}}}, \ and\ \bibinfo {author} {\bibfnamefont {D.}~\bibnamefont
  {{Shahar}}},\ }\bibfield  {title} {\enquote {\bibinfo {title} {{Continuous
  quantum phase transitions}},}\ }\href {\doibase 10.1103/RevModPhys.69.315}
  {\bibfield  {journal} {\bibinfo  {journal} {Reviews of Modern Physics}\
  }\textbf {\bibinfo {volume} {69}},\ \bibinfo {pages} {315} (\bibinfo {year}
  {1997})},\ \Eprint {http://arxiv.org/abs/cond-mat/9609279}
  {arXiv:cond-mat/9609279 [cond-mat]} \BibitemShut {NoStop}%
\bibitem [{\citenamefont {{Finkel'stein}}(1987)}]{Finkelstein87}%
  \BibitemOpen
  \bibfield  {author} {\bibinfo {author} {\bibfnamefont {A.~M.}\ \bibnamefont
  {{Finkel'stein}}},\ }\href
  {http://www.jetpletters.ac.ru/ps/1235/article_18665.shtml} {\bibfield
  {journal} {\bibinfo  {journal} {JETP Lett.}\ }\textbf {\bibinfo {volume}
  {45}},\ \bibinfo {pages} {46} (\bibinfo {year} {1987})}\BibitemShut {NoStop}%
\bibitem [{\citenamefont {{Sachdev}}\ and\ \citenamefont {{Ye}}(1993)}]{SY92}%
  \BibitemOpen
  \bibfield  {author} {\bibinfo {author} {\bibfnamefont {S.}~\bibnamefont
  {{Sachdev}}}\ and\ \bibinfo {author} {\bibfnamefont {J.}~\bibnamefont
  {{Ye}}},\ }\bibfield  {title} {\enquote {\bibinfo {title} {{Gapless
  spin-fluid ground state in a random quantum Heisenberg magnet}},}\ }\href
  {\doibase 10.1103/PhysRevLett.70.3339} {\bibfield  {journal} {\bibinfo
  {journal} {Phys. Rev. Lett.}\ }\textbf {\bibinfo {volume} {70}},\ \bibinfo
  {pages} {3339} (\bibinfo {year} {1993})},\ \Eprint
  {http://arxiv.org/abs/cond-mat/9212030} {cond-mat/9212030} \BibitemShut
  {NoStop}%
\bibitem [{\citenamefont {{Kitaev}}(2015)}]{kitaev2015talk}%
  \BibitemOpen
  \bibfield  {author} {\bibinfo {author} {\bibfnamefont {A.~Y.}\ \bibnamefont
  {{Kitaev}}},\ }\bibfield  {title} {\enquote {\bibinfo {title} {{Talks at
  KITP, University of California, Santa Barbara}},}\ }\href
  {http://online.kitp.ucsb.edu/online/entangled15/} {\bibfield  {journal}
  {\bibinfo  {journal} {Entanglement in Strongly-Correlated Quantum Matter}\ }
  (\bibinfo {year} {2015})}\BibitemShut {NoStop}%
\bibitem [{\citenamefont {Sachdev}(2015)}]{SS15}%
  \BibitemOpen
  \bibfield  {author} {\bibinfo {author} {\bibfnamefont {S.}~\bibnamefont
  {Sachdev}},\ }\bibfield  {title} {\enquote {\bibinfo {title}
  {{Bekenstein-Hawking Entropy and Strange Metals}},}\ }\href {\doibase
  10.1103/PhysRevX.5.041025} {\bibfield  {journal} {\bibinfo  {journal} {Phys.
  Rev. X}\ }\textbf {\bibinfo {volume} {5}},\ \bibinfo {pages} {041025}
  (\bibinfo {year} {2015})},\ \Eprint {http://arxiv.org/abs/1506.05111}
  {arXiv:1506.05111 [hep-th]} \BibitemShut {NoStop}%
\bibitem [{\citenamefont {Gu}\ \emph {et~al.}(2020)\citenamefont {Gu},
  \citenamefont {Kitaev}, \citenamefont {Sachdev},\ and\ \citenamefont
  {Tarnopolsky}}]{GKST}%
  \BibitemOpen
  \bibfield  {author} {\bibinfo {author} {\bibfnamefont {Y.}~\bibnamefont
  {Gu}}, \bibinfo {author} {\bibfnamefont {A.}~\bibnamefont {Kitaev}}, \bibinfo
  {author} {\bibfnamefont {S.}~\bibnamefont {Sachdev}}, \ and\ \bibinfo
  {author} {\bibfnamefont {G.}~\bibnamefont {Tarnopolsky}},\ }\bibfield
  {title} {\enquote {\bibinfo {title} {{Notes on the complex Sachdev-Ye-Kitaev
  model}},}\ }\href {\doibase 10.1007/JHEP02(2020)157} {\bibfield  {journal}
  {\bibinfo  {journal} {JHEP}\ }\textbf {\bibinfo {volume} {02}},\ \bibinfo
  {pages} {157} (\bibinfo {year} {2020})},\ \Eprint
  {http://arxiv.org/abs/1910.14099} {arXiv:1910.14099 [hep-th]} \BibitemShut
  {NoStop}%
\bibitem [{\citenamefont {Kurland}\ \emph {et~al.}(2000)\citenamefont
  {Kurland}, \citenamefont {Aleiner},\ and\ \citenamefont {Altshuler}}]{KAA00}%
  \BibitemOpen
  \bibfield  {author} {\bibinfo {author} {\bibfnamefont {I.~L.}\ \bibnamefont
  {Kurland}}, \bibinfo {author} {\bibfnamefont {I.~L.}\ \bibnamefont
  {Aleiner}}, \ and\ \bibinfo {author} {\bibfnamefont {B.~L.}\ \bibnamefont
  {Altshuler}},\ }\bibfield  {title} {\enquote {\bibinfo {title} {{Mesoscopic
  magnetization fluctuations for metallic grains close to the Stoner
  instability}},}\ }\href {\doibase 10.1103/PhysRevB.62.14886} {\bibfield
  {journal} {\bibinfo  {journal} {Phys. Rev. B}\ }\textbf {\bibinfo {volume}
  {62}},\ \bibinfo {pages} {14886} (\bibinfo {year} {2000})}\BibitemShut
  {NoStop}%
\bibitem [{\citenamefont {Alhassid}(2000)}]{Alhassid00}%
  \BibitemOpen
  \bibfield  {author} {\bibinfo {author} {\bibfnamefont {Y.}~\bibnamefont
  {Alhassid}},\ }\bibfield  {title} {\enquote {\bibinfo {title} {{The
  Statistical theory of quantum dots}},}\ }\href {\doibase
  10.1103/RevModPhys.72.895} {\bibfield  {journal} {\bibinfo  {journal} {Rev.
  Mod. Phys.}\ }\textbf {\bibinfo {volume} {72}},\ \bibinfo {pages} {895}
  (\bibinfo {year} {2000})},\ \Eprint {http://arxiv.org/abs/cond-mat/0102268}
  {arXiv:cond-mat/0102268} \BibitemShut {NoStop}%
\bibitem [{\citenamefont {{Aleiner}}\ \emph {et~al.}(2002)\citenamefont
  {{Aleiner}}, \citenamefont {{Brouwer}},\ and\ \citenamefont
  {{Glazman}}}]{Aleiner02}%
  \BibitemOpen
  \bibfield  {author} {\bibinfo {author} {\bibfnamefont {I.~L.}\ \bibnamefont
  {{Aleiner}}}, \bibinfo {author} {\bibfnamefont {P.~W.}\ \bibnamefont
  {{Brouwer}}}, \ and\ \bibinfo {author} {\bibfnamefont {L.~I.}\ \bibnamefont
  {{Glazman}}},\ }\bibfield  {title} {\enquote {\bibinfo {title} {{Quantum
  effects in Coulomb blockade}},}\ }\href {\doibase
  10.1016/S0370-1573(01)00063-1} {\bibfield  {journal} {\bibinfo  {journal}
  {Physics Reports}\ }\textbf {\bibinfo {volume} {358}},\ \bibinfo {pages}
  {309} (\bibinfo {year} {2002})},\ \Eprint
  {http://arxiv.org/abs/cond-mat/0103008} {arXiv:cond-mat/0103008
  [cond-mat.mes-hall]} \BibitemShut {NoStop}%
\bibitem [{\citenamefont {{Cha}}\ \emph {et~al.}(2020)\citenamefont {{Cha}},
  \citenamefont {{Wentzell}}, \citenamefont {{Parcollet}}, \citenamefont
  {{Georges}},\ and\ \citenamefont {{Kim}}}]{Cha19}%
  \BibitemOpen
  \bibfield  {author} {\bibinfo {author} {\bibfnamefont {P.}~\bibnamefont
  {{Cha}}}, \bibinfo {author} {\bibfnamefont {N.}~\bibnamefont {{Wentzell}}},
  \bibinfo {author} {\bibfnamefont {O.}~\bibnamefont {{Parcollet}}}, \bibinfo
  {author} {\bibfnamefont {A.}~\bibnamefont {{Georges}}}, \ and\ \bibinfo
  {author} {\bibfnamefont {E.-A.}\ \bibnamefont {{Kim}}},\ }\bibfield  {title}
  {\enquote {\bibinfo {title} {{Linear resistivity and Sachdev-Ye-Kitaev (SYK)
  spin liquid behavior in a quantum critical metal with spin-$1/2$
  fermions}},}\ }\href {\doibase 10.1073/pnas.2003179117} {\bibfield  {journal}
  {\bibinfo  {journal} {Proc. Nat. Acad. Sci.}\ }\textbf {\bibinfo {volume}
  {117}},\ \bibinfo {pages} {18341} (\bibinfo {year} {2020})},\ \Eprint
  {http://arxiv.org/abs/2002.07181} {arXiv:2002.07181 [cond-mat.str-el]}
  \BibitemShut {NoStop}%
\bibitem [{\citenamefont {Joshi}\ \emph {et~al.}(2020)\citenamefont {Joshi},
  \citenamefont {Li}, \citenamefont {Tarnopolsky}, \citenamefont {Georges},\
  and\ \citenamefont {Sachdev}}]{Joshi2019}%
  \BibitemOpen
  \bibfield  {author} {\bibinfo {author} {\bibfnamefont {D.~G.}\ \bibnamefont
  {Joshi}}, \bibinfo {author} {\bibfnamefont {C.}~\bibnamefont {Li}}, \bibinfo
  {author} {\bibfnamefont {G.}~\bibnamefont {Tarnopolsky}}, \bibinfo {author}
  {\bibfnamefont {A.}~\bibnamefont {Georges}}, \ and\ \bibinfo {author}
  {\bibfnamefont {S.}~\bibnamefont {Sachdev}},\ }\bibfield  {title} {\enquote
  {\bibinfo {title} {{Deconfined critical point in a doped random quantum
  Heisenberg magnet}},}\ }\href {\doibase 10.1103/PhysRevX.10.021033}
  {\bibfield  {journal} {\bibinfo  {journal} {Phys. Rev. X}\ }\textbf {\bibinfo
  {volume} {10}},\ \bibinfo {pages} {021033} (\bibinfo {year} {2020})},\
  \Eprint {http://arxiv.org/abs/1912.08822} {arXiv:1912.08822
  [cond-mat.str-el]} \BibitemShut {NoStop}%
\bibitem [{\citenamefont {Tarnopolsky}\ \emph {et~al.}(2020)\citenamefont
  {Tarnopolsky}, \citenamefont {Li}, \citenamefont {Joshi},\ and\ \citenamefont
  {Sachdev}}]{Tarnopolsky20}%
  \BibitemOpen
  \bibfield  {author} {\bibinfo {author} {\bibfnamefont {G.}~\bibnamefont
  {Tarnopolsky}}, \bibinfo {author} {\bibfnamefont {C.}~\bibnamefont {Li}},
  \bibinfo {author} {\bibfnamefont {D.~G.}\ \bibnamefont {Joshi}}, \ and\
  \bibinfo {author} {\bibfnamefont {S.}~\bibnamefont {Sachdev}},\ }\bibfield
  {title} {\enquote {\bibinfo {title} {{Metal-insulator transition in a random
  Hubbard model}},}\ }\href {\doibase 10.1103/PhysRevB.101.205106} {\bibfield
  {journal} {\bibinfo  {journal} {Phys. Rev. B}\ }\textbf {\bibinfo {volume}
  {101}},\ \bibinfo {pages} {205106} (\bibinfo {year} {2020})},\ \Eprint
  {http://arxiv.org/abs/2002.12381} {arXiv:2002.12381 [cond-mat.str-el]}
  \BibitemShut {NoStop}%
\bibitem [{\citenamefont {{Joshi}}\ and\ \citenamefont
  {{Sachdev}}(2020)}]{Joshi2020}%
  \BibitemOpen
  \bibfield  {author} {\bibinfo {author} {\bibfnamefont {D.~G.}\ \bibnamefont
  {{Joshi}}}\ and\ \bibinfo {author} {\bibfnamefont {S.}~\bibnamefont
  {{Sachdev}}},\ }\bibfield  {title} {\enquote {\bibinfo {title} {{Anomalous
  density fluctuations in a random t -J model}},}\ }\href {\doibase
  10.1103/PhysRevB.102.165146} {\bibfield  {journal} {\bibinfo  {journal}
  {Phys. Rev. B}\ }\textbf {\bibinfo {volume} {102}},\ \bibinfo {eid} {165146}
  (\bibinfo {year} {2020})},\ \Eprint {http://arxiv.org/abs/2006.13947}
  {arXiv:2006.13947 [cond-mat.str-el]} \BibitemShut {NoStop}%
\bibitem [{\citenamefont {{Shackleton}}\ \emph {et~al.}(2020)\citenamefont
  {{Shackleton}}, \citenamefont {{Wietek}}, \citenamefont {{Georges}},\ and\
  \citenamefont {{Sachdev}}}]{Shackleton20}%
  \BibitemOpen
  \bibfield  {author} {\bibinfo {author} {\bibfnamefont {H.}~\bibnamefont
  {{Shackleton}}}, \bibinfo {author} {\bibfnamefont {A.}~\bibnamefont
  {{Wietek}}}, \bibinfo {author} {\bibfnamefont {A.}~\bibnamefont {{Georges}}},
  \ and\ \bibinfo {author} {\bibfnamefont {S.}~\bibnamefont {{Sachdev}}},\
  }\bibfield  {title} {\enquote {\bibinfo {title} {{Quantum phase transition at
  non-zero doping in a random $t$-$J$ model}},}\ }\href@noop {} {\  (\bibinfo
  {year} {2020})},\ \Eprint {http://arxiv.org/abs/2012.06589} {arXiv:2012.06589
  [cond-mat.str-el]} \BibitemShut {NoStop}%
\bibitem [{\citenamefont {Guo}\ \emph {et~al.}(2020)\citenamefont {Guo},
  \citenamefont {Gu},\ and\ \citenamefont {Sachdev}}]{GGS20}%
  \BibitemOpen
  \bibfield  {author} {\bibinfo {author} {\bibfnamefont {H.}~\bibnamefont
  {Guo}}, \bibinfo {author} {\bibfnamefont {Y.}~\bibnamefont {Gu}}, \ and\
  \bibinfo {author} {\bibfnamefont {S.}~\bibnamefont {Sachdev}},\ }\bibfield
  {title} {\enquote {\bibinfo {title} {Linear in temperature resistivity in the
  limit of zero temperature from the time reparameterization soft mode},}\
  }\href {\doibase https://doi.org/10.1016/j.aop.2020.168202} {\bibfield
  {journal} {\bibinfo  {journal} {Annals of Physics}\ }\textbf {\bibinfo
  {volume} {418}},\ \bibinfo {pages} {168202} (\bibinfo {year}
  {2020})}\BibitemShut {NoStop}%
\bibitem [{\citenamefont {Georges}\ \emph {et~al.}(1996)\citenamefont
  {Georges}, \citenamefont {Kotliar}, \citenamefont {Krauth},\ and\
  \citenamefont {Rozenberg}}]{GeorgesRMP}%
  \BibitemOpen
  \bibfield  {author} {\bibinfo {author} {\bibfnamefont {A.}~\bibnamefont
  {Georges}}, \bibinfo {author} {\bibfnamefont {G.}~\bibnamefont {Kotliar}},
  \bibinfo {author} {\bibfnamefont {W.}~\bibnamefont {Krauth}}, \ and\ \bibinfo
  {author} {\bibfnamefont {M.~J.}\ \bibnamefont {Rozenberg}},\ }\bibfield
  {title} {\enquote {\bibinfo {title} {{Dynamical mean-field theory of strongly
  correlated fermion systems and the limit of infinite dimensions}},}\ }\href
  {\doibase 10.1103/RevModPhys.68.13} {\bibfield  {journal} {\bibinfo
  {journal} {Rev. Mod. Phys.}\ }\textbf {\bibinfo {volume} {68}},\ \bibinfo
  {pages} {13} (\bibinfo {year} {1996})}\BibitemShut {NoStop}%
\bibitem [{\citenamefont {{Georges}}\ \emph {et~al.}(2000)\citenamefont
  {{Georges}}, \citenamefont {{Parcollet}},\ and\ \citenamefont
  {{Sachdev}}}]{GPS00}%
  \BibitemOpen
  \bibfield  {author} {\bibinfo {author} {\bibfnamefont {A.}~\bibnamefont
  {{Georges}}}, \bibinfo {author} {\bibfnamefont {O.}~\bibnamefont
  {{Parcollet}}}, \ and\ \bibinfo {author} {\bibfnamefont {S.}~\bibnamefont
  {{Sachdev}}},\ }\bibfield  {title} {\enquote {\bibinfo {title} {{Mean Field
  Theory of a Quantum Heisenberg Spin Glass}},}\ }\href {\doibase
  10.1103/PhysRevLett.85.840} {\bibfield  {journal} {\bibinfo  {journal} {Phys.
  Rev. Lett.}\ }\textbf {\bibinfo {volume} {85}},\ \bibinfo {pages} {840}
  (\bibinfo {year} {2000})},\ \Eprint {http://arxiv.org/abs/cond-mat/9909239}
  {arXiv:cond-mat/9909239 [cond-mat.dis-nn]} \BibitemShut {NoStop}%
\bibitem [{\citenamefont {{Georges}}\ \emph {et~al.}(2001)\citenamefont
  {{Georges}}, \citenamefont {{Parcollet}},\ and\ \citenamefont
  {{Sachdev}}}]{GPS01}%
  \BibitemOpen
  \bibfield  {author} {\bibinfo {author} {\bibfnamefont {A.}~\bibnamefont
  {{Georges}}}, \bibinfo {author} {\bibfnamefont {O.}~\bibnamefont
  {{Parcollet}}}, \ and\ \bibinfo {author} {\bibfnamefont {S.}~\bibnamefont
  {{Sachdev}}},\ }\bibfield  {title} {\enquote {\bibinfo {title} {{Quantum
  fluctuations of a nearly critical Heisenberg spin glass}},}\ }\href {\doibase
  10.1103/PhysRevB.63.134406} {\bibfield  {journal} {\bibinfo  {journal} {Phys.
  Rev. B}\ }\textbf {\bibinfo {volume} {63}},\ \bibinfo {eid} {134406}
  (\bibinfo {year} {2001})},\ \Eprint {http://arxiv.org/abs/cond-mat/0009388}
  {arXiv:cond-mat/0009388 [cond-mat.str-el]} \BibitemShut {NoStop}%
\bibitem [{\citenamefont {Kotliar}\ and\ \citenamefont
  {Ruckenstein}(1986)}]{KR86}%
  \BibitemOpen
  \bibfield  {author} {\bibinfo {author} {\bibfnamefont {G.}~\bibnamefont
  {Kotliar}}\ and\ \bibinfo {author} {\bibfnamefont {A.~E.}\ \bibnamefont
  {Ruckenstein}},\ }\bibfield  {title} {\enquote {\bibinfo {title} {{New
  Functional Integral Approach to Strongly Correlated Fermi Systems: The
  Gutzwiller Approximation as a Saddle Point}},}\ }\href {\doibase
  10.1103/PhysRevLett.57.1362} {\bibfield  {journal} {\bibinfo  {journal}
  {Phys. Rev. Lett.}\ }\textbf {\bibinfo {volume} {57}},\ \bibinfo {pages}
  {1362} (\bibinfo {year} {1986})}\BibitemShut {NoStop}%
\bibitem [{\citenamefont {{Sachdev}}(2012)}]{HouchesSS}%
  \BibitemOpen
  \bibfield  {author} {\bibinfo {author} {\bibfnamefont {S.}~\bibnamefont
  {{Sachdev}}},\ }\bibfield  {title} {\enquote {\bibinfo {title} {{Quantum
  phase transitions of antiferromagnets and the cuprate superconductors}},}\
  }\href {\doibase 10.1007/978-3-642-10449-7_1} {\bibfield  {journal} {\bibinfo
   {journal} {Lect. Notes Phys.}\ }\textbf {\bibinfo {volume} {843}},\ \bibinfo
  {pages} {1} (\bibinfo {year} {2012})},\ \Eprint
  {http://arxiv.org/abs/1002.3823} {arXiv:1002.3823 [cond-mat.str-el]}
  \BibitemShut {NoStop}%
\bibitem [{\citenamefont {Fu}\ \emph {et~al.}(2018)\citenamefont {Fu},
  \citenamefont {Gu}, \citenamefont {Sachdev},\ and\ \citenamefont
  {Tarnopolsky}}]{Fu2018}%
  \BibitemOpen
  \bibfield  {author} {\bibinfo {author} {\bibfnamefont {W.}~\bibnamefont
  {Fu}}, \bibinfo {author} {\bibfnamefont {Y.}~\bibnamefont {Gu}}, \bibinfo
  {author} {\bibfnamefont {S.}~\bibnamefont {Sachdev}}, \ and\ \bibinfo
  {author} {\bibfnamefont {G.}~\bibnamefont {Tarnopolsky}},\ }\bibfield
  {title} {\enquote {\bibinfo {title} {{$\mathbb Z_2$ fractionalized phases of
  a solvable, disordered, $t$-$J$ model}},}\ }\href {\doibase
  10.1103/PhysRevB.98.075150} {\bibfield  {journal} {\bibinfo  {journal} {Phys.
  Rev. B}\ }\textbf {\bibinfo {volume} {98}},\ \bibinfo {pages} {075150}
  (\bibinfo {year} {2018})},\ \Eprint {http://arxiv.org/abs/1804.04130}
  {arXiv:1804.04130 [cond-mat.str-el]} \BibitemShut {NoStop}%
\bibitem [{\citenamefont {Tikhanovskaya}\ \emph
  {et~al.}(2020{\natexlab{a}})\citenamefont {Tikhanovskaya}, \citenamefont
  {Guo}, \citenamefont {Sachdev},\ and\ \citenamefont
  {Tarnopolsky}}]{Tikhanovskaya:2020elb}%
  \BibitemOpen
  \bibfield  {author} {\bibinfo {author} {\bibfnamefont {M.}~\bibnamefont
  {Tikhanovskaya}}, \bibinfo {author} {\bibfnamefont {H.}~\bibnamefont {Guo}},
  \bibinfo {author} {\bibfnamefont {S.}~\bibnamefont {Sachdev}}, \ and\
  \bibinfo {author} {\bibfnamefont {G.}~\bibnamefont {Tarnopolsky}},\
  }\bibfield  {title} {\enquote {\bibinfo {title} {{Excitation spectra of
  quantum matter without quasiparticles I: Sachdev-Ye-Kitaev models}},}\
  }\href@noop {} {\  (\bibinfo {year} {2020}{\natexlab{a}})},\ \Eprint
  {http://arxiv.org/abs/2010.09742} {arXiv:2010.09742 [cond-mat.str-el]}
  \BibitemShut {NoStop}%
\bibitem [{\citenamefont {Tikhanovskaya}\ \emph
  {et~al.}(2020{\natexlab{b}})\citenamefont {Tikhanovskaya}, \citenamefont
  {Guo}, \citenamefont {Sachdev},\ and\ \citenamefont
  {Tarnopolsky}}]{Tikhanovskaya:2020zcw}%
  \BibitemOpen
  \bibfield  {author} {\bibinfo {author} {\bibfnamefont {M.}~\bibnamefont
  {Tikhanovskaya}}, \bibinfo {author} {\bibfnamefont {H.}~\bibnamefont {Guo}},
  \bibinfo {author} {\bibfnamefont {S.}~\bibnamefont {Sachdev}}, \ and\
  \bibinfo {author} {\bibfnamefont {G.}~\bibnamefont {Tarnopolsky}},\
  }\bibfield  {title} {\enquote {\bibinfo {title} {{Excitation spectra of
  quantum matter without quasiparticles II: random $t$-$J$ models}},}\
  }\href@noop {} {\  (\bibinfo {year} {2020}{\natexlab{b}})},\ \Eprint
  {http://arxiv.org/abs/2012.14449} {arXiv:2012.14449 [cond-mat.str-el]}
  \BibitemShut {NoStop}%
\bibitem [{\citenamefont {{Vojta}}\ and\ \citenamefont
  {{Fritz}}(2004)}]{VojtaFritz04}%
  \BibitemOpen
  \bibfield  {author} {\bibinfo {author} {\bibfnamefont {M.}~\bibnamefont
  {{Vojta}}}\ and\ \bibinfo {author} {\bibfnamefont {L.}~\bibnamefont
  {{Fritz}}},\ }\bibfield  {title} {\enquote {\bibinfo {title} {{Upper critical
  dimension in a quantum impurity model:Critical theory of the asymmetric
  pseudogap Kondo problem}},}\ }\href {\doibase 10.1103/PhysRevB.70.094502}
  {\bibfield  {journal} {\bibinfo  {journal} {Phys. Rev. B}\ }\textbf {\bibinfo
  {volume} {70}},\ \bibinfo {eid} {094502} (\bibinfo {year} {2004})},\ \Eprint
  {http://arxiv.org/abs/cond-mat/0309262} {arXiv:cond-mat/0309262
  [cond-mat.str-el]} \BibitemShut {NoStop}%
\bibitem [{\citenamefont {{Vojta}}\ \emph {et~al.}(2000)\citenamefont
  {{Vojta}}, \citenamefont {{Buragohain}},\ and\ \citenamefont
  {{Sachdev}}}]{vbs}%
  \BibitemOpen
  \bibfield  {author} {\bibinfo {author} {\bibfnamefont {M.}~\bibnamefont
  {{Vojta}}}, \bibinfo {author} {\bibfnamefont {C.}~\bibnamefont
  {{Buragohain}}}, \ and\ \bibinfo {author} {\bibfnamefont {S.}~\bibnamefont
  {{Sachdev}}},\ }\bibfield  {title} {\enquote {\bibinfo {title} {{Quantum
  impurity dynamics in two-dimensional antiferromagnets and
  superconductors}},}\ }\href {\doibase 10.1103/PhysRevB.61.15152} {\bibfield
  {journal} {\bibinfo  {journal} {Phys. Rev. B}\ }\textbf {\bibinfo {volume}
  {61}},\ \bibinfo {pages} {15152} (\bibinfo {year} {2000})},\ \Eprint
  {http://arxiv.org/abs/cond-mat/9912020} {arXiv:cond-mat/9912020
  [cond-mat.str-el]} \BibitemShut {NoStop}%
\bibitem [{\citenamefont {{Sachdev}}(2001)}]{SS2001}%
  \BibitemOpen
  \bibfield  {author} {\bibinfo {author} {\bibfnamefont {S.}~\bibnamefont
  {{Sachdev}}},\ }\bibfield  {title} {\enquote {\bibinfo {title} {{Static hole
  in a critical antiferromagnet: field-theoretic renormalization group}},}\
  }\href {\doibase 10.1016/S0921-4534(01)00198-8} {\bibfield  {journal}
  {\bibinfo  {journal} {Physica C Superconductivity}\ }\textbf {\bibinfo
  {volume} {357}},\ \bibinfo {pages} {78} (\bibinfo {year} {2001})},\ \Eprint
  {http://arxiv.org/abs/cond-mat/0011233} {arXiv:cond-mat/0011233
  [cond-mat.str-el]} \BibitemShut {NoStop}%
\bibitem [{\citenamefont {{Fritz}}\ and\ \citenamefont
  {{Vojta}}(2004)}]{FritzVojta04}%
  \BibitemOpen
  \bibfield  {author} {\bibinfo {author} {\bibfnamefont {L.}~\bibnamefont
  {{Fritz}}}\ and\ \bibinfo {author} {\bibfnamefont {M.}~\bibnamefont
  {{Vojta}}},\ }\bibfield  {title} {\enquote {\bibinfo {title} {{Phase
  transitions in the pseudogap Anderson and Kondo models:{\quad} Critical
  dimensions, renormalization group, and local-moment criticality}},}\ }\href
  {\doibase 10.1103/PhysRevB.70.214427} {\bibfield  {journal} {\bibinfo
  {journal} {Phys. Rev. B}\ }\textbf {\bibinfo {volume} {70}},\ \bibinfo {eid}
  {214427} (\bibinfo {year} {2004})},\ \Eprint
  {http://arxiv.org/abs/cond-mat/0408543} {arXiv:cond-mat/0408543
  [cond-mat.str-el]} \BibitemShut {NoStop}%
\bibitem [{\citenamefont {Whitsitt}\ and\ \citenamefont
  {Sachdev}(2017)}]{Whitsitt17}%
  \BibitemOpen
  \bibfield  {author} {\bibinfo {author} {\bibfnamefont {S.}~\bibnamefont
  {Whitsitt}}\ and\ \bibinfo {author} {\bibfnamefont {S.}~\bibnamefont
  {Sachdev}},\ }\bibfield  {title} {\enquote {\bibinfo {title} {{Critical
  behavior of an impurity at the boson superfluid–Mott-insulator
  transition}},}\ }\href {\doibase 10.1103/PhysRevA.96.053620} {\bibfield
  {journal} {\bibinfo  {journal} {Phys. Rev. A}\ }\textbf {\bibinfo {volume}
  {96}},\ \bibinfo {pages} {053620} (\bibinfo {year} {2017})},\ \Eprint
  {http://arxiv.org/abs/1709.04919} {arXiv:1709.04919 [cond-mat.quant-gas]}
  \BibitemShut {NoStop}%
\end{thebibliography}%

\end{document}